\documentclass{aa}  
\usepackage{graphicx}
\usepackage{txfonts}
\usepackage{CJKutf8}
\usepackage{longtable}
\usepackage{lscape} 
\usepackage{pdflscape} 
\usepackage{hyperref}
\usepackage{xcolor}
\usepackage{comment}
\usepackage{placeins}
\definecolor{burntorange}{rgb}{0.8, 0.33, 0.0}

\usepackage[nameinlink]{cleveref}
\usepackage{orcidlink}

\begin{document} 

   \title{Preferential alignment of Class 0, Class I protostellar disks in multiple systems across nine nearby molecular clouds}

   \author{{Cheng-Han Hsieh \begin{CJK}{UTF8}{bsmi}(謝承翰)\end{CJK} \orcidlink{0000-0003-2803-6358}}
          \inst{1,2}
          \and
          Aleksey Generozov \inst{2,3} \orcidlink{0000-0001-9261-0989}
          \and 
          Stella S. R. Offner \inst{2,3} \orcidlink{0000-0003-1252-9916}
          \and
          H{\'e}ctor G. Arce\inst{4} \orcidlink{0000-0001-5653-7817}
          \and 
          Jaime E. Pineda\inst{5} \orcidlink{0000-0002-3972-1978}
          \and 
          Michael M. Dunham\inst{6} \orcidlink{0000-0003-0749-9505}
          \and 
          Diego Mardones\inst{7} \orcidlink{0000-0002-5065-9175}
          \and 
          Dominique Segura-Cox\inst{8} \orcidlink{0000-0003-3172-6763}
          \and 
          Bethany Grimm\inst{2}\orcidlink{0009-0006-2010-5928}
          }

   \institute{The NASA Hubble Fellowship Program Sagan Fellow\\
              \email{chenghan.hsieh@utexas.edu}
        \and 
        Department of Astronomy, The University of Texas at Austin, 2515 Speedway, Stop C1400, Austin, Texas 78712, USA
        \and Oden Institute, The University of Texas at Austin, 201 E 24th St, Austin, Texas 78712, USA
        \and
        Department of Astronomy, Yale University, New Haven, CT 06511, USA
        \and 
        Max Planck Institute for Extraterrestrial Physics, Gießenbachstraße 1, 85748, Garching bei München, Germany
        \and Department of Physics, Middlebury College, Middlebury, VT 05753, USA
        \and Departamento de Astronomía, Universidad de Chile, Camino El Observatorio 1515, Las Condes, Chile 
        \and Department of Physics and Astronomy, University of Rochester, Rochester, NY 14609, USA
             }

   \date{Submitted 2026.6.9; accepted 2026.7.15} 
\abstract{ 
Protostellar disk orientations in multiple systems provide critical insights into the primary mechanisms that govern the formation of multiple-star systems, their subsequent dynamical evolution, and their impact on planet-forming disks. We present a disk alignment study of 512 Class 0, Class I, and flat-spectrum protostars across nine nearby molecular clouds within 500\,pc, utilizing data from the CAMPOS and VANDAM surveys. Our sample includes 74 binaries and 31 high-order multiple systems.
We find that multiple systems with projected pair separations up to 6000 au exhibit preferential disk alignment with respect to each other across all evolutionary classes, deviating significantly from the random distribution predicted by turbulent fragmentation models. This suggests that the formation of multiple systems cannot be explained by turbulent fragmentation alone. Disk alignment on scales of a few thousand au is also difficult to explain by disk fragmentation as the dominant origin. We further find that the degree of nearest-neighbor disk alignment in higher-order multiples is comparable to that in binaries. Finally, we identify a significant deficit of flat-spectrum protostellar disks in high-order multiple systems as compared to younger Class 0 and Class I phases. The decline is consistent with rapid dynamical evolution, in which most higher-order systems dissolve by the end of the Class I phase.
}


\titlerunning{Preferential alignment of Class 0/I protostellar disks in multiple systems}

   \maketitle

\section{IntroductioFn}

Most young stellar objects form in multiple systems \citep{2017ApJS..230...15M,2010ApJS..190....1R,2023ASPC..534..275O}. Both models and observations indicate that the initially high multiplicity of newborn stars decreases during the protostellar phase \citep{2010ApJS..190....1R,2022ApJ...925...39T}, implying that the primordial initial conditions of these systems are progressively erased in more evolved stages. Consequently, observations of extremely young systems are required to reliably probe their initial conditions. 

Multiple systems and proto-clusters are primarily formed from fragmentation of disks \citep{2010ApJ...708.1585K, 2016ARA&A..54..271K,2021PASJ...73L..25T}, turbulent cores \citep{2010ApJ...725.1485O,2020ApJ...897L..22S}, and filaments \citep{1992ApJ...388..392I}. If multiple systems formed via disk fragmentation, their disk angular momenta are expected to be aligned \citep{2018MNRAS.475.5618B}. In contrast, if turbulence sets the angular momentum (turbulent fragmentation), disks' angular momenta should become less correlated with wider separations. Different formation mechanisms leave traces on how angular momentum is distributed and transported within the multiple systems.

Angular momentum plays a crucial role in regulating protostellar accretion, outflows, and the formation and evolution of planet and star-forming disks \citep{2017A&A...602A..17H}. Stars inherit angular momentum from their birth environment; however, the natal gas in the parent dense core has orders of magnitude more angular momentum than the final star \citep{2007ARA&A..45..565M}. A complex set of turbulent, magnetic, and dynamical processes act to regulate angular momentum transport and determine the stellar angular momentum at birth.  Understanding exactly how the angular momentum of protostars, circumstellar disks, and dense cores evolves remains one of the biggest problems in star formation \citep{2023ASPC..534..233P}.  

The study of angular momentum in young systems usually focuses on four aspects: outflow-filament orientation, core rotation, stellar spins, and disk inclination angle studies. However, directly measuring angular momentum is difficult in the absence of full three-dimensional velocity information. For filaments and dense cores, velocity gradients are therefore commonly used as proxies for the direction of projected angular momentum \citep{1993ApJ...406..528G,2019MNRAS.490..527C,2019ApJ...886..119C}.
The discovery of rotating filaments poses interesting questions on how much angular momentum of cores or disks is inherited from the initial angular momentum of filaments \citep[e.g.,][]{2016A&A...590A...2S,2021ApJ...908...92H,2021ApJ...908...86A}. The connection between the angular momentum of small-scale disks and large-scale filaments is shown in the recent Atacama Large Millimeter Array (ALMA) CO J = 2--1 observations toward a massive infrared dark cloud G28.37+0.07, where strong orthogonal outflow-filament alignment is found \citep{2019ApJ...874..104K}. Protostellar outflows are generally perpendicular to the disk plane and have been used as a tracer for disk angular momentum. However, the CARMA-NRO outflow study of 45 outflows in Orion A shows that the distribution of outflow position angle is randomly oriented with respect to their parent filaments \citep{2020ApJ...896...11F}. Similarly, relative angles between protostellar outflows and filaments in the Perseus molecular cloud are also found to be randomly aligned \citep{2017ApJ...846...16S}. A study of rotational axes of cores in the Orion Molecular Cloud 2/3 region by \citet{2020ApJ...894L..20X} also shows no correlation with the filament direction. 

The most surprising result comes from the study of stellar spins. \citet{2017NatAs...1E..64C} used asteroseismology and found the strong alignment of stellar inclinations of 48 red giants in two open clusters. The alignment of stellar spins suggested that the global angular momentum of cluster-forming clouds was effectively transferred to each star in NGC 6791 and NGC 6819 \citep{2017NatAs...1E..64C}. Their results, however, are challenged by \citet{2018A&A...618A.109M} where no alignment of stellar spin is found using the same data. Studies of stellar spins in Pleiades and Alpha Per clusters also show no strong evidence of the alignment \citep{2010MNRAS.402.1380J,2018MNRAS.476.3245J}. Moreover, the discovery of the transiting multi-planetary system, Kepler-56, with a stellar inclination angle of 45$^\circ$ shows that the stellar spin and disk rotation can be significantly different either due to primordial or dynamical interactions \citep{2013Sci...342..331H}. 

Protostellar disks regulate the transport of angular momentum and accretion onto protostars and protoplanets. The degree of their alignment provides clues about the formation process. A recent ALMA and Hubble Space Telescope (HST) study of the protoplanetary disk in five nearby star-forming regions: Lupus, Taurus, Upper Scorpius, $\rho$ Ophiuchus, and the Orion Nebula Cluster (ONC) region has shown that disks are randomly aligned except for the Lupus III cloud \citep{2020ApJ...899...55A}. The random alignment is consistent with turbulent fragmentation models. Interestingly, the perpendicular alignment of disk position angles in Lupus III with the magnetic field directions may suggest the link between the disk-specific angular momentum vector and the large-scale collapse dynamics influenced by the magnetic fields \citep{2020ApJ...899...55A}.

While the study of protoplanetary disk alignment gives us the current angular momentum direction, we do not know whether the detected disk angular momentum vector direction is primarily set by the initial conditions of formation or by subsequent dynamic interactions. To study the primordial disk alignment in multiple systems, it is important to look at the youngest Class 0 and Class I protostars where the disk angular momentum vectors are pristine, not significantly affected by the dynamical process. We surveyed all the Class 0, Class I, and Flat-spectrum sources in 7 molecular clouds: Chamaeleon I, Chamaeleon II, Ophiuchus, Ophiuchus North, Aquila, Cor Australis, and Serpens from \citet{2015ApJS..220...11D} with ALMA (hereafter the CAMPOS survey) \citep{2024ApJ...973..138H}. For Orion A and B, we include data from the VANDAM survey \citep{2020ApJ...890..130T}, which covered most of the Class 0, Class I, Flat-spectrum protostars from the Herschel Orion Protostar Survey (HOPS) with ALMA and VLA (HOPS: \citet{2010A&A...518L.122F,2013ApJ...767...36S,2016ApJS..224....5F}). In total, our sample comprises 512 Class 0 and Class I protostellar disks, of which 282 are members of multiple systems. These 282 disks form 105 independent multiple systems, including binaries, triples, quadruples, quintuples, and octuples. This constitutes the largest systematic study to date of disk orientations (as a proxy for three-dimensional angular momentum) in Class 0 and Class I protostellar disks. 

This paper is organized as follows. In Section~\ref{sec:definitions}, we introduce the definitions adopted throughout this work. Section~\ref{sec:observation} describes the ALMA observational data used in this study. In Section~\ref{sec:analysis}, we outline the methodology for determining the three-dimensional angular differences between pairs of protostellar disks. Section~\ref{sec:Obs_Results} presents our ALMA observational results. In Section~\ref{sec:STARFORGE_discussion}, we compare the observations with existing \textsc{Starforge simulations}. We interpret and discuss our findings in the context of multiple-system formation in Section~\ref{sec:discussion}, and summarize our conclusions in Section~\ref{sec:conclusion}.

\section{Definition}
\label{sec:definitions}
Following \citet{2024ApJ...963..164R} and for consistency, we define the following:

\begin{enumerate}
    \item Source: A single protostar that has a compact or extended dust disk.
    \item Binary system: There are 2 sources within a 6000 au projected separation. 
    \item High-order multiple system: At least 3 sources, where each one is within 6000 au (projected) of at least one of the others. 
    \item Inclination angle ($i$): This is the angle between the disk angular momentum vector, which is defined to be perpendicular to the disk surface, and the line of sight. We define $0^\circ$ as a face-on disk, and $90^\circ$ as an edge-on disk. The inclination angle in this study has a range between $-90^\circ$ and $90^\circ$. 
    \item Position angle ($PA$):  This is the angle measured 
    counterclockwise in the plane of the sky, between the North Celestial Pole and the angular momentum vector.
    \item Disk angle difference ($\theta$): The angle between two protostellar disk angular momentum vectors. $\theta = \arccos(|\mathbf{n}_1 \cdot \mathbf{n}_2|)$, where $\mathbf{n}_1 = (x_1, y_1, z_1) = \langle \sin i_1 \cos \text{PA}_1, \, \sin i_1 \sin \text{PA}_1, \, \cos i_1 \rangle$. For sources with degenerate inclination angle sign, the angular momentum vector can be expressed as $\mathbf{n}_1(i_1,{PA}_1)$ or $\mathbf{n}_1'(-i_1,{PA}_1)$. We assign equal weight for the two corresponding angle difference $\theta_1 = \arccos(|\mathbf{n}_1 \cdot \mathbf{n}_2|)$ and $\theta_2 = \arccos(|\mathbf{n}'_1 \cdot \mathbf{n}_2|)$.
    Note that we cannot determine the disk rotation direction (clockwise vs. counter-clockwise), such that $\mathbf{n}_1$ and $-\mathbf{n}_1$ are indistinguishable. For that, we would need disk molecular line data, which we do not possess at this moment. Thus, the disk angle difference ranges between $0^\circ$ and $90^\circ$ (alignment to perpendicular). The true disk angular momentum angle difference should range between $0^\circ$ and $180^\circ$ (parallel to anti-parallel). We refer to the observationally derived three-dimensional disk vectors as disk orientations rather than true disk angular momentum directions.
  
\end{enumerate}

We note that binaries and higher-order multiples in our sample are identified based on projected separation. Because protostellar masses are unknown for the majority of sources, we cannot confirm whether these systems are gravitationally bound. Therefore, some of the identified systems may be chance projections rather than truly bound.

\section{Observation}
\label{sec:observation}

\begin{table*}[tbh!] 
\centering
\caption{Number of protostars in multiple systems in each molecular cloud}
\begin{tabular}{ l rrrrrrrr}
\hline \hline
 Molecular Cloud & Binary & Triple & Quadruple & Quintuples & Sextuple & Septuple & Octuple & Total\\
 \hline 
 Aquila Serpens & 26 & 9 & 4 & 20 & 0 & 0 & 0 & 59\\
 Chanmaeleon I and II & 2 & 0 & 0 & 0 & 0 & 0 & 0 & 2 \\
 Corona Australis & 6 & 3 & 0 & 0 & 0 & 0 & 8 & 17 \\
 Ophiuchus and Ophiuchus North & 12 & 6 & 8 & 0 & 0 & 0 & 0 & 26\\
 Orion A and B & 102 & 18 & 12 & 30 & 0 & 0 & 16 & 178 \\
 \hline
 Total & 148 & 36 & 24 & 50 & 0 & 0 & 24 & 282\\
\hline
\end{tabular}
\label{Table:Multiplicity_stat}
\tablefoot{The corresponding systems are shown as diagrams in  \autoref{fig:Flow_chart_Aql_Serp} to \ref{fig:Flow_chart_Orion4}. For systems with compact disks, the disk inclination and position angles of the disk orientation cannot be determined. Thus, we exclude them in this study. }
\end{table*}

We compile the list of target multiple systems from the CAMPOS survey \citep{2024ApJ...973..138H} and from the VLA Nascent Disk and Multiplicity (VANDAM) Survey for Orion \citep{2020ApJ...890..130T}. The combined CAMPOS and VANDAM survey represents a uniform high-resolution (0.1\arcsec) Atacama Large Millimeter/submillimeter Array (ALMA) survey of 512 Class 0, Class I, and flat-spectrum protostellar disks in 9 nearby molecular clouds. The two surveys cover the majority of the known protostars within 500\,pc.

CAMPOS surveyed nearly all the embedded protostars in seven nearby molecular clouds, Corona Australis, Aquila, Chamaeleon I and II, Ophiuchus North, Ophiuchus, and Serpens, at a wavelength of 1.3\,mm. The survey included all the protostars with bolometric temperature ($T_{\rm bol}$) less than 1900\,K that are detected in submillimeter or millimeter continuum from the \citet{2015ApJS..220...11D} catalog, which compiles two {\it Spitzer Space Telescope} Legacy Projects: ``From Molecular Cores to Planet-forming Disks" (c2d) survey and the ``Spitzer Gould Belt (GB) Legacy Survey" \citep{2003PASP..115..965E,2009ApJS..181..321E}. In total, CAMPOS detected 39 multiple systems (binary and higher-order multiple systems) out of 184 detected protostellar disks with projected separations less than 6000 au. The detailed data reduction and the survey design for the CAMPOS survey are described in \citet{2024ApJ...973..138H}. 

The VANDAM survey characterized the multiplicity of nearly the entire protostellar population in the Orion A and Orion B molecular clouds \citep{2020ApJ...890..130T}. The survey included all the Class 0, I and flat spectrum protostars with reliable measurements of bolometric temperature ($T_{\rm bol}$), bolometric luminosity ($L_{\rm bol}$) and a 70\,$\mu m$ detection from the {\it Herschel} Orion Protostar Survey (HOPS: \citet{2010A&A...518L.122F,2013ApJ...767...36S,2016ApJS..224....5F}). In addition, the few sources that are not part of the HOPS sample in Orion B are also included in the VANDAM survey \citep{2020ApJ...890..130T}. In Orion A and Orion B, \citet{2020ApJ...890..130T} detected 66 total multiple systems with projected separations less than 6000 au out of 328 sources. 

In total, 282 out of 512 Class 0, Class I, and flat-spectrum sources in 9 nearby clouds are in multiple systems. This includes 74 binaries, 12 triples, 6 quadruples, 10 quintuples, and 3 octuples. The octuples can be considered as small clusters. The multiplicity statistics of each molecular cloud are summarized in \autoref{Table:Multiplicity_stat}.

In this study, we use the ALMA Band 6 continuum data from the CAMPOS survey \citep{2024ApJ...973..138H} and Band 7 continuum data from the VANDAM Orion survey \citep{2020ApJ...890..130T} to determine disk orientations, including position angles and absolute values of the inclination angles. We additionally use CO $J =2-1$ and CO $J=3-2$ molecular line data from the CAMPOS and VANDAM surveys, respectively, to construct protostellar outflow integrated-intensity maps to break the inclination-angle degeneracy.

\section{Analysis}
\label{sec:analysis}

\subsection{Position Angles of protostellar disks}

The major and minor axes, as well as the position angles of the disks in the CAMPOS and VANDAM surveys, were previously reported by \citet{2024ApJ...973..138H} and \citet{2020ApJ...890..130T}, respectively. In both the CAMPOS and VANDAM Orion surveys, protostellar disk radii were derived by fitting two-dimensional Gaussian models to the continuum emission using the CASA \emph{imfit}\footnote{Another alternative approach to obtain the protostellar disk radii is to fit models directly in the visibility plane using the CASA \emph{uvmodelfit} task. Direct fitting in the visibility plane can reduce both the fitting uncertainty and the systematic bias associated with the image-plane fitting using \emph{imfit}. However, \emph{uvmodelfit} is not commonly applied to Class 0/I protostars because many of these systems contain multiple sources or complicated structures (e.g., streamers) within the same field of view, whereas the current implementation of \emph{uvmodelfit} can fit only a single component near the phase center. Support for multiple-component fitting is currently under development in CASA.} task \citep{2024ApJ...973..138H,2020ApJ...890..130T}.

\subsection{Inclination angles of protostellar disks}
\label{Sec:derive_i}

\begin{figure*}[tbh!]
    \includegraphics[width=0.99\textwidth]{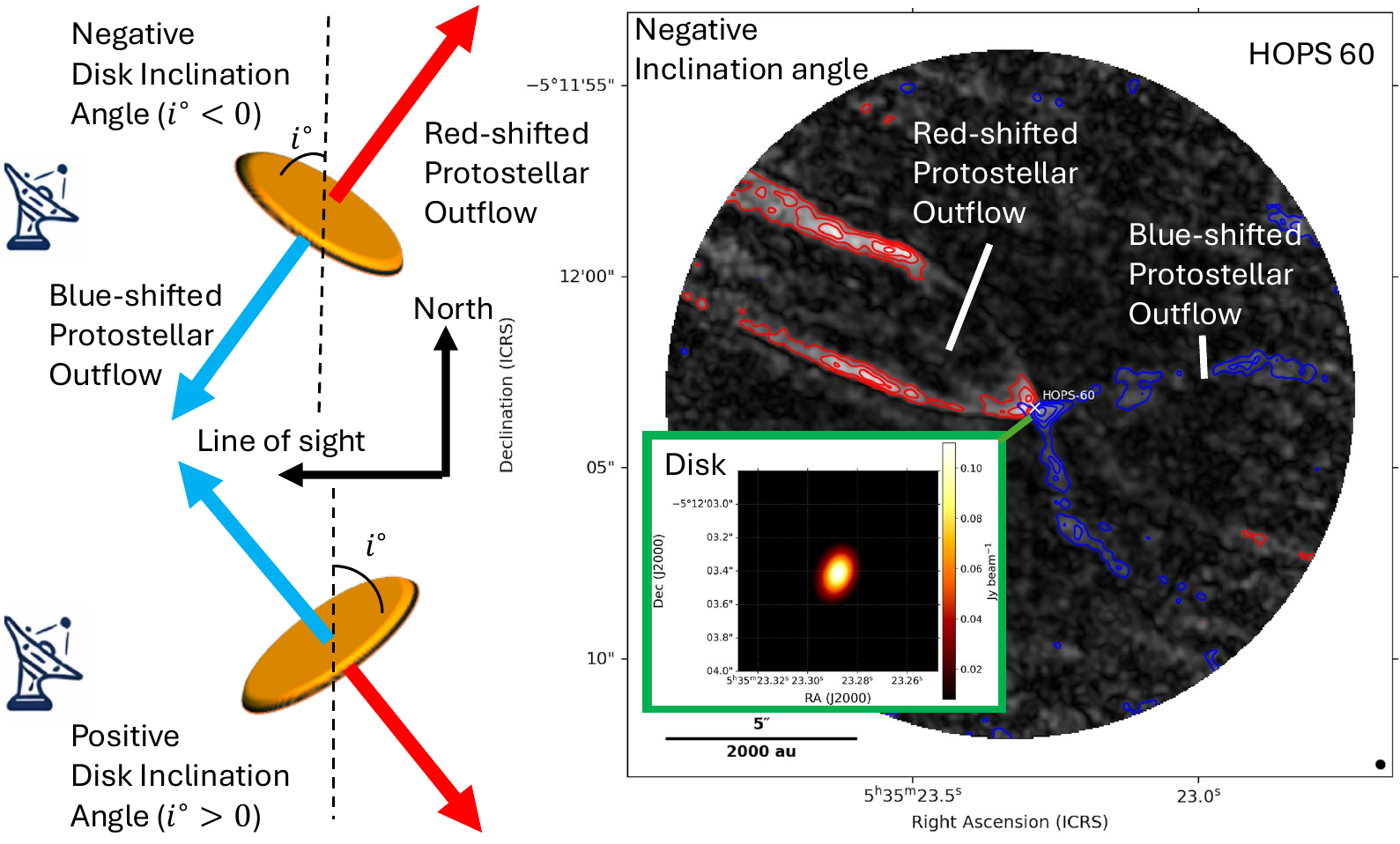}
    \caption{Diagram illustrating how the sign of the disk inclination angle is determined by protostellar outflows. The inclination is positive if the disk normal vector is pointing towards the observer (the radio telescope) with increasing declination. The right panel shows the source HOPS-60 from the VANDAM survey. The red and blue contours show the red-shifted and blue-shifted protostellar outflow lobes overlaid on the CO $J=3-2$ integrated intensity maps (grayscale). The small, green panel shows the disk dust continuum of HOPS-60 \citep{2020ApJ...890..130T}. 
   } 
\label{fig:Cartoon}
\end{figure*} 

We assume that the disks are geometrically thin and intrinsically circular, and we use the deconvolved major-to-minor axis ratio of the dust continuum emission to estimate the disk inclination angle ($i$). The inclination is given by:
\begin{equation}
        i = 90^\circ-\arcsin(R_{\rm min}/R_{\rm maj})\times 180^\circ/\pi,
        \label{eq:inc}
\end{equation}
where $R_{\rm maj}$ and $R_{\rm min}$ are the deconvolved major and minor axes of the protostellar disk. We define $i=90^\circ$ as an edge-on disk, and $i=0^\circ$ as a face-on disk. 

The thin-disk approximation introduces a bias against edge-on systems because finite disk thickness, when viewed edge-on, can be misinterpreted as a lower inclination angle. For a Class I protostellar disk with a typical dust scale height of $\sim$6 au at a radius of 100 au from the central star \citep{2023ApJ...951....9L}, the thin-disk approximation maps disks with true inclination angles $>86^\circ$ to an apparent inclination angle of $\sim86^\circ$. This effect is minimal for our study, as it introduces only a small systematic error of $\sim4^\circ$ for edge-on disks, which comprise approximately 6.7\% of the population. Class 0 disks likely have larger scale heights. However, even for a dust scale height of $\sim 10$ au at a radius of 100 au, the effect remains small, corresponding to a systematic error of only $\sim 6^\circ$ for edge-on disks. Similarly, if the beam size is greater than the scale height, the systematic error would instead be set by the beam size. The beam size affecting the derived inclination angles will be quantified in Section ~\ref{sec:analysis:CASA}.

The derived inclination angle from the disk major-to-minor axis ratio is the absolute value of the disk inclination angle \citep{2014Natur.511..567J}. We can break the sign degeneracy of the inclination angle by determining the location of the red-shifted and blue-shifted protostellar outflow lobe. As illustrated in \autoref{fig:Cartoon}, we define the protostellar disk to have a positive disk inclination angle ($i$) if the disk normal vector is pointing towards the observer (the radio telescope in \autoref{fig:Cartoon}) with increasing declination. Under this convention, we expect the protostellar outflows to be blue-shifted to the north of the disk. If the protostellar outflow is red-shifted to the north of the disk, as shown in the case for HOPS-60, the sign of the inclination angle is negative. We inspect the protostellar outflows for all sources in the VANDAM and CAMPOS surveys; out of the 512 protostars, we can reliably determine the sign of the inclination angle for 250 protostars, or around 48\%.

\subsection{Three-dimensional angle difference between disks}

For each protostellar disk, we specify the three-dimensional orientation of the protostellar disk with the inclination and position angle. The three-dimensional protostellar disk orientation derived in this study is closely related to, but not equivalent to, the true disk angular momentum unit vector. Because we lack high angular resolution molecular line observations comparable to the $\sim0.1''$ dust continuum data, we cannot determine whether the disks rotate clockwise or counterclockwise. As a result, the inferred three-dimensional disk orientation vector ($\vec{v}$) is degenerate, such that $\vec{v}$ and $-\vec{v}$ are indistinguishable. Thus, we refer to the observationally derived three-dimensional disk vectors as disk orientations rather than angular momentum directions.

The inclination angle and the position angle form an orthogonal basis. We compute the three-dimensional acute angle between all disk orientations in a molecular cloud with a dot product. For example, for a sample of $n$ disks, there are $n (n-1)/2$ distance and disk angle difference pairs. The angle difference ($\Delta \theta$) spans 0$^\circ$ to 90$^\circ$, where 0$^\circ$ indicates alignment and 90$^\circ$ indicates polar alignment. For pairs with degenerate inclinations, there are two possible angles. We include both in our analysis with equal weight.

\subsection{Identifying the multiple systems}

\begin{figure}[tbh!]
    \includegraphics[width=\columnwidth]{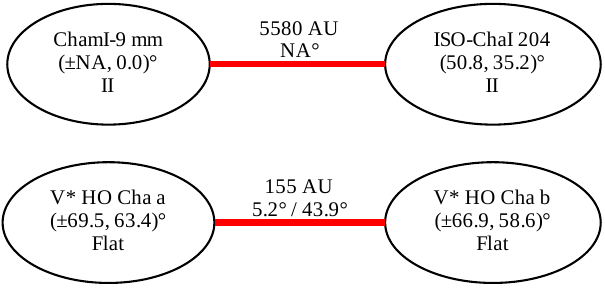}
    \caption{Protostellar disks in multiple systems within the Chamaeleon molecular cloud are represented as a network, where each disk is shown as a node labeled with the source name, inclination, position angle, and evolutionary class. An edge is drawn between two nodes if their projected separation is less than 6000 au. Each edge is labeled with the projected separation and the three-dimensional angle between the disk orientations. For each disk (node), the edge connecting to its nearest neighbor is highlighted in red. For sources with a sign degeneracy in the inclination angle, the three-dimensional angle between the disk orientations will also be degenerate. 
    } 
\label{fig:Flow_chart_Cham}
\end{figure} 

We first compute the pairwise projected distance between all protostellar disks given their RA (right ascension) and Dec (declination) within each molecular cloud. We adopt the distances of 436 pc for the Serpens and Aquila molecular clouds \citep{2018ApJ...869L..33O}, 144 pc for Ophiuchus and Ophiuchus North \citep{2019ApJ...879..125Z}, 149 pc for Corona Australis \citep{2020A&A...634A..98G}, 179 pc for Chamaeleon I \citep{2018A&A...610A..64V}, and 400 pc for Orion \citep{2020ApJ...890..130T}. 

We group all the protostellar disks in Class 0, Class I, and Flat-spectrum sources within a projected distance of 6000 au of each other into multiple systems. Dense protostellar cores with densities at least $10^4-10^6$\,cm$^{-3}$ have a typical size around 0.1 pc  (20,000 au; e.g., \citealt{2007ApJ...666..982E,2007ARA&A..45..339B,2014prpl.conf...27A}). We adopt a projected distance cutoff of $\sim1/3$ of a protostellar core to increase the likelihood that the sources are located in the same core.

To identify and visualize the multiple systems, we construct graphs for all systems in the nine molecular clouds; an example for the Chamaeleon molecular cloud is shown in \autoref{fig:Flow_chart_Cham}. We represent each protostellar disk as a node with the source name, inclination, position angle, and evolutionary class. We add an edge if the disks are within a projected distance of 6000$\,$au.  Each edge is labeled by the separation between the disks as well as the three-dimensional angle between the disk orientations. For each disk (node), we highlight the line (edge) connecting the nearest neighbor in red. If the multiple system has only 2 nodes connected by 1 edge, then it is a binary system as shown in \autoref{fig:Flow_chart_Cham}. The most complicated high-order protostellar systems are clusters in the  Orion and Corona Australis molecular clouds, which are composed of 8 Class 0 and Class I protostars. The graphs of the Class 0, Class I, flat-spectrum, and early Class II ($T_{\rm bol}\le1900$\,K) multiple systems in Chamaeleon I and Chamaeleon II are shown in \autoref{fig:Flow_chart_Cham}.  The graphs of the rest of the multiple systems are in Appendix \ref{Appendix}.

\subsection{Observational bias for deriving the inclination angle}
\label{sec:analysis:CASA}

\begin{figure*}[tbh!]
    \includegraphics[width=0.99\textwidth]{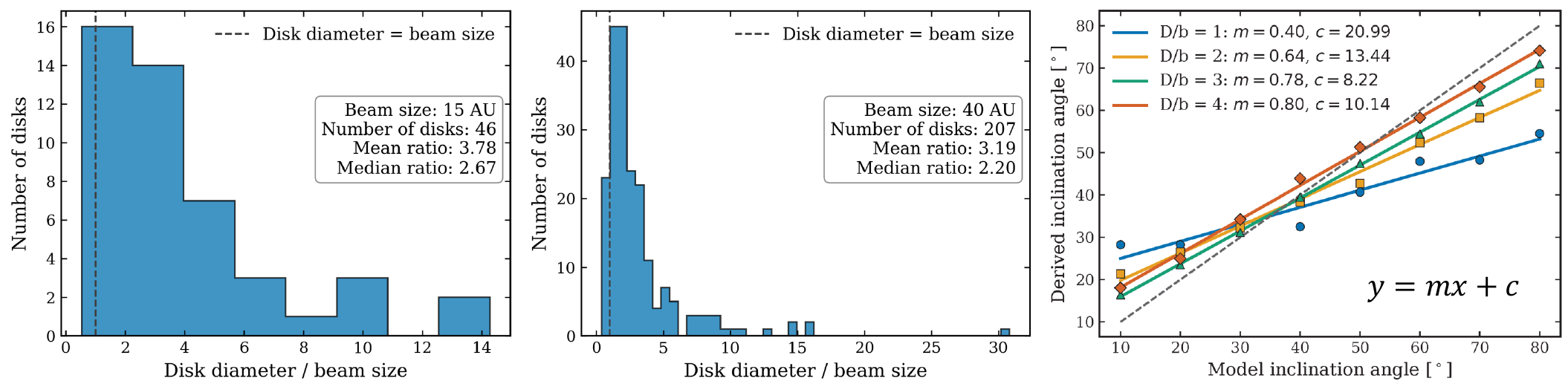}
    \caption{Systematic bias in deriving disk inclination angles. \emph{Left and Center}: Distribution of disk diameter-to-beam size ratios for this study. Compact, unresolved disks are excluded. \emph{Right}: The right panel shows the relationship between the inclination angle of the flat-disk model and the inclination derived from mock observations generated with CASA SimObserve and SimAnalyze. The inclination is estimated from the deconvolved major-to-minor axis ratio of the disk (\autoref{eq:inc}).  
    }
\label{fig:Correction}
\end{figure*} 

For both the CAMPOS and VANDAM survey, the beam size is comparable to or greater than the disk scale height for most of the sample; thus, the observational bias due to the thin-disk approximation would instead be set by the beam
size.

To quantify this observational bias, we first simulate circular, flat disks with inclination angles of $10^\circ$--$80^\circ$ and a range of disk diameters corresponding to 1--4 telescope beam sizes ($D/b$). We then generate mock observations using the same observational setup as the CAMPOS survey \citep{2024ApJ...973..138H}, including the integration time and array configurations, with the CASA tasks \emph{simobserve} and \emph{simanalyze}. Next, we fit a deconvolved two-dimensional Gaussian model to derive the protostellar disk radii, following procedures similar to those in \citet{2020ApJ...890..130T} and \citet{2024ApJ...973..138H}. Finally, we derive the corresponding inclination angles as described in Section~\ref{Sec:derive_i}.

\autoref{fig:Correction} compares the true model inclination angle with the derived inclination angle. The upper left panel shows the linear fit between the true model inclination angle ($\theta_{\text {true}}$) and the derived observed inclination angle ($\theta_{\text {obs }}$) for different disk diameter to beam size ratios ($D/b$). 

This observational bias depends only weakly on the signal-to-noise ($S/N$) ratio for well-detected disks. The VANDAM Orion disk sample typically has $S/N$ ratios between 20 and 100 \citep{2020ApJ...890..130T}, and all CAMPOS disks are detected with $S/N > 5$ \citep{2024ApJ...973..138H}. Therefore, the adopted inclination-bias correction is largely insensitive to the $S/N$ ratios of the disks in our sample.

The correction equation ($y=mx+c$) for the inclination angle is shown in the right panel of \autoref{fig:Correction}. In the left and center panels, we plot the distribution of the disk diameter-to-beam size ratio, $D/b$, for all resolved disks in this study. We find that the majority of disks have $D/b>2$. We therefore adopt the $D/b=2$ correction curve (orange squares) to the expected distribution for isotropically oriented disks to account for observational bias. The choice of $D/b=2$ provides a conservative estimate when comparing with observations with the random distribution. As shown in the following sections, we found that including this observational bias has only a minor effect on the expected $1-\cos(\theta)$ distribution.

\section{Observational Results}
\label{sec:Obs_Results}

We combine all identified multiple systems across nine molecular clouds into a single, unified dataset to achieve a statistically robust sample (see \autoref{fig:Flow_chart_Cham} and \autoref{fig:Flow_chart_Aql_Serp}–\autoref{fig:Flow_chart_Orion4}). Using this combined sample, we investigate disk alignment in both binary systems and higher-order multiple systems, examining its dependence on separation ($\leq$1000 au versus 1000--6000 au) and evolutionary stage (Class 0, Class I, and flat-spectrum sources).

\subsection{Protostellar disks in binaries and high-order multiples exhibit preferential alignment}
\label{sec:result_disk_alignment}

\begin{figure*}[tbh!]
    \includegraphics[width=.99\textwidth]{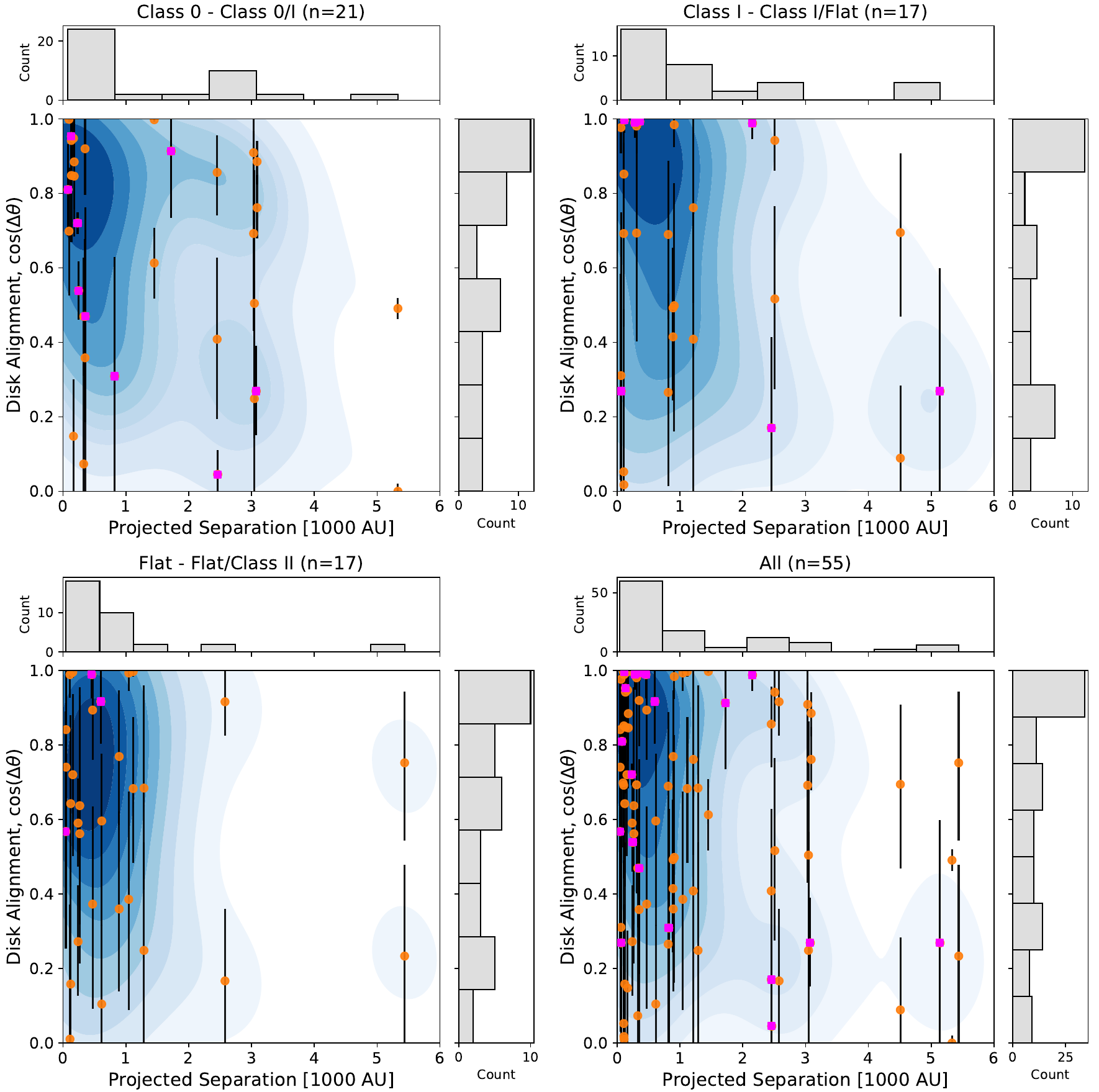}
    \caption{Evolution of the cosine of the angle ($\theta$) between two disk angular momentum vectors versus projected separation for binary systems. For randomly oriented disks, $\cos(\theta)$ is uniformly distributed between 0 and 1. Here, $\cos(\theta)=1$ corresponds to perfectly parallel disk orientations, whereas $\cos(\theta)=0$ corresponds to perpendicular orientations. Anti-parallel orientations are not distinguished because the disk rotation axis has a 180° ambiguity. Blue contours show kernel density estimates using a Gaussian kernel, weighted to account for data degeneracy. Contours mark intervals of cumulative probability, with each consecutive ring enclosing an additional 10\% of the stellar pairs moving outward from the peak density. The orange circles represent data with degenerate inclination angle signs (weight of 0.5), and magenta squares represent the non-degenerate data points (weight of 1). Each panel shows data for different protostellar classes, as indicated by the subplot titles. The last panel shows the combined sample with Class 0, Class I, Flat spectrum, and early Class II sources ($T_{\rm bol} \le 1900$\,K). } 
\label{fig:KED_binary_nearest_neighbor}
\end{figure*}

\begin{figure*}[tbh!]
    \includegraphics[width=.99\textwidth]{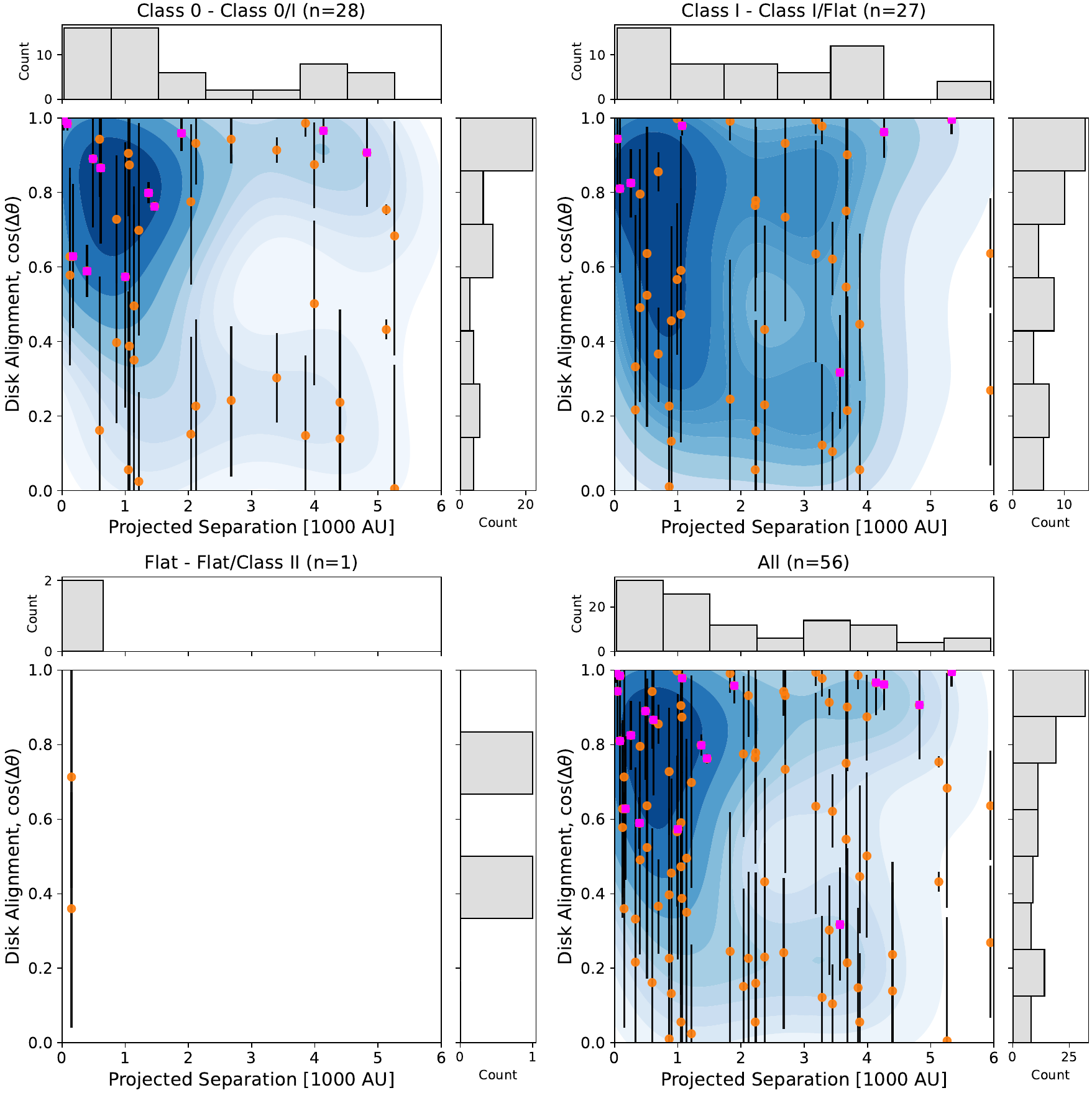}
    \caption{Evolution of the cosine of the angle ($\theta$) between two disk angular momentum vectors versus projected separation for the nearest neighbors in a high-order multiple system. Blue contours show kernel density estimates using a Gaussian kernel, weighted to account for data degeneracy. Contours mark intervals of cumulative probability, with each consecutive ring enclosing an additional 10\% of the stellar pairs moving outward from the peak density. The orange circles show data with degenerate inclination angle signs (weight of 0.5), and magenta squares show the non-degenerate data points (weight of 1). Each panel shows data for different protostellar classes, as indicated by the subplot titles. The last panel shows the combined samples in Class 0, Class I, Flat spectrum sources, and early Class II ($T_{\rm bol} \le 1900$\,K). } 
\label{fig:KED_high_order_nearest_neighbor}
\end{figure*}

\begin{figure*}[tbh!]
    \includegraphics[width=.99\textwidth]{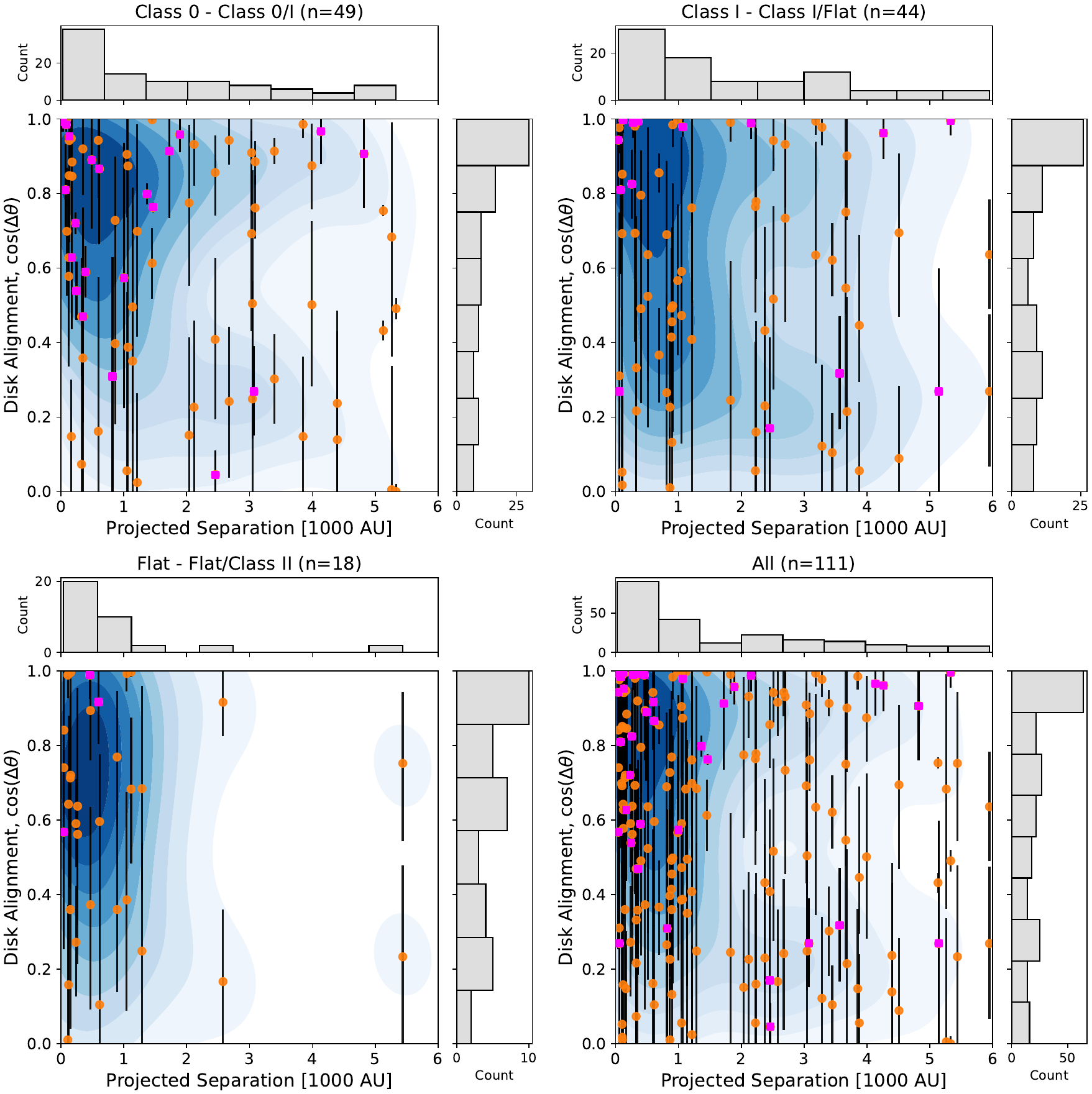}
    \caption{ Evolution of the cosine of the angle ($\theta$) between two disk angular momentum vectors versus projected separation for the combined sample of all the binaries and the nearest neighbors in high-order multiple systems. Blue contours show kernel density estimates using a Gaussian kernel, weighted to account for data degeneracy. Contours mark intervals of cumulative probability, with each consecutive ring enclosing an additional 10\% of the stellar pairs moving outward from the peak density. The scatter plots represent each data point, with orange circles representing data with the degenerate inclination angle signs (weight of 0.5), and magenta squares representing the non-degenerate data points (weight of 1). The Class of each protostellar disk pair and the number of samples are shown in the subplot title. The last panel shows the combined samples in Class 0, Class I, Flat spectrum sources, and early Class II ($T_{\rm bol} \le 1900$\,K). } 
\label{fig:KED_All}
\end{figure*}

\autoref{fig:KED_binary_nearest_neighbor}, \autoref{fig:KED_high_order_nearest_neighbor}, and \autoref{fig:KED_All} show the evolution of the cosine of the angle ($\cos\theta$) between two disk vectors versus projected separation for binaries, high-order protostellar systems, and the combined sample, respectively. For randomly oriented disks, $\cos(\theta)$ is uniformly distributed between 0 and 1. Here, $\cos(\theta)=1$ corresponds to perfectly parallel disk orientations, whereas $\cos(\theta)=0$ corresponds to perpendicular orientations. Anti-parallel orientations are not distinguished because the disk rotation axis has a 180° ambiguity. For high-order systems, we only include the `nearest-neighbor' pairs. That is, we only include each source and its closest neighbor (e.g., the red edges in \autoref{fig:Flow_chart_Aql_Serp}). The upper left panels show the youngest protostellar phase, with Class 0 - Class 0, and Class 0 - Class I pairs. They are followed by Class I - Class I, Class I - flat-spectrum pairs (upper right panels), and flat-spectrum - flat-spectrum and flat-spectrum - Class II pairs (bottom left). The last panel shows the combined sample of Class 0, Class I, flat-spectrum, and early Class II sources.

Pairs without reliably associated protostellar outflows have degenerate inclinations. In these cases, both possible angles are plotted. Uncertainties are estimated via bootstrapping, assuming Gaussian measurement errors for the inclination and position angles derived from the CASA \emph{imfit} task.\footnote{We thank Dr. John Tobin for providing the CASA \emph{imfit} uncertainties for the VANDAM survey.} We apply a weighted kernel density estimation (KDE) with a Gaussian kernel to highlight the underlying distributions. For source pairs with two degenerate angle measurements, each measurement is assigned a weight of 0.5, while source pairs with a single, well-defined three-dimensional angle difference are assigned a weight of 1.

The binaries shown in \autoref{fig:KED_binary_nearest_neighbor} exhibit a preference for disk alignment, with the cosine of the angle between the two disk angular momentum vectors clustering near unity (parallel to each other). There is a range of binary separations, but binaries in all Classes peak at $<1000$\,au. In addition, there is a deficiency of Flat/Class II binaries at wide separations ($> 1000$\,au), indicating evolution in binary separation.

The nearest neighbors in high-order multiple systems ($n>3$), shown in \autoref{fig:KED_high_order_nearest_neighbor}, also exhibit a preference for disk alignment and are clustered near 1. Their disk separations are more spread out than binaries, but still peak at $<2000$\,au. There is a dearth of Flat/Class II disks in high-order multiples, suggesting that the majority of high-order multiples disintegrate by the end of the Class I phase due to dynamical interactions.

\begin{figure*}[tbh!]
\centering
    \includegraphics[width=.83\textwidth]{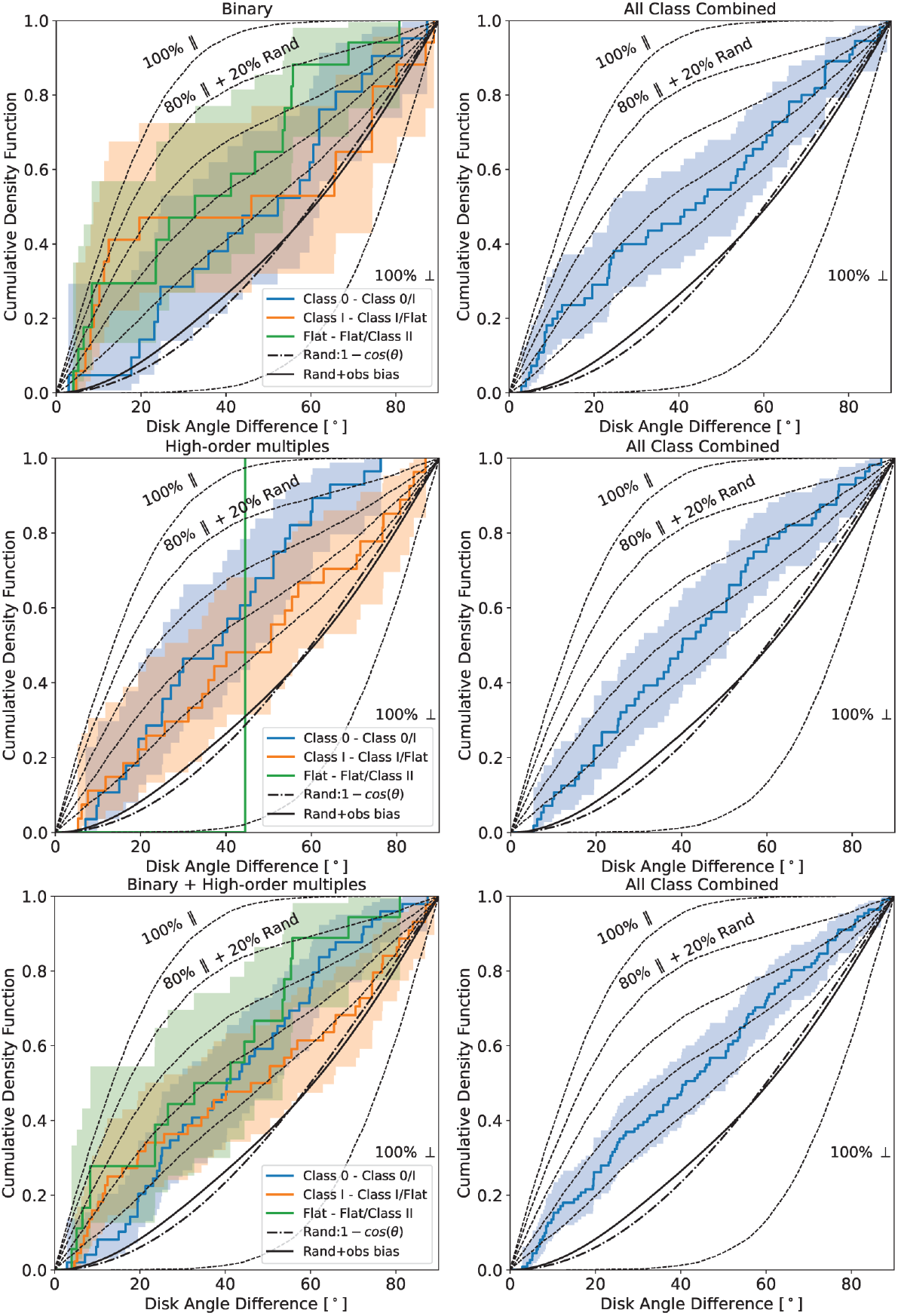}
    \caption{Cumulative distribution plot of protostellar disk angle differences in binary systems (top panel), the nearest neighbor of the higher-order multiple system (center panel), and combined binary and the higher-order multiple system (bottom panel). The cumulative plots are computed from separations within 6000$\,$au. The dashed line represents the $1-\cos(\theta)$ random distribution. The black solid line represents the random distribution including the observational bias (obs bias), and the dotted lines represent a bimodal mixture model consisting of parallel and random + obs bias components. 
    } 
\label{fig:Cumulative_all_dist}
\end{figure*}

\autoref{fig:Cumulative_all_dist} compares the cumulative distributions of disk angle differences for different classes (colored lines) with the random distribution (black lines). The former are computed using the Python package \emph{lifelines} \citep{cameron_davidson_pilon_2019_2652543}. We employ a weighted Kaplan–Meier estimator to properly account for source pairs with degenerate angle measurements. To ensure a fair comparison, we include observational bias in the random distribution via Monte Carlo simulations.  We generate one million three-dimensional random vectors, with the position angle and inclination angle sampled isotropically. We then apply the $D/b=2$ correction to the sampled inclination angles. Finally, we compute the angle between each pair of vectors to obtain the expected distribution for random orientations including observational bias (solid black line in \autoref{fig:Cumulative_all_dist}).

With the exception of the Class 0-Class 0/I binaries and the Flat–Flat/Class II nearest-neighbor sample in high-order multiple systems, disk-orientation distributions in both binaries and higher-order multiples deviate significantly from the random plus observational bias line, with $p$-values $<0.01$ according to two-sample Kolmogorov–Smirnov tests. The $p$-values, sample sizes ($n$), and median disk-orientation angle differences for all samples are summarized in \autoref{table:pvalue}.

We find that protostellar disks in both binaries and high-order multiples exhibit preferential alignment across all evolutionary classes, as shown in \autoref{fig:Cumulative_all_dist}. By comparison with a simple mixture model combining a parallel-orientation component and a random distribution, the observed data are best described by a population that is approximately 40\% three-dimensionally parallel and 60\% randomly oriented for the combined sample of all classes.

The presence of preferential alignment across all evolutionary classes, both in binaries and among nearest neighbors in higher-order multiple systems, is unexpected. At larger separations ($>1000$ au), turbulent core fragmentation is expected to dominate \citep{2022ApJ...925...39T}, which should lead to random disk orientations (see the \textsc{Starforge} simulation results discussed in the next section). In the following subsections, we describe how the degree of preferential alignment depends on separation and evolutionary stage.

\subsection{Evolution of protostellar disk alignment}
\label{sec:result_evolution}

The current sample is not large enough to establish statistically significant trends in the evolution of protostellar disk alignment. However, it reveals several intriguing patterns that warrant follow-up observations with a larger sample.

We find a possible trend suggesting that protostellar disks in binary systems become increasingly aligned over time, whereas disks in higher-order multiple systems become more misaligned as they evolve. As shown in \autoref{fig:Cumulative_all_dist}, the Flat–Flat/Class II binary population (green line) exhibits a higher degree of alignment than the younger Class 0–Class 0/I population (blue line) at a statistical p-value of 0.1. Additional observations are needed to increase the sample size and determine whether this apparent trend is statistically robust.

In contrast, for nearest-neighbor pairs in higher-order multiple systems, the Class I–Flat spectrum population (orange line) is more misaligned than the younger Class 0–Class 0/I population (blue line). However, it is important to note that this alignment evolution trend is not statistically significant, with p-values of approximately 0.4. We therefore present it only as a tentative evolutionary trend that warrants future follow-up.

An explanation for the evolutionary trends is that different sub-populations contain different fractions of sources with degenerate inclination angle signs. As shown in the following section, source pairs with unknown inclination signs are approximately randomly distributed. We therefore examine the fraction of sources with degenerate inclination angles. For binaries, these fractions are 0.43 and 0.42 for the Class 0–Class 0/I and Class I–Class I/Flat pairs, respectively. Thus, the observed evolutionary trend in disk alignment cannot be attributed to different fractions of sources with degenerate inclination angle signs. With the caveat of limited statistics, we speculate that the disk alignment in binaries may be a natural consequence of continued accretion from a common mass reservoir with a coherent angular-momentum direction. In contrast, for the nearest neighbors of higher-order multiples, the corresponding fractions are 0.68 and 0.33 for the Class 0–Class 0/I and Class I–Class I/Flat pairs, respectively. Therefore, for the nearest neighbors of higher-order multiples, the apparent misalignment may be explained by the inclusion of a larger fraction of disks with degenerate inclination angle signs.

\subsection{Comparison of nearest-neighbor protostellar disk alignment in high-order multiples and binary systems}
\label{sec:result_high_order_vs_binary}

\begin{figure*}[tbh!]
\centering
    \includegraphics[width=1\textwidth]{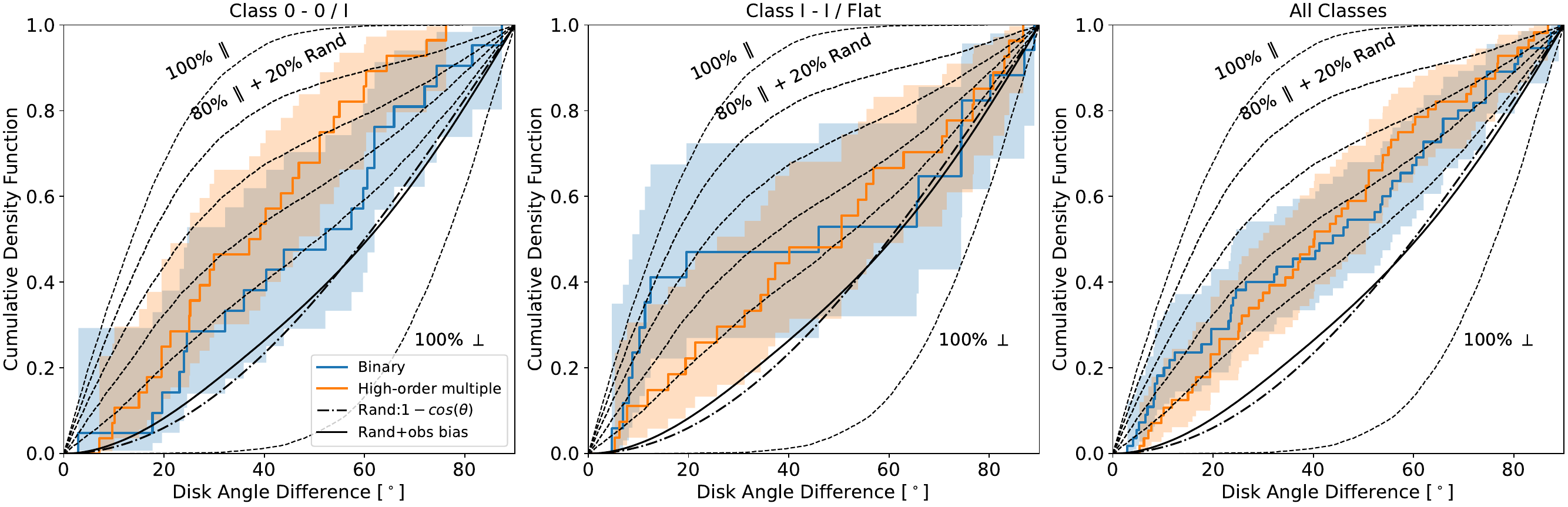}
    \caption{Cumulative distributions of protostellar disk-orientation angle differences for binary systems and nearest-neighbor pairs in higher-order multiple systems. The dashed line represents the $1-\cos(\theta)$ random distribution. The black solid line represents the random distribution including the observational bias (obs bias), and the dotted lines represent a bimodal mixture model consisting of parallel and random + obs bias components.} 
\label{fig:Binary_vs_high_order_Cumulative_dist}
\end{figure*}

We found that the degree of nearest-neighbor protostellar disk alignment in higher-order multiple systems is comparable to that in binary systems. \autoref{fig:Binary_vs_high_order_Cumulative_dist} compares the cumulative disk-alignment distributions of binary systems and higher-order multiple systems. Combining all evolutionary classes, with a sample size of 111, the cumulative distributions of disk alignments for both binary systems and nearest-neighbor higher-order multiple systems largely overlap.

Although the overall disk alignment in the nearest-neighbor of higher-order multiple systems is statistically indistinguishable from that in binary systems, the data reveal several tentative trends worthy of future investigation. Among the youngest Class 0 - Class 0/I sources, we found a tentative trend that disks in nearest-neighbor pairs within high-order multiples tend to be more aligned as compared to those in binary systems. This distinction between binary and higher-order systems is only present in the youngest Class 0–Class 0/I pairs and disappears for more evolved Class I–flat-spectrum pairs. This difference is visible in the cumulative distributions shown in \autoref{fig:Binary_vs_high_order_Cumulative_dist}. However, this difference is not statistically significant, with p-values of approximately 0.2 (see \autoref{table:pvalue}). Future follow-up observations are needed to increase the sample size and determine whether this apparent difference is statistically robust.

\subsection{Protostellar disk alignment showed little variation with increasing separation}
\label{sec:alignment_distance}

\begin{figure*}[tbh!]
\centering
    \includegraphics[width=.99\textwidth]{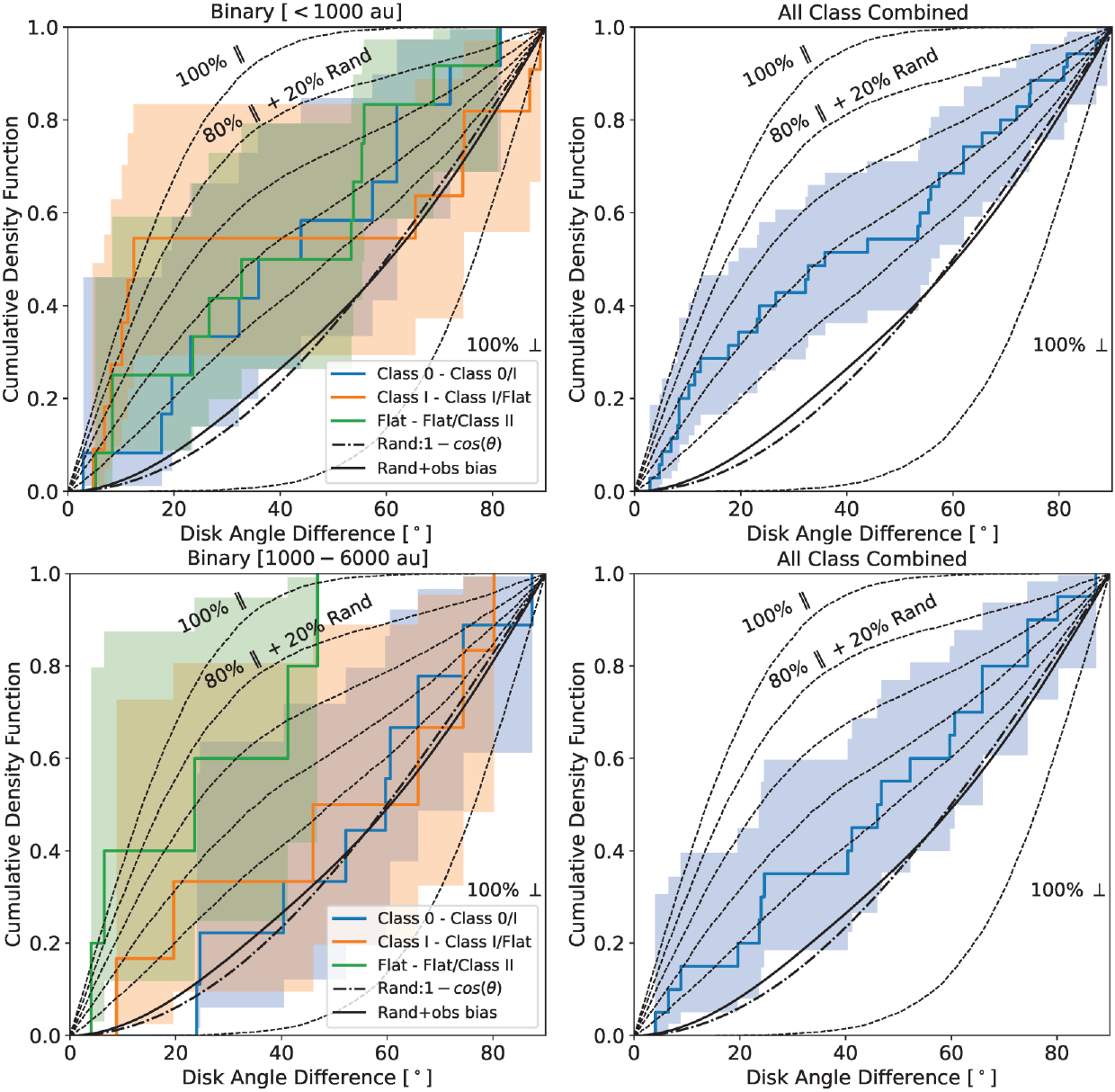}
    \caption{ Cumulative distribution plot of protostellar disk angle differences in close binary systems with separation less than 1000 au (top panel), and wide binary systems with separation between 1000--6000 au (bottom panel). The dashed line represents the $1-cos(\theta)$ random distribution. The black solid line represents the random distribution including the observational bias (obs bias), and the dotted lines represent a bimodal mixture model consisting of parallel and random + obs bias components. } 
\label{fig:Cumulative_dist_1000_Binary}
\end{figure*} 

\begin{figure*}[tbh!]
\centering
    \includegraphics[width=.99\textwidth]{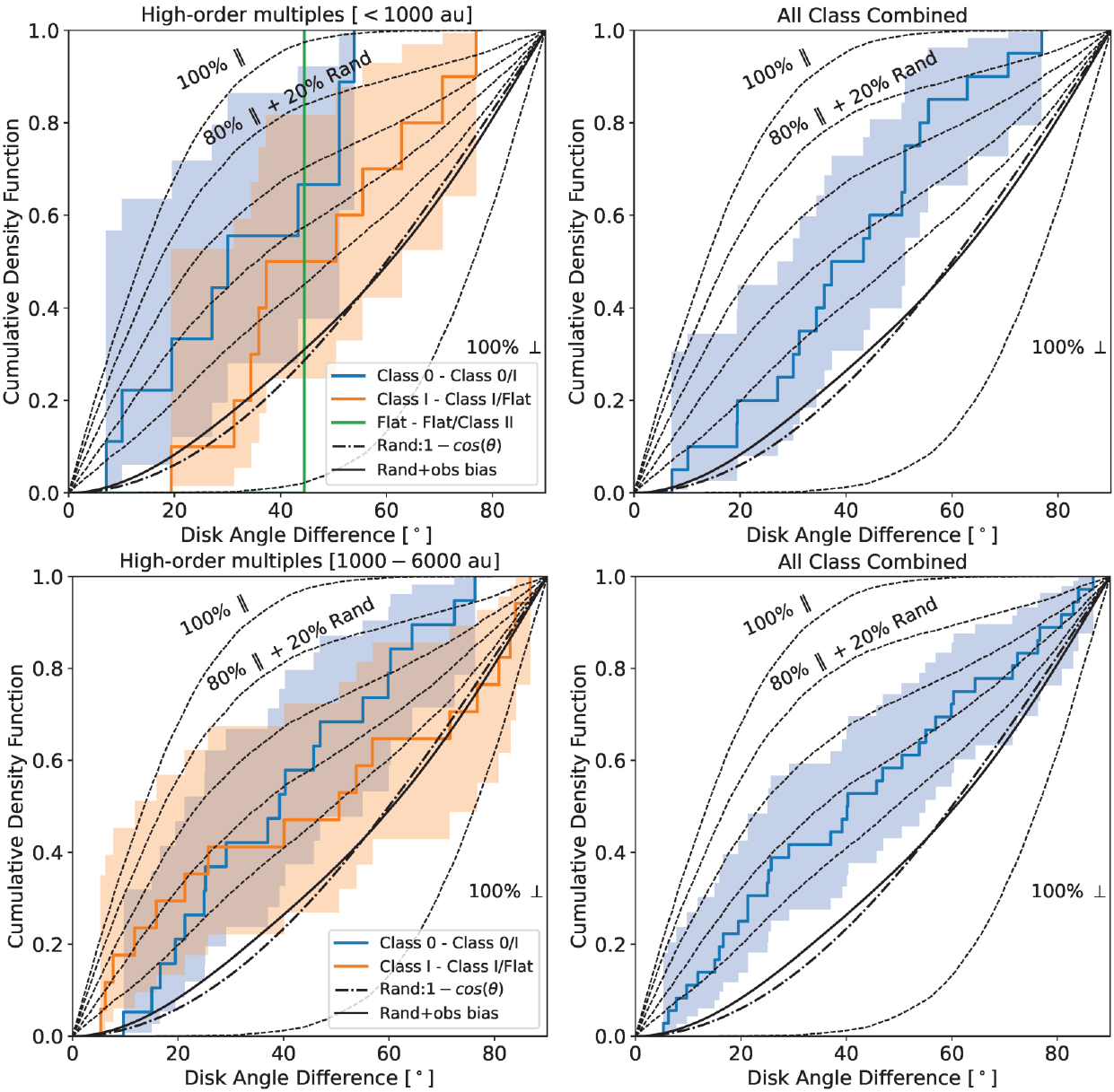}
    \caption{Cumulative distribution plot of protostellar disk angle differences in the nearest-neighbor high-order multiple system with separation less than 1000 au (top panel), and wide binary with separation between 1000--6000 au (bottom panel). The dashed line represents the $1-cos(\theta)$ random distribution. The black solid line represents the random distribution, including the observational bias (obs bias), and the dotted lines represent a bimodal mixture model consisting of parallel and random + obs bias components.} 
\label{fig:Cumulative_dist_1000_high_order}
\end{figure*}

Multiple systems or proto-clusters are primarily formed from fragmentation of disks \citep{2010ApJ...708.1585K, 2016ARA&A..54..271K,2021PASJ...73L..25T}, turbulent cores \citep{2010ApJ...725.1485O,2020ApJ...897L..22S}, and filaments \citep{1992ApJ...388..392I}. These mechanisms operate on different spatial scales: disk fragmentation typically occurs on scales of $\sim$10–100 au, turbulent core fragmentation on scales of a few thousand au, and filament fragmentation produces separations of approximately four times the filament diameter ($\sim10^4$–$10^5$ au; \citealt{1992ApJ...388..392I}). The disk and turbulent core fragmentation mechanisms can together produce two populations of multiple systems at wide ($>1000$ au) and close-in distances \citep{2016ApJ...818...73T}. Based on the observed bimodal separation peaks for Class 0 sources at $\sim$100 au and $\sim$10$^3$ au, \citet{2022ApJ...925...39T} conclude that multiples within 500 au may arise from both disk fragmentation and turbulent fragmentation followed by migration, while systems at separations $\gtrsim$1000 au are predominantly formed via turbulent fragmentation. However, simulations indicate that multiple systems formed via turbulent core fragmentation can migrate by several hundred au within $\sim10^4$ yr \citep{2010ApJ...725.1485O,2019ApJ...887..232L,guszejnov+2023,2023A&A...674A.196K}. The recent study by \citet{2023A&A...674A.196K} shows that a bimodal distribution of protostellar separation in multiple systems can be reproduced with turbulent fragmentation with migration. Thus, the distribution of companion separations alone is insufficient to determine the dominant formation pathway of multiple systems.

If multiple systems form within the same massive, fragmenting disk, their disk angular momenta are expected to be aligned \citep{2018MNRAS.475.5618B}. In contrast, turbulent fragmentation should produce randomly oriented disks within a system. In this section, we investigate disk alignment in multiple systems by separating them into close multiples ($\leq$1000 au) and wide multiples (1000--6000 au).

In contrast to the expectation of a transition from aligned to randomly oriented disks at larger separations, we find that protostellar disk alignment shows small variation with increasing separation, out to $\sim6000$ au. \autoref{fig:Cumulative_dist_1000_Binary} shows the cumulative distributions of protostellar disk-orientation angle differences for close ($\leq1000$ au) and wide (1000--6000 au) binary systems. To improve the statistical robustness, we combine all evolutionary classes, as shown in the right column of the figure. While disks in close binaries ($\leq1000$ au) are more strongly aligned than those in wide binaries (1000--6000 au), the wide-binary disk orientations are not consistent with a completely random distribution. The combined wide-binary sample deviates from the random expectation at the $2\sigma$ level.

Similarly, for the nearest neighbor in the high-order multiple system, we do not observe a significant change from alignment to random distribution (see \autoref{fig:Cumulative_dist_1000_high_order}). This is in contrast with the \textsc{Starforge} simulation (see next section), where at larger separation (1000--6000 au), the disks are randomly aligned as predicted by the turbulent fragmentation model. These observations suggest that fragmentation within a core is not entirely random, but instead retains a preferential direction, potentially influenced by magnetic fields, large-scale accretion flows, or the angular momentum of the parent core.

\subsection{High-order multiple systems are unstable, and the majority disintegrate by the end of the Class I phase}
\label{sec:res_high-order}

In \autoref{fig:KED_high_order_nearest_neighbor}, we found that there are significantly fewer nearest neighbor pairs for more evolved flat-spectrum /Class II protostellar disks (1 pair with degenerate measurement) as compared to the younger Class 0/I protostellar disks (28 pairs) in the high-order multiple system. The lack of flat-spectrum /Class II disks in the high-order multiple system indicates that high-order multiple systems are unstable, and the majority of the high-order multiple systems would disintegrate by the end of the Class I phase. Another possibility is that migration has happened and one of the companions has migrated to close separations and is not resolved ($<20-40$\,au).

\begin{table*}[tbh!] 
\centering
\caption{KS test p-values of disk angle difference distribution }
\begin{tabular}{ l r  r r r r}
\hline \hline
   & $n_1$ & $n_2$ & median 1 & median 2 & p value  \\
 \hline 
 Binary \\
 \hline
 Class 0 - 0/I vs. Random + obs bias  & 21 & -- & 52.2& -- &  $ 0.139$ \\
 Class I - I/Flat vs. Random + obs bias  & 17 & -- & 46.0&-- &   $0.008$\\
 Flat - Flat/Class II vs. Random + obs bias  & 17 & -- & 32.7& --&  $0.001$ \\
 All Class vs. Random + obs bias  & 55 & -- & 43.9 &-- & $8\times10^{-4}$ \\
 Class 0 - 0/I vs. Class I - Class I/Flat  & 21 & 17 & 52.2 & 46.0& 0.121\\
 Class I - I/Flat vs. Flat - Flat/Class II & 17 & 17 & 46.0 & 32.7 & 0.245\\
 Class 0 - 0/I vs. Flat - Flat/Class II  & 21& 17& 52.2 & 32.7 & 0.139\\
\hline
\multicolumn{6}{p{0.95\linewidth}} {High-order multiple system (nearest neighbor, NN)}\\
 \hline
 Class 0 - 0/I vs. Random + obs bias  & 28&  --& 38.1 & --&  $ 2\times10^{-4}$ \\
 Class I - I/Flat vs. Random + obs bias  & 27 & -- & 50.5&-- &  $0.137$ \\
 Flat - Flat/Class II vs. Random + obs bias  & 1 &--  & NA &-- & NA \\
 All Class vs. Random + obs bias  & 56 & -- & 40.2 & --& $< 1\times10^{-4}$ \\
 Class 0 - 0/I vs. Class I - I/Flat  & 28 & 27 & 38.1 & 50.5& 0.406\\
 Class I - I/Flat vs. Flat - Flat/Class II & 27& 1 & 50.5 & NA & NA\\
 Class 0 - 0/I vs. Flat - Flat/Class II  & 28 & 1 & 38.1 & NA & NA\\
\hline
\multicolumn{6}{p{0.95\linewidth}} {Binary + High-order multiple system (nearest neighbor)}\\
 \hline
 Class 0 - 0/I vs. Random + obs bias  & 49 & -- & 40.4 & --& $< 1\times10^{-4}$ \\
 Class I - I/Flat vs. Random + obs bias  & 44 & -- & 48.2 & --&  $0.007$ \\
 Flat - Flat/Class II vs. Random + obs bias & 18 & -- & 37.0 & -- & $ 7\times10^{-4}$ \\
 All Class vs. Random+ obs bias  & 111 & -- & 41.2 & --&  $< 1\times10^{-4}$ \\
 Class 0 - 0/I vs. Class I - I/Flat  & 49 & 44 & 40.4 & 48.2 &  0.165 \\
 Class I - I/Flat vs. Flat - Flat/Class II & 44 &  18 & 48.2 & 37.0 & 0.165 \\
 Class 0 - 0/I vs. Flat - Flat/Class II  & 49 & 18 & 40.4 & 37.0 & 0.377 \\

 \hline 
 \multicolumn{6}{p{0.95\linewidth}} {Binary vs. High-order multiple system (nearest neighbor)}\\
  \hline
 Class 0 - 0/I  & 21 &  28& 52.2& 38.1& 0.202 \\
 Class I - I/Flat  &17 & 27 & 46.0& 50.5& 0.378 \\
 All Class & 55 & 56 & 43.9& 39.7 & 0.799 \\
 \hline
\multicolumn{6}{p{0.95\linewidth}} {$<1000$ au vs. 1000--6000 au }\\
\hline
Binary Class 0 - 0/I  & 12 & 9 & 39.9 & 59.7 & 0.547\\
Binary Class I - I/Flat  & 11 & 6 & 12.4 & 56.0 &  0.525\\
Binary Flat - Flat/Class II & 12 & 5 &  43.0 & 23.6& 0.267\\
Binary All Class & 35 & 20 & 43.9 & 46.4 & 0.621\\
high-order multiple system (NN) Class 0 - 0/I & 9&  19 &30.0 &39.2 & 0.449\\
high-order multiple system (NN) Class I - I/Flat & 10 & 17 & 43.9 & 50.5& 0.461\\
high-order multiple system (NN) Flat - Flat/Class II & 1 & 0 & NA & NA & NA\\
high-order multiple system (NN) All Class & 20 &  36 &  40.3 &  40.2 & 0.678 \\
Binary + high-order multiple system (NN) Class 0 - 0/I & 21&  28 & 35.9 & 43.0& 0.848\\
Binary + high-order multiple system (NN) Class I - I/Flat & 21 & 23 & 37.3 & 50.5& 0.964\\
Binary + high-order multiple system (NN) Flat - Flat/Class II & 13 & 5 & 44.5 & 23.6 & 0.319 \\
Binary + high-order multiple system (NN) All Class & 55 & 56 & 37.3& 43.4 & 0.914 \\

\hline
\hline
\end{tabular}
\label{table:pvalue}
\end{table*}

\subsection{Effects of inclination degeneracy on disk alignment}
\label{sec:result_disk_alignment}

\begin{figure*}[tbh!]
\centering
    \includegraphics[width=1\textwidth]{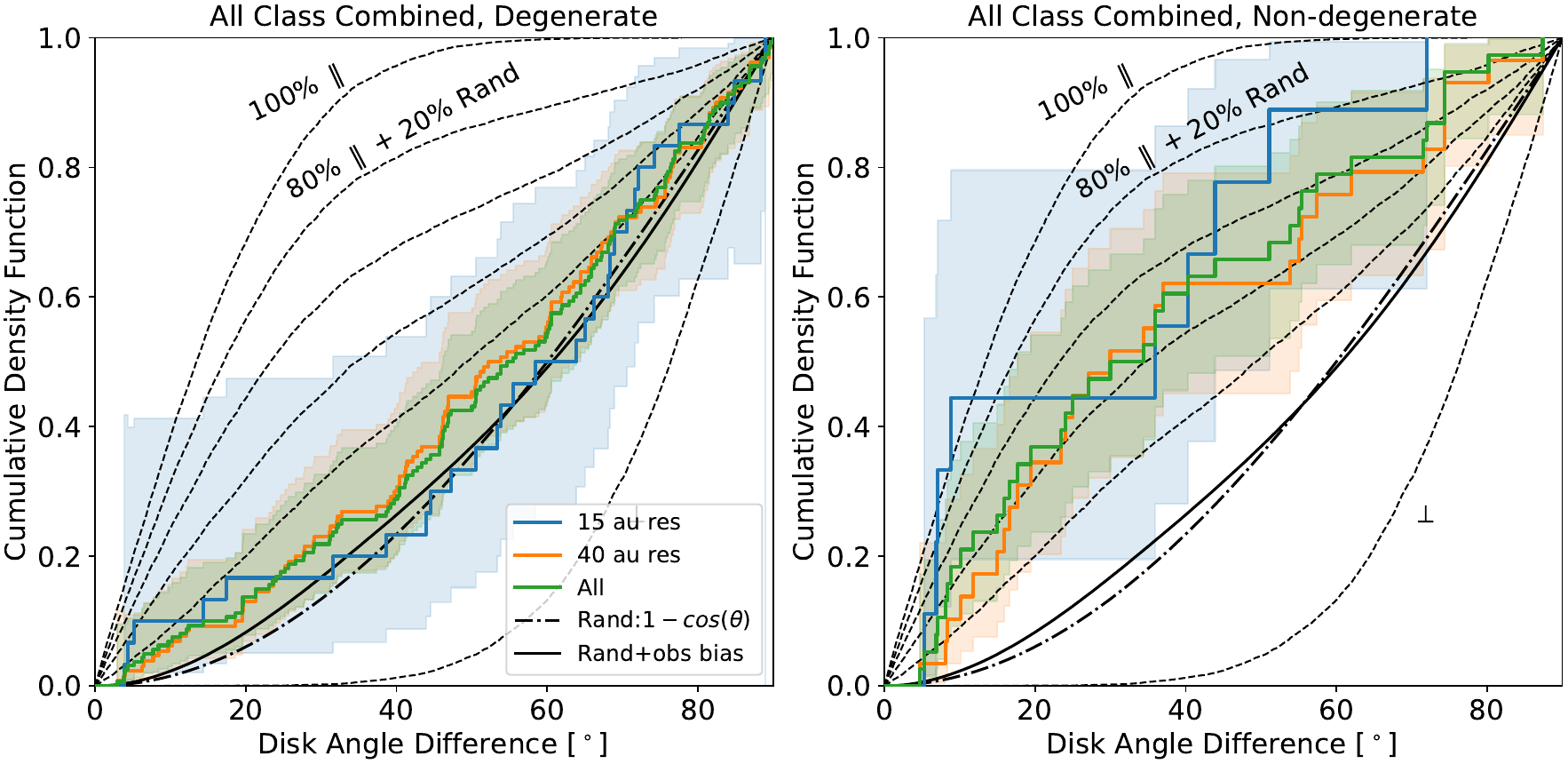}
    \caption{Cumulative distribution of protostellar disk-orientation angle differences for the combined binary and higher-order cluster sample across all evolutionary classes. \emph{Left:} For source pairs with a degenerate inclination angle ($\pm i^{\circ}$). \emph{Right:} For source pairs with a non-degenerate inclination angle ($i^{\circ}$). The blue curve represents sources observed at $\sim$15 au linear resolution (Chamaeleon I and Chamaeleon II, Corona Australis, Ophiuchus, and Ophiuchus North). The orange curve corresponds to sources observed at $\sim$40 au linear resolution (Orion A, Orion B, Serpens, and Aquila). The green curve shows the combined sample. The dashed line represents the $1-\cos(\theta)$ random distribution. The black solid line represents the random distribution, including the observational bias (obs bias), and the dotted lines represent a bimodal mixture model consisting of parallel and random + obs bias components.} 
\label{fig:cdf_degenerate_inclination}
\end{figure*}

\begin{figure*}[tbh!]
\centering
    \includegraphics[width=1\textwidth]{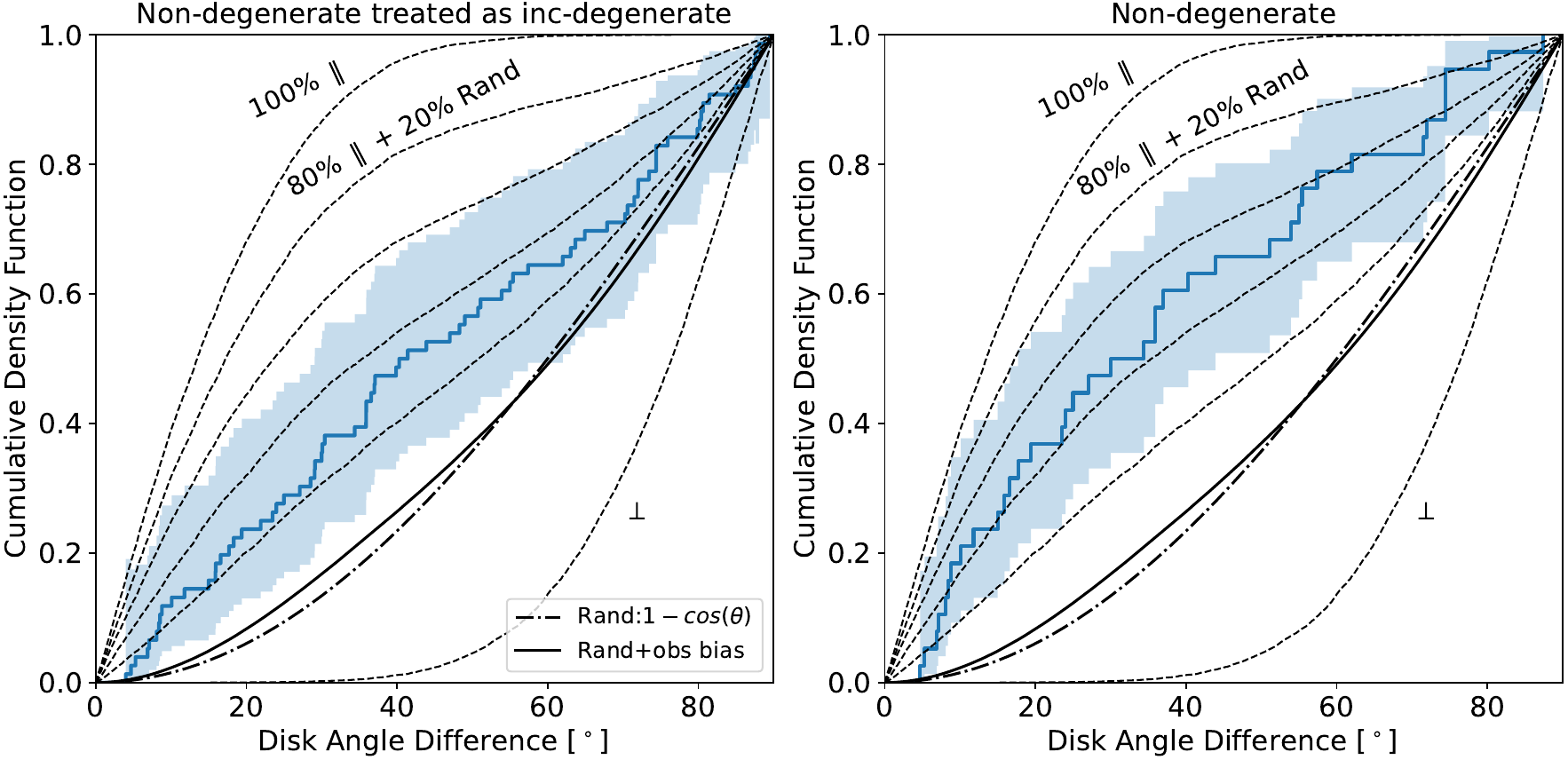}
    \caption{Cumulative distribution of protostellar disk-orientation angle differences for the combined binary and higher-order cluster sample across all evolutionary classes. \emph{Left:} For source pairs with a non-degenerate inclination angle ($i^{\circ}$), we removed the sign information and treated it as a degenerate sample. \emph{Right:} For same source pairs with a non-degenerate inclination angle ($i^{\circ}$). The dashed line represents the $1-\cos(\theta)$ random distribution. The black solid line represents the random distribution, including the observational bias (obs bias), and the dotted lines represent a bimodal mixture model consisting of parallel and random + obs bias components.}
\label{fig:degenerate_inclination_test}
\end{figure*} 

We find that the sign of the protostellar disk inclination angle plays a crucial role in recovering disk alignment in multiple systems. We inspected the protostellar outflow data for all 512 protostars and obtained reliable inclination-angle sign determinations for 250 sources, corresponding to approximately 48\% of the sample. \autoref{fig:Cartoon} illustrates the schematic used to determine the sign of the protostellar disk inclination from the associated protostellar outflows. For source pairs with degenerate inclination-angle signs, we assign equal weights to both possible solutions in the kernel density estimation plots, cumulative distribution functions, and Kolmogorov–Smirnov (KS) tests presented above.

To assess the impact of inclination-angle degeneracy on the inferred disk alignment, we divide the source pairs into those with non-degenerate inclination signs and those with degenerate signs. The resulting cumulative distributions of protostellar disk-orientation angle differences, for the combined binary and higher-order cluster sample across all evolutionary classes, are shown in \autoref{fig:cdf_degenerate_inclination}. For the non-degenerate sample, where the inclination-angle sign is known for both members of the source pair, the distribution shows strong alignment and is well described by a simple binomial mixture model consisting of 50\% aligned and 50\% randomly oriented disks (green line). A one-sample KS test yields a $p$-value of $3\times10^{-5}$, indicating a highly significant deviation from a random distribution.

In contrast, the distribution for source pairs with degenerate inclination-angle signs is statistically indistinguishable from random, with a KS $p$-value of $\sim$0.24. The striking difference between the degenerate and non-degenerate samples highlights that properly accounting for inclination-angle sign degeneracy is essential. When this degeneracy is resolved, the intrinsic degree of protostellar disk alignment is substantially higher (green line), approaching $\sim$50\% in a simple binomial mixture model of parallel and random orientations.  

However, the subsample with and without resolved inclination angle sign degeneracies may differ in other systematic ways that contribute to the contrast. For example, misaligned disks in multiple systems are expected to drive misaligned outflows, leading to more complex interactions between the outflows and the surrounding core. Consequently, it becomes more difficult to distinguish individual protostellar outflows from the ambient cloud emission or from one another, especially given the relatively shallow CO observations in both the VANDAM and CAMPOS surveys. In short, misaligned sources may be more likely to fall into the degenerate sample, due to the difficulty of extracting outflows.

Separately from the intrinsic alignment of each sample, the contrast also depends on the joint distribution of angle differences and inclinations, which determines the difference between two degenerate inclinations. For two intrinsically aligned disks with inclinations of $45^\circ$, the inclination-sign degeneracy can cause the inferred angle between their angular momentum vectors to be either $0^\circ$ or $90^\circ$. In contrast, for disks with inclination angles of $0^\circ$ or $90^\circ$, the degeneracy would not affect the measured alignment.

To study the effect of inclination angle sign degeneracy, we take the sign-determined systems and repeat the same analysis after intentionally removing sign information. This allows us to directly isolate the impact of sign determination within the same set of systems. In \autoref{fig:degenerate_inclination_test}, we find that intentionally removing the inclination-sign information shifts the distribution closer to the random $1-\cos(\theta)$ expectation. However, it remains less isotropic than the degenerate sample shown in the left panel of \autoref{fig:cdf_degenerate_inclination}.

\subsection{Effects of the linear resolution on disk alignment}
\label{sec:result_disk_alignment_linear_res}

Although all protostellar disks are observed at a uniform angular resolution of 0.1\arcsec, differences in cloud distances result in varying linear resolutions across the sample. To assess the impact of linear resolution on disk alignment, we divide the sample into two sub-samples: sources observed at $\sim$15 au linear resolution (Chamaeleon I and II, Corona Australis, Ophiuchus, and Ophiuchus North), shown in blue, and sources observed at $\sim$40 au linear resolution (Orion A and B, Serpens, and Aquila), shown in orange in \autoref{fig:cdf_degenerate_inclination}.

For both the inclination-angle degenerate and non-degenerate cases, the distributions of disk-orientation angle differences for the $\sim$15 au and $\sim$40 au samples are statistically consistent. This result indicates that the lower linear-resolution sub-sample ($\sim$40 au) still reliably recovers disk orientations using the deconvolved radii derived from the CASA \emph{imfit}, comparable to the higher-resolution ($\sim$15 au) sub-sample.

\section{STARFORGE Simulation Results}
\label{sec:STARFORGE_discussion}
We will now compare the observations to detailed simulations from the {\sc starforge} project \citep{GrudicGuszejnov2022a,GuszejnovGrudic2022a}. {\sc starforge} is a state-of-the-art framework built around the {\sc gizmo}  Lagrangian meshless finite-mass MHD code \citep{HopkinsRaives2016a,GrudicGuszejnov2021a} to model star formation physics from giant molecular clouds to stars.

{\sc{starforge}}, like previous Lagrangian 3D star formation simulations, integrates the MHD equations to follow the evolution of discrete mass elements \citep{klessen&burkert2000,bate+2003}. For computational efficiency, sink particles are placed in gravitationally collapsing regions that exceed the Jeans criterion \cite{bate+1995}, i.e., $\rho \sim 3 \times 10^{-14}$~g~cm$^{-3}$.\footnote{See \cite{grudic+2021Starforge} for the full list of sink insertion criteria.} The sink particles correspond to forming stars and interact with surrounding gas through gravity, accretion, and feedback, as determined by a sub-grid model for (proto-)star evolution \citep{offner+2009feedback,grudic+2021Starforge}. \textsc{starforge} includes all the most important feedback mechanisms: multi-band radiation, stellar winds, protostellar outflows, and supernovae. These simulations are among the first that incorporate all of this physics while predicting the IMF self-consistently, rather than using a sub-grid prescription for the IMF. They are also the first to be able to do this in massive GMCs, in which most stars actually form. 

Stellar dynamics is handled with a high-order integrator, which is essential for following binaries over many orbits. Gravitational interactions between sinks within $\sim$20 au are softened. 

We focus on clouds with properties typical of a Milky Way GMC, adopting the \emph{M2e4\_mu1.3} simulations from \citet{guszejnov+2022a} as our fiducial simulation (see their Table 1). The initial mass and radius are 20,000 M$_\odot$ and 10 pc, respectively, matching the mean surface density of GMCs in the Solar neighborhood (e.g., \citealp{lada&dame2020}). The virial parameter (the ratio of twice the kinetic to gravitational energy) is $\alpha_{\rm vir}=2$, comparable to that of massive ($\gtrsim 10^4 M_{\odot}$) Milky Way GMCs \citep{larson1981,chevance+2023}. The cloud is initially threaded by a uniform magnetic field with $B_z \approx 6~\mu$G, corresponding to a mass-to-flux ratio, $\mu$, of 1.3. This field is similar to that found by Zeeman surveys of molecular clouds \citep{crutcher+2010}. We analyze three realizations of this cloud with different initial turbulent seeds. Each seed corresponds to a separate simulation with a different random realization of the initial turbulent velocity field. 

To test the robustness of our results, we also include clouds with $B_z=2~\mu$G and $20~\mu$G ($\mu=4.2$ and 0.42) in our analysis (with three seeds per magnetic field). For $\mu=0.42$, we also include simulations with $\alpha_{\rm vir}=1$ and 4 (with two seeds per virial parameter). In total, we use 13 full physics simulations.

We analyze the alignment and separations of close stellar pairs in {\sc starforge} data, closely following the observational analysis of the previous sections. First, we identify close sets of stars with projected separations $<6000$ au, taking the z-axis to be the line of sight. The {\sc starforge} simulations do not contain disks. Instead, we use two different proxies for disk orientation: (i) the spin vector of the stars, (ii) the change in the spin vector between consecutive snapshots (excluding cases where the spin has not changed). The spin vector depends on the total angular momentum accreted. As the angular momentum of infalling gas can change over time, the total spin may not accurately reflect what the disk orientation is, especially in cases where the disk is truncated by dynamical interactions, as discussed in \citep{bate+2010}. The change in spin reflects recent accretion and may be a better proxy. For a pair of nearest neighbors within 6000 au, we calculate the acute angle between the stellar spins. We only include pairs in associations with five stars or less, as larger associations are rare in the observational data. Finally, we mimic the observational inclination bias and modify the inclination of each spin according to the linear fit in the upper left panel of \autoref{fig:Correction} (assuming $D/b=2$). We don't include any inclination degeneracy in our modeling (see \autoref{fig:Cartoon}), and only compare to the non-degenerate observed sample.

Finally, to compare to observations, we must assign a Class to stars in the simulations. We use \autoref{tab:classConv} to convert between stellar age and Class \citep{dunham+2015, kristensen&dunham2018}. These timescales are consistent with the main accretion phase and protostellar core phase measured for low-mass ($< 2 M_{\odot}$) stars in \textsc{Starforge} \citep{offner+2025, kaalva+2026}. The age of the star is measured from the time it reaches 0.07 $M_{\odot}$, and we exclude any lower-mass sink particles from the analysis. We also exclude Class II and older sources from our analysis, as older sources are rare in the observations.

For each simulation, we have hundreds of snapshots corresponding to different times.
We stack the simulation snapshots until the end of each simulation (between $\sim5$ and 16 Myr, depending on simulation parameters). In other words, we analyze each simulation snapshot independently and then combine the results. Snapshots are spaced by $\sim 25$ kyr. We take every fourth snapshot to avoid repeated pairs within a Class.

\begin{table}[h!]
    \centering
    \caption{Mapping of stellar age to protostar class in {\sc starforge}  \label{tab:classConv}}
    \begin{tabular}{l|c|c}
    Class & Start time [Myr] & End time [Myr] \\
    \hline
    0     & 0 & 0.1\\
    I     & 0.1 & 0.2 \\
    Flat    & 0.2 & 0.3 \\
    II  & 0.3 & 2 \\
    \hline
    \end{tabular}
\end{table}

\begin{figure*}[h!]
\includegraphics[width=\textwidth]{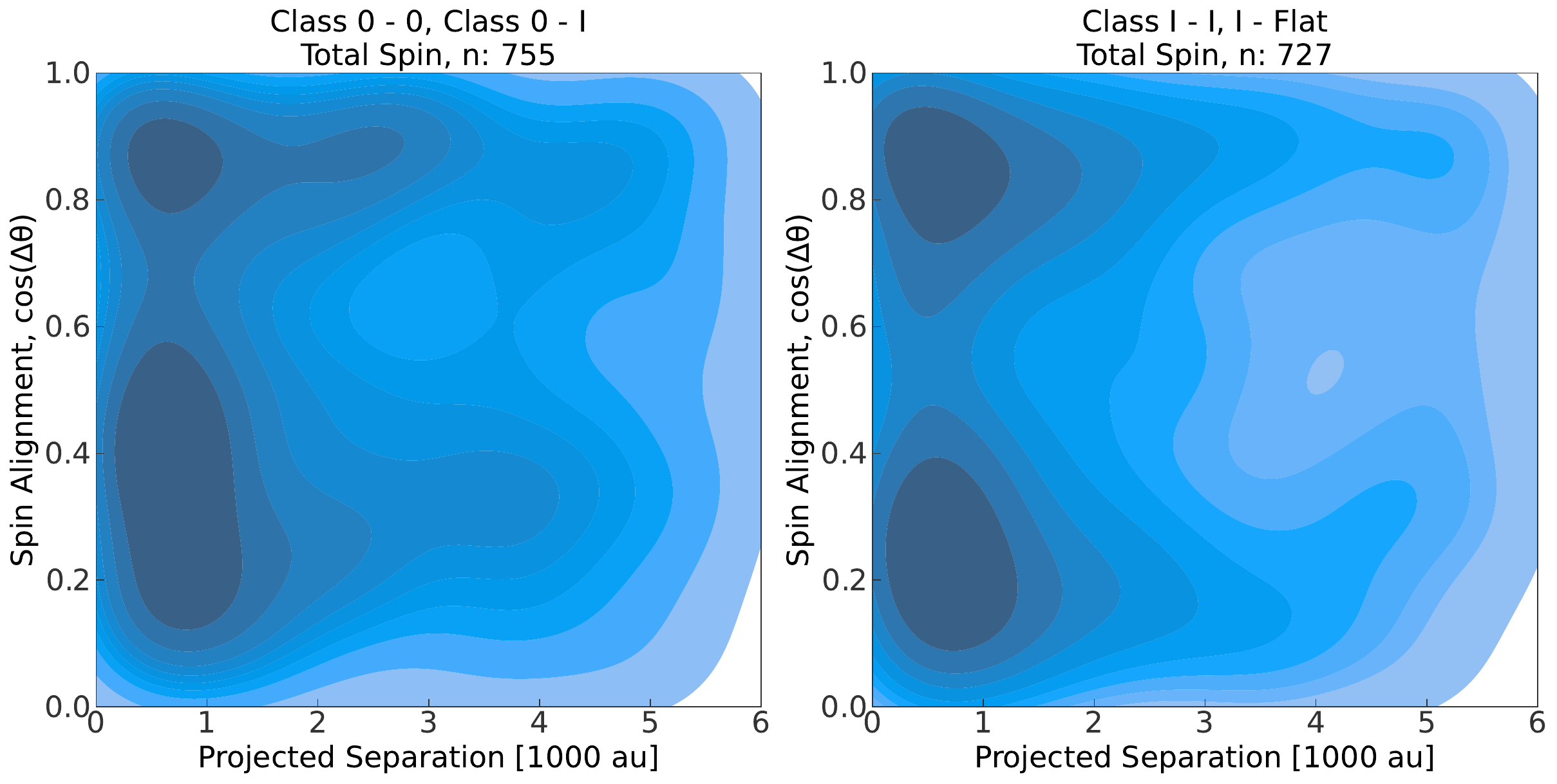}
\includegraphics[width=\textwidth]{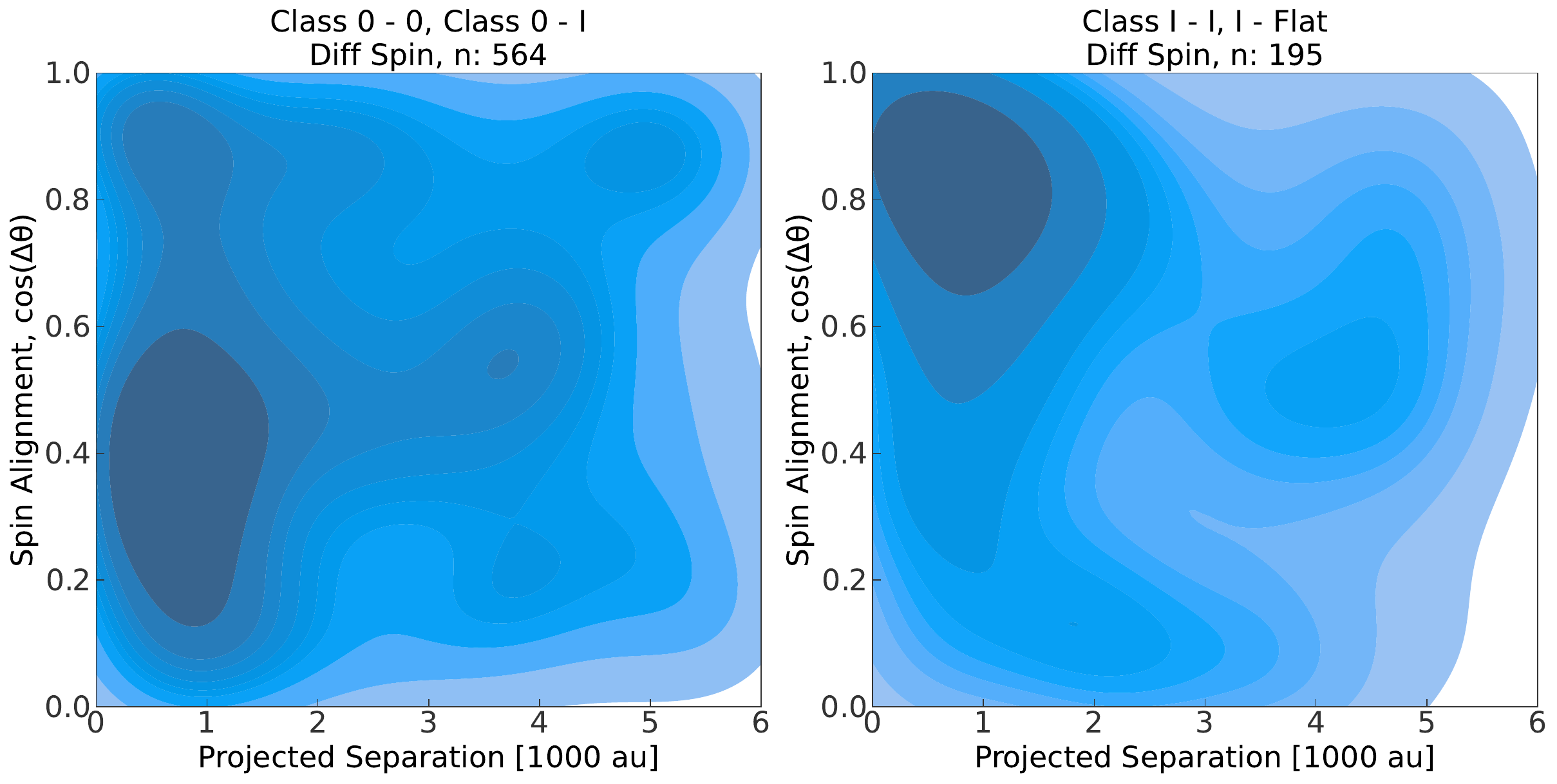}
\caption{Cosine of spin-spin angle versus projected separation for the nearest-neighbor stars in our fiducial {\sc starforge} simulation. Contours mark intervals of cumulative probability, with each consecutive ring enclosing an additional 10\% of the stellar pairs moving outward from the peak density. The bottom row shows the angle between the change in spin over a single snapshot. We include all pairs of stars with projected separations between 40 and 6000 au and (individual) masses $\geq 0.07 M_{\odot}$, stacking data from different snapshots (see text for details). The different columns correspond to different protostellar Classes, with the mapping between stellar age and Class defined by \autoref{tab:classConv}.
\label{fig:fidSimEvolve}}
\end{figure*}

\begin{figure*}[tbh!]
\includegraphics[width=\columnwidth]{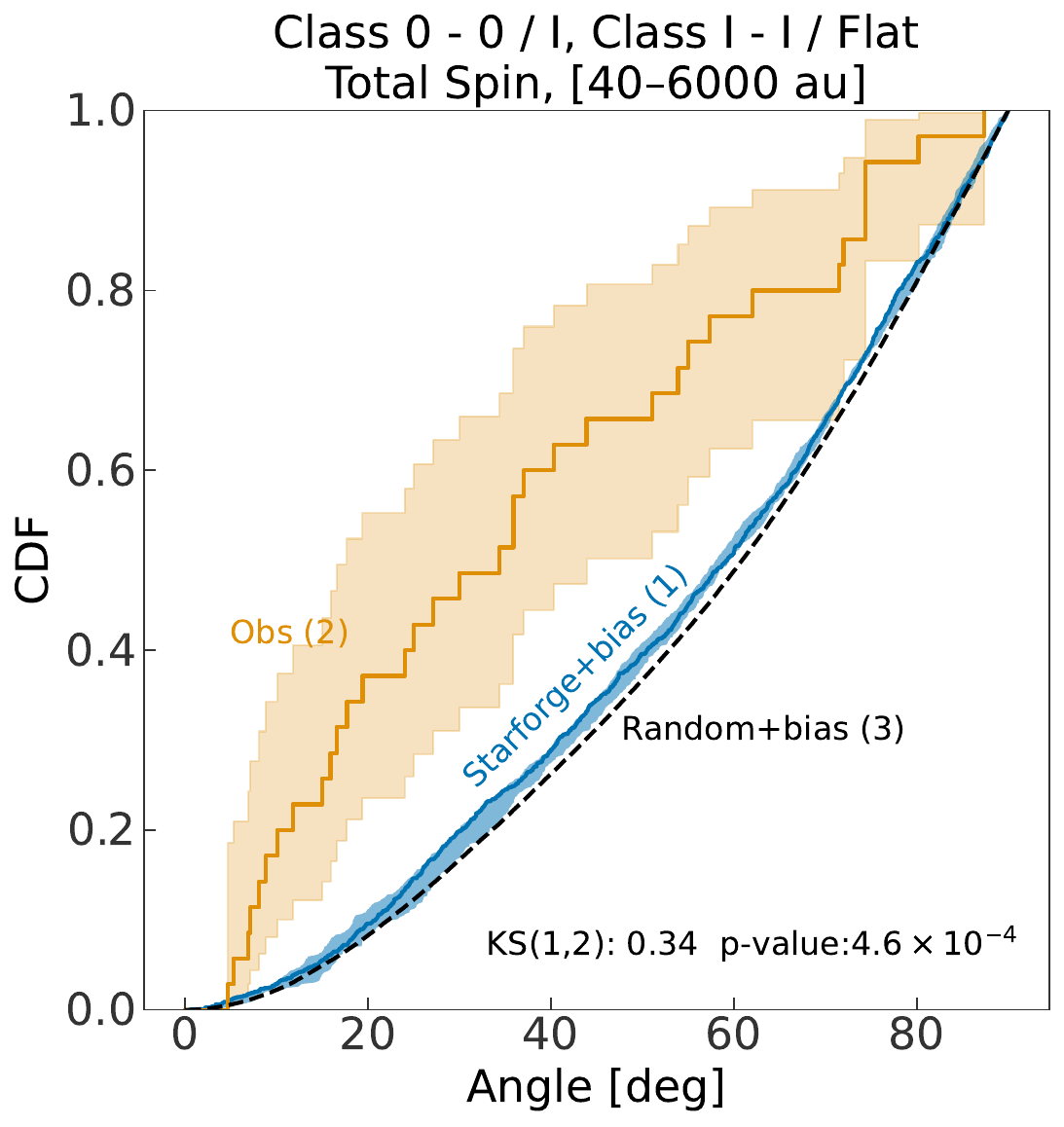}
\includegraphics[width=\columnwidth]{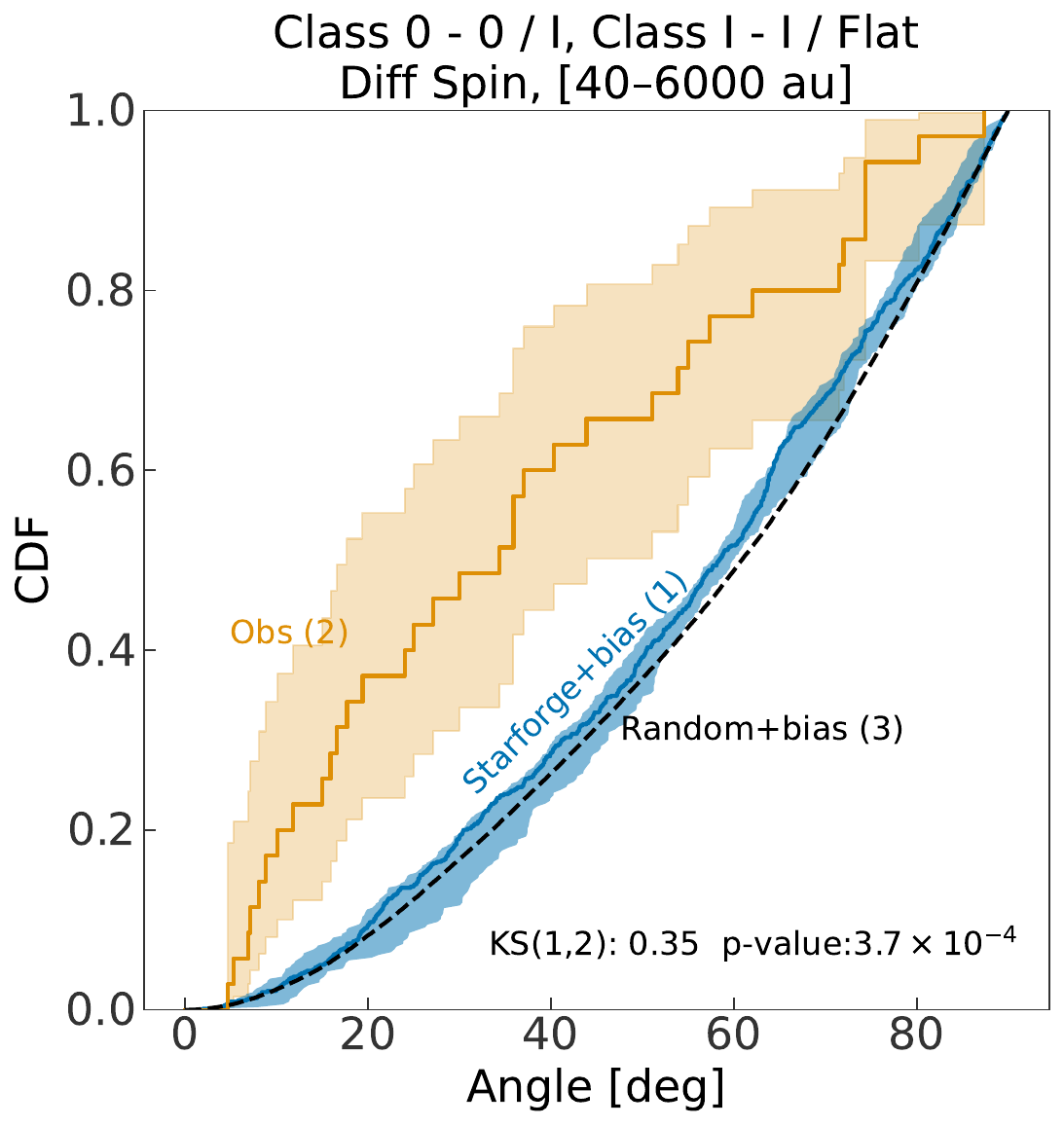}
\caption{Cumulative distributions of spin-spin angles from the fiducial simulation (blue line) and non-degenerate disk-disk angles from observations (orange line) for nearest neighbor pairs between 40 and 6000 au, with one additional observed source at 30 au. We only include pairs in associations of five or fewer stars in the simulation for consistency with the observations. The left panel shows the angle between the total stellar spins, while the right panel shows the angle between the changes in spin over one snapshot ($\sim 25$ kyr). The spin-spin angle distribution differs significantly from the observed distribution of disk orientations and is close to random (dashed, black line). The shaded region around the blue line shows how the spin-spin distribution varies with the initial magnetic field and virial parameter. The shaded region around the orange curve shows the 95\% confidence interval.} 
\label{fig:fidObs1D}
\end{figure*}

\autoref{fig:fidSimEvolve} shows the spin-spin angle versus projected separation between the closest neighbors in the fiducial simulation. Only pairs with projected separations between 40 and 6000 au are included, as closer pairs are unlikely to be resolved observationally. The spin-spin angle distribution is close to random, and far from the observations, as shown in \autoref{fig:fidObs1D}. This result is not sensitive to the initial magnetic field or virial parameter. We also found similar results in a simulation without any stellar feedback (outflows or radiation). \citet{guszejnov+2023} found a similar distribution of angles between binary spins at formation, though there are differences due to sample selection (e.g., we select pairs based on separation without requiring them to be bound to match the observational analysis). This is expected as we are analyzing the same underlying \textsc{Starforge} simulations.

\begin{figure*}[h!]
\includegraphics[width=\columnwidth]{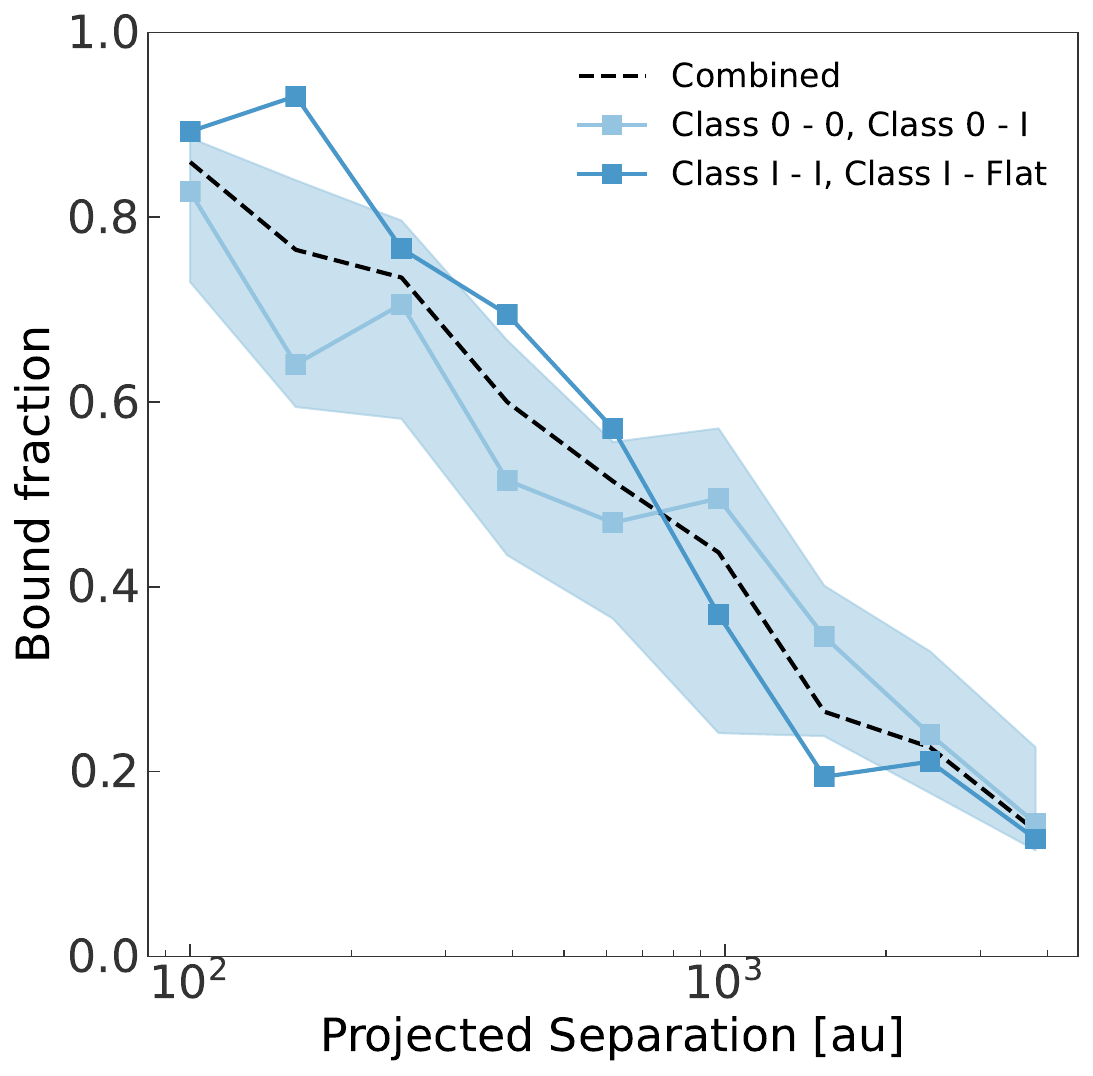}
\includegraphics[width=\columnwidth]{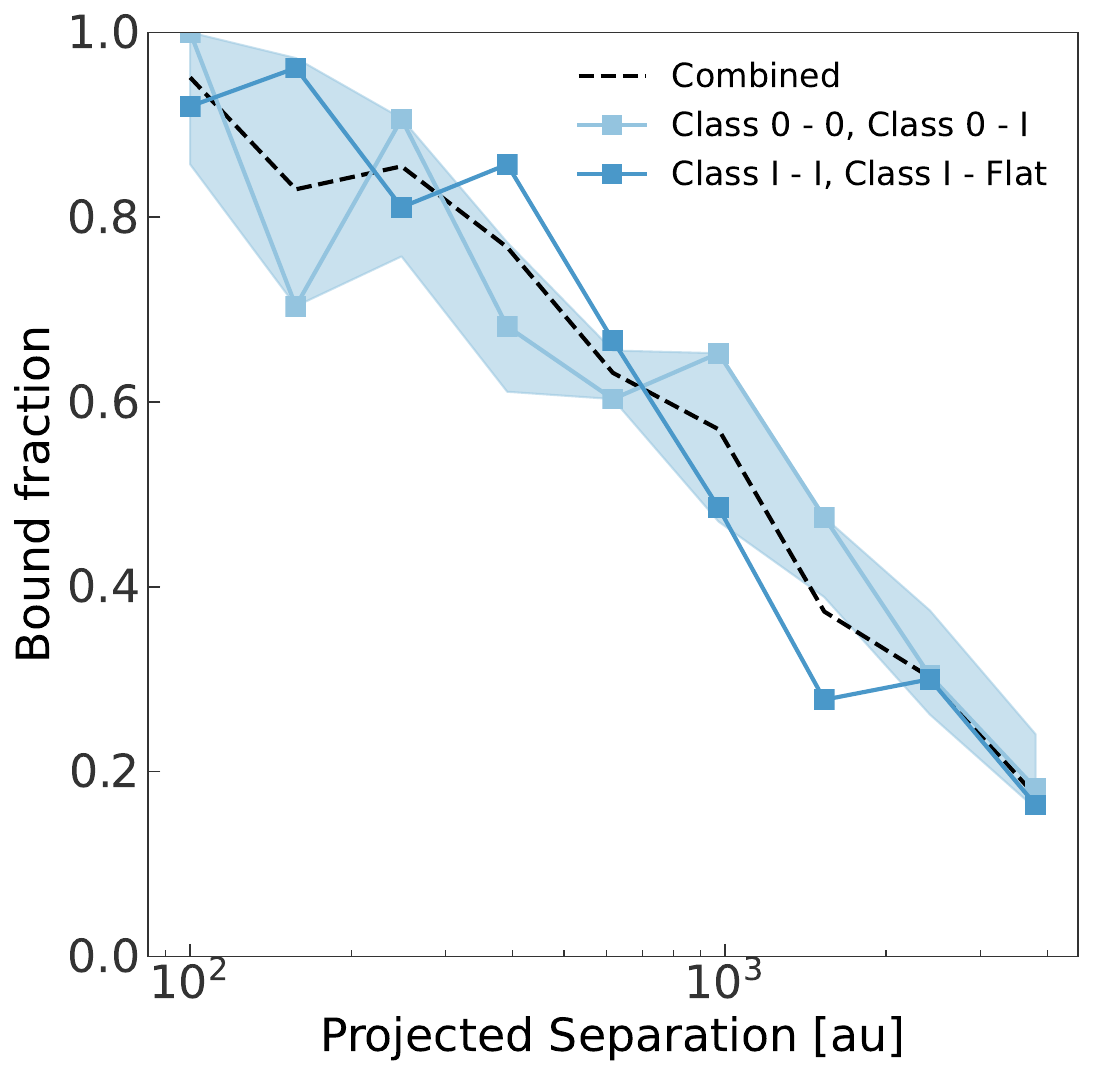}
\caption{\label{fig:boundProb} The lines show the fraction of nearest neighbors bound in a persistent multiple as a function of projected separation in the fiducial simulation for all pairs (left) and for pairs in associations of five or fewer stars (right). The latter is likely more relevant, as larger associations are rare in our observations. The lighter line corresponds to Class 0 - 0 and Class 0 - I pairs, while the darker line corresponds to Class I - I and Class I - Flat pairs. The dashed black line shows the fraction combining these classes. The shaded areas around the lighter lines show the variation in the bound fraction with the initial magnetic field and virial parameter. Pairs with projected separations $\gtrsim$1000 au are most likely unbound chance alignments.}
\end{figure*}

The observed signal is all the more striking, considering many of the wider pairs are likely unbound. \autoref{fig:boundProb} shows the probability that close pairs are bound as a function of projected separation. Here stars are bound if they are in the same persistent multiple system, according to the definitions in \cite{generozov+2025}. In short, multiples are considered persistent if they are bound together (i) for at least $\sim$50,000 yr and (ii) for at least one period. Most pairs outside of $\sim$1000 au are not bound.

\begin{figure*}[tbh!]
\includegraphics[width=\columnwidth]{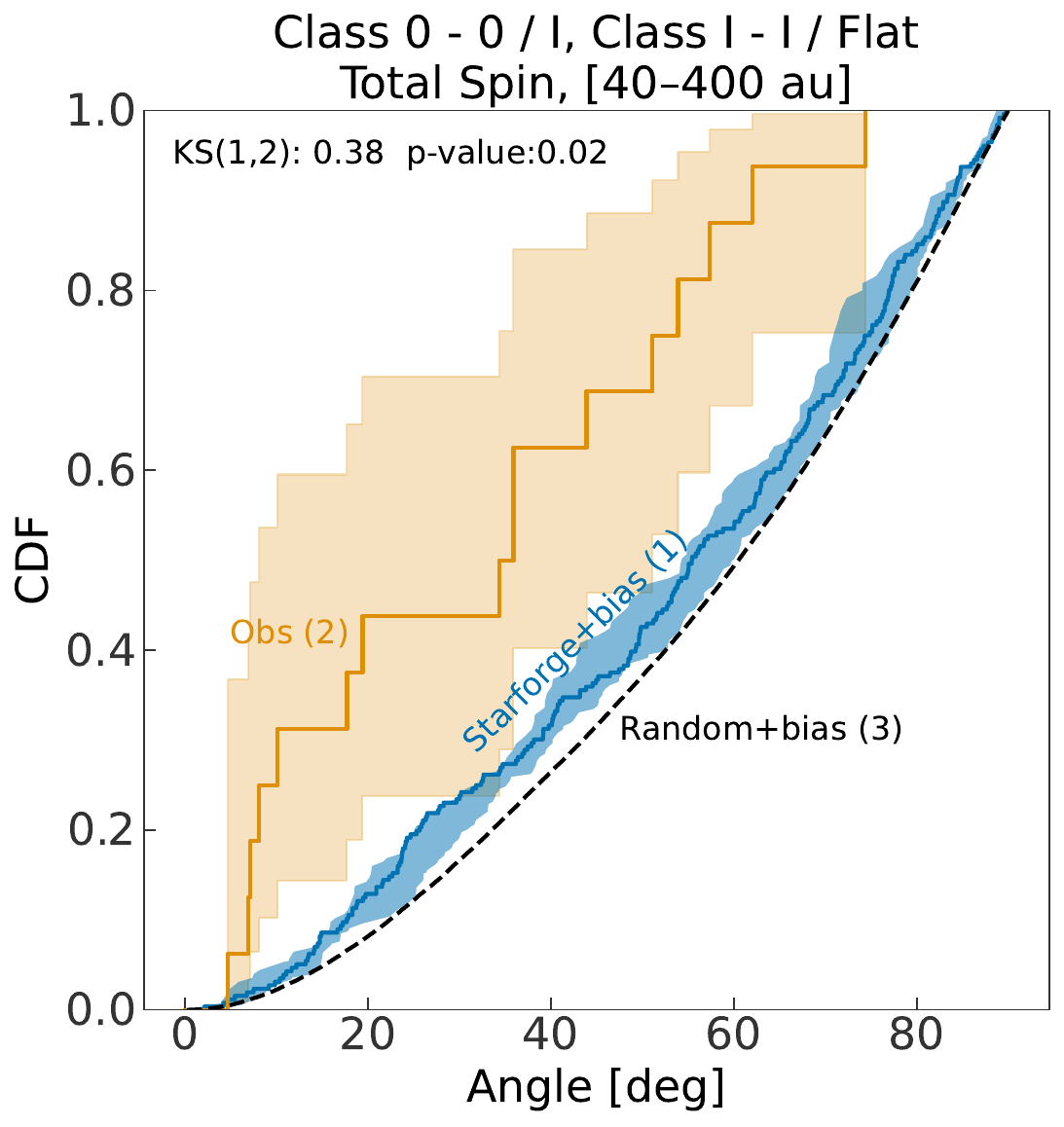}
\includegraphics[width=\columnwidth]{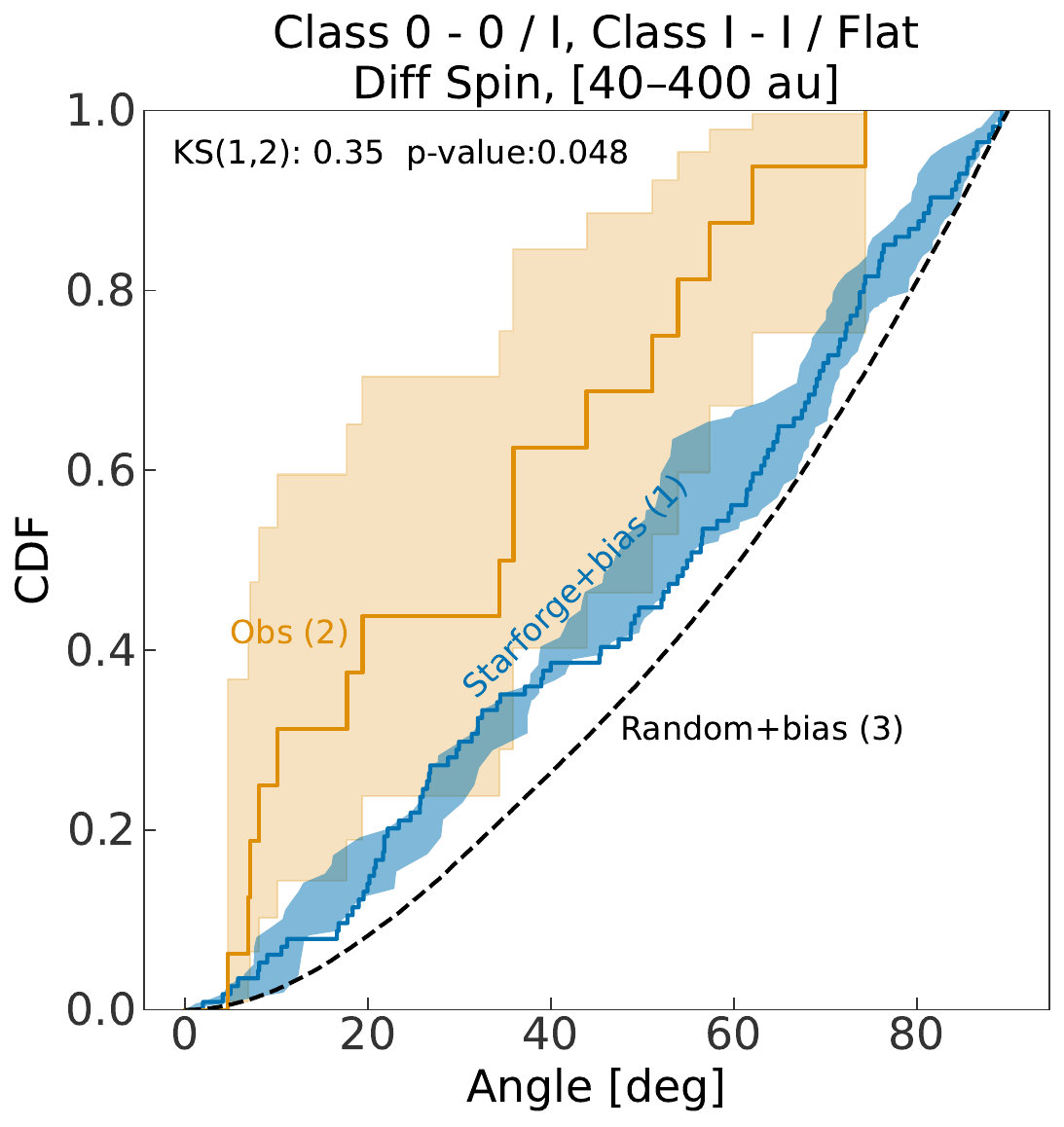}
\caption{Same as \autoref{fig:fidObs1D}, except excluding pairs with projected separations greater than 400 au, removing the majority of chance alignments. Observed disks in close pairs are more aligned than the spins of close stars in our simulations.} 
\label{fig:fidObs1DCut}
\end{figure*}

\autoref{fig:fidObs1DCut}
compares the observed and simulated 1D spin-spin distribution of closer pairs (between 40 and 400 au) that are most likely bound. The simulation appears slightly more aligned with the distance cut, though the distributions are not significantly different from those in \autoref{fig:fidObs1D}, and the tension with observations remains. 

The alignment signal is suppressed by the exclusion of pairs within a projected separation of 40 au, which are generally not resolved in observations. We find a much stronger alignment signal without this cut (see Fig.~\ref{fig:theoryNoCut}), though we note the pairs within $\sim$20 au in the simulations would be affected by gravitational softening.

Finally, caution is warranted in interpreting the comparisons between theory and observations. First of all, we are comparing observed \emph{disk} alignments with \emph{spin} alignments in the simulations. As pointed out by \citet{bate+2010}, disks can be truncated such that the final disk orientation likely reflects the angular momentum of the most recently accreted material, which will in general not be aligned with the stellar spin. Furthermore, the simulation spins can evolve in non-physical ways, considering the sinks accrete from much larger radii than stars and are not allowed to spin down due to outflows or magnetic braking \citep{guszejnov+2023}. Finally, close binaries (within $\sim$20 au) are affected by gravitational softening, which may also affect their spin evolution.

\section{Discussion}
\label{sec:discussion}

Our results suggest that turbulent fragmentation alone is insufficient to explain the formation of wide protostellar multiples (separations of $\sim$10$^3$ au). We show that protostellar disks in binaries and high-order multiples with projected separations up to 6000 au exhibit preferential alignment across all evolutionary classes, deviating significantly from the random or nearly random distributions predicted by turbulent fragmentation models (see \autoref{fig:fidObs1D}, \citealt{offner+2016}). While disk fragmentation can result in strong alignment \citep{bate2018}, this cannot account for pairs separated by thousands of au. The persistence of alignment at wide separations (1000–6000 au) indicates that the formation of multiple systems is more coherent than expected under fully stochastic turbulence; instead, the fragmentation and accretion processes retain a substantial degree of angular momentum coherence.

In contrast to our observation, \citet{2016ApJ...820L...2L} found misaligned protostellar outflow position angles in nine of the youngest binary/multiple systems with separations $a > 1000$\,au in Perseus. This apparent discrepancy may be partly due to the small sample size in \citet{2016ApJ...820L...2L}. More recently, \citet{2026ApJ..1001..134S} examined 51 Class 0/I protostellar systems with close companions in Perseus and Orion molecular clouds and found that the companion position angles are preferentially orthogonal to molecular outflow position angles. If protostellar outflows are perpendicular to disks, this result implies preferential disk alignment in multiple systems in the plane of the sky \citep{2026ApJ..1001..134S}, in agreement with our work.

Based on these findings, \citet{2026ApJ..1001..134S} suggested that disk fragmentation is the dominant formation pathway for close-companion protostellar systems. However, our finding that disks are preferentially aligned in multiple systems with separations of a few $10^3$ au cannot be explained by disk fragmentation. Instead, it suggests that disk alignment may be established at wide separations and later preserved as systems migrate inward \citep[e.g.,][]{2023A&A...674A.196K}. Such a scenario could provide an alternative or complementary explanation for the alignment of protostellar outflow position angles in close-companion protostellar systems reported by \citet{2026ApJ..1001..134S}.

Our finding is also consistent with recent high-resolution ($\sim8$\,au) disk-orientation studies in the Perseus molecular cloud \citep{2024ApJ...963..164R}. That study found that protostellar disks with projected separations $\lesssim 10{,}000$ au exhibit preferential alignment, with approximately 40–80\% of sources aligned. The strong alignment observed in Perseus multiple systems likewise disfavors a purely random disk-orientation distribution. 

The relatively high fraction (up to 80\%) of preferentially aligned disks reported by \citet{2024ApJ...963..164R} may be partially attributable to their treatment of the inclination-angle degeneracy. Specifically, they do not resolve the sign ambiguity and instead restrict inclination angles to be positive, which can artificially enhance the inferred degree of alignment. Nevertheless, this inclination-angle degeneracy cannot erase the intrinsic alignment signal. As shown in \autoref{fig:cdf_degenerate_inclination}, comparing source pairs with and without non-degenerate inclination angles reveals that the distribution of inclination-angle signs is not random; many disks within multiple systems share the same sign.

Extrapolating from our results (e.g., as shown in \autoref{fig:cdf_degenerate_inclination}), we believe that \citet{2024ApJ...963..164R} would still find preferential alignment in their sample if they were to account for the sign ambiguity, as we do. Taken together, these studies of disk alignments in multiple systems support a scenario in which fragmentation and disk accretion preserve a certain degree of coherent core angular momentum, rather than being fully randomized by turbulence or dynamics. 

A growing number of recent observations raise questions about whether turbulent fragmentation alone can account for the formation of wide protostellar multiple systems. In the L1448 region of Perseus, \citet{2015ApJ...814..114L} compared multiplicity with local core properties and found no correlation between the non-thermal velocity dispersion and the level of fragmentation (turbulence), although their analysis was limited by small-number statistics.
This lack of correlation deviates from a simple turbulent-fragmentation picture, in which stronger turbulent motions are expected to seed larger density perturbations and produce a higher degree of fragmentation.

The Submillimeter Array (SMA) Large Program ($\sim$600 observing hours over 3–4 years), Mass Assembly of Stellar Systems and Their Evolution with the SMA (MASSES), targeted all 73 known protostars in the Perseus molecular cloud in dust continuum and spectral line emission at 230 and 345 GHz. Using the MASSES dataset, \citet{2018ApJ...853....5P} conducted a detailed multiplicity analysis tracing the hierarchical structure of Perseus from cloud scales ($\sim$10 pc) to clumps ($\sim$1 pc), cores ($\sim$0.05–0.1 pc), envelopes ($\sim$300–3000 au), and individual protostars ($\sim$15 au). On small scales, classical thermal plus non-thermal Jeans fragmentation predicts an excess of protostellar objects compared to observations. Instead, the observed fragmentation level is more consistent with inefficient thermal Jeans fragmentation \citep{2018ApJ...853....5P}.

Our finding of preferential alignment among protostellar disks in wide binaries and higher-order multiple systems supports a coherent formation scenario for multiple systems. The observed alignment implies that the parent cores possess low levels of turbulence and maintain a coherent velocity structure during collapse. Evidence for such coherent cores is provided by \citet{2010ApJ...712L.116P} and \citet{2019ApJ...877...93C}. \citet{2010ApJ...712L.116P} conducted NH$_3$ observations of the B5 region in the Perseus molecular cloud using the Green Bank Telescope and identified a sharp transition from the surrounding turbulent gas to a coherent dense core. 

Rather than being driven purely by turbulent fragmentation, the collapse of a protostellar core and the formation of multiple systems may be influenced by magnetic-field–guided collapse, which can impose preferred directions during fragmentation. \citet{2013ApJ...770..151C} found that, in four Class 0 low-mass cores, the core-scale magnetic fields are aligned with the pseudodisk symmetry axes, supporting a scenario of magnetically regulated infall. More recently, an SMA study of the magnetic fields in 20 Class 0 protostars, probing envelope scales of $\sim$1000 au, identified a strong correlation between the misalignment of the magnetic field relative to the outflow axis and the envelope angular momentum \citep{2020A&A...644A..47G}. On even larger scales ($0.2$--$0.6$\,pc),\citet{2022ApJ...941...81X} used machine learning to identify 200 protostellar outflows in Ophiuchus, Taurus, Perseus, and Orion, and found that the outflow position angles are aligned with the large-scale magnetic-field directions traced by Planck 353\,GHz dust polarization. However, some numerical simulations do not find a trend toward stronger alignment with increasing cloud magnetic-field strength, and may even find the opposite trend \citep{2019ApJ...887..232L,guszejnov+2023}.

In addition to magnetic fields, large-scale gas flows or accretion may also help explain the preferential alignment of protostellar disks in multiple systems. Recent ALMA CO (2–1) observations toward the massive infrared dark cloud G28.37+0.07 reveal a strong orthogonal alignment between outflows and filamentary structures \citep{2019ApJ...874..104K}. If protostellar outflows trace the disk angular momentum axis, this orthogonal outflow–filament alignment implies coherent disk orientations on even larger, parsec filamentary scales. Such large-scale coherence is further supported by the recent James Webb Space Telescope Near Infrared Camera (JWST NIRCam) imaging of the Serpens Main star-forming regions, which reveals that nearly all the protostellar outflows are aligned in the plane-of-sky over scales of $\sim$1\,pc \citep{2024ApJ...972....5G}.

In summary, our results reveal preferential alignment of protostellar disks in multiple systems, significantly deviating from the random distributions predicted by turbulent fragmentation models. These observations indicate that additional physical mechanisms, such as strong magnetic fields or large-scale accretion within fibers or filaments \citep{2017A&A...606A.123H}, are required to explain the formation of multiple systems.

\section{Conclusion}
\label{sec:conclusion}

We combine the CAMPOS survey \citep{2024ApJ...973..138H} and the VLA Nascent Disk and Multiplicity (VANDAM) survey \citep{2020ApJ...890..130T}, which together cover the vast majority of known protostellar disks within 500 pc of the solar neighborhood at angular resolutions down to $\sim$0.1\arcsec. This combined dataset enables a comprehensive study of disk alignment in multiple systems. In total, 282 out of 512 Class 0, Class I, and flat-spectrum protostars across nine nearby molecular clouds are found in multiple systems (projected separation $\le 6000$\,au). When classified by multiplicity, the sample comprises 74 binaries, 12 triples, 6 quadruples, 10 quintuples, and 3 octuple systems.  We summarize our major conclusions below: 

\begin{enumerate}

    \item Our results suggest that turbulent fragmentation alone is insufficient to explain the formation of wide protostellar multiples (separations of $\sim$10$^3$ au). We show that protostellar disks in binaries and high-order multiples with projected separations up to 6000 au exhibit preferential alignment across all evolutionary classes, deviating significantly from the random distributions predicted by turbulent fragmentation models (see \autoref{fig:fidObs1D}). When the inclination angle sign degeneracy is resolved, the intrinsic degree of protostellar disk alignment is substantially higher, approaching $\sim50$\% in a simple binomial mixture model of parallel and random orientations.\\

    \item Comparing protostellar disk alignment in close ($<1000$\,au) separation versus wide (1000--6000\,au) separation shows little variation with increasing separation, in strong contrast to the randomly oriented disks at larger separations predicted by turbulent fragmentation models (See \autoref{fig:fidObs1D} for comparison). \\

    \item We found that the degree of nearest-neighbor protostellar disk alignment in higher-order multiple systems is comparable to that in binary systems. \\ 

    \item We find that there are significantly fewer nearest-neighbor pairs among more evolved flat-spectrum protostellar disks than among younger Class 0/I disks within high-order multiple systems. This trend indicates that most high-order multiple systems disintegrate or migrate to unresolved separations ($<20-40$\,au) and appear as binaries by the end of the Class I phase.  \\ 

    \item We find a tentative trend (p-value of 0.1) suggesting that protostellar disks in binary systems become increasingly aligned over time. Additional observations are needed to increase the sample size and determine whether this apparent trend is statistically robust.

    \item We compare the disk alignment distribution from observations with spin alignments in \textsc{Starforge} simulations of molecular clouds similar to those in the Solar neighborhood. Close pairs in the \textsc{Starforge} simulations show spin alignment, especially within 40 au. However, such close pairs are not resolved in observations, and there is no spin alignment on 1000 au scales and beyond. These simulations do not have disks and therefore cannot say anything directly about disk alignment. The spin may simply be a poor proxy for disk orientation, especially considering that the spin is not allowed to evolve due to outflows and magnetic braking. \\

    \item Pairs beyond $\sim$1000 au in projection in the \textsc{Starforge} simulations are more likely to be chance alignments rather than bound systems. This suggests that many of the observed wide pairs show spin alignment despite being unbound.

\end{enumerate}

\section*{Data Availability}
All data used in this study are publicly available and can be downloaded from the ALMA Science Archive. Table A.1 is only available in electronic form at the CDS via anonymous ftp to cdsarc.u-strasbg.fr (130.79.128.5) or via http://cdsweb.u-strasbg.fr/cgi-bin/qcat?J/A+A/.

\begin{acknowledgements}
We thank Dr. Kaitlin Kratter for helpful conversations. Special thanks to Dr. John Tobin for providing the CASA \emph{imfit} uncertainties for the VANDAM survey. We thank the anonymous referee for the comments and suggestions. C.H.H. is supported by the NASA Hubble Fellowship Program under award HF2-51556. A.G. and  S.S.R.O. acknowledge support from NSF AAG 2407522. S.S.R.O. acknowledges support from a Peter O'Donnell Distinguished Researcher Fellowship and a Donald Harrington Fellowship.
 This paper makes use of the following ALMA data: ADS/JAO. ALMA \#2015.1.00041.
S., \#2019.1.01792.S. ALMA is a partnership of the ESO (representing its member states),NSF(USA), and NINS (Japan), together with the NRC (Canada), NSC and ASIAA(Taiwan), and KASI(Republic of Korea), in cooperation with the Republic of Chile. The Joint ALMA Observatory is operated by the ESO, AUI/NRAO, and NAOJ. The National Radio Astronomy Observatory and Green Bank Observatory are facilities of the National Science Foundation operated under cooperative agreement by Associated Universities, Inc. The authors acknowledge the Texas Advanced Computing Center (TACC) at The University of Texas at Austin for providing computational resources that have contributed to the research results reported within this paper. URL:
http://www.tacc.utexas.edu This paper used the following software packages: Astropy: \citet{2013ascl.soft04002G,2018AJ....156..123A,2022ApJ...935..167A}, CASA: \citet{2007ASPC..376..127M} SciPy:  \citet{2020NatMe..17..261V}.

\end{acknowledgements}

\clearpage
\bibliographystyle{aa}
\bibliography{Disk.bib}

\begin{appendix}

\section{Full data of multiple systems in 9 nearby molecular clouds}
\label{Appendix}

\begin{figure*}[tbh!]
    \includegraphics[width=.99\textwidth]{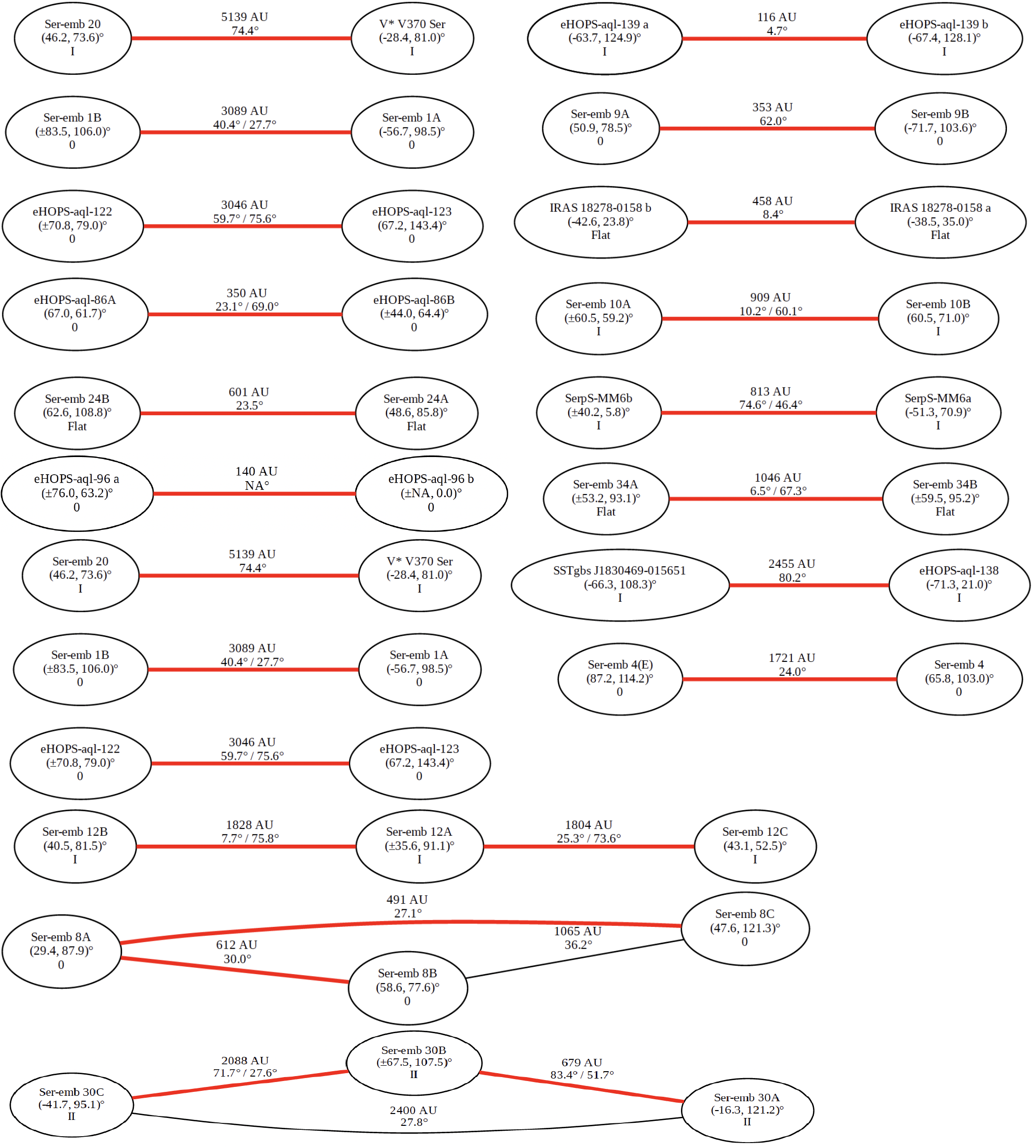}
    \caption{The protostellar disks in the binary and triple systems in the Serpens and Aquila molecular cloud. We represent each protostellar disk as a node with the source name, inclination,  position angle, and evolutionary class labeled. We add an edge if the disks are within a projected distance of 6000$\,$au. Each edge is labeled by the separation between the disks as well as the three-dimensional angle between the disk orientations. For each disk (node), we highlighted the edges connecting the nearest neighbor in red. Data from the CAMPOS survey \citep{2024ApJ...973..138H}. } 
\label{fig:Flow_chart_Aql_Serp} 
\end{figure*}

\begin{figure*}[tbh!]
    \includegraphics[width=.99\textwidth]{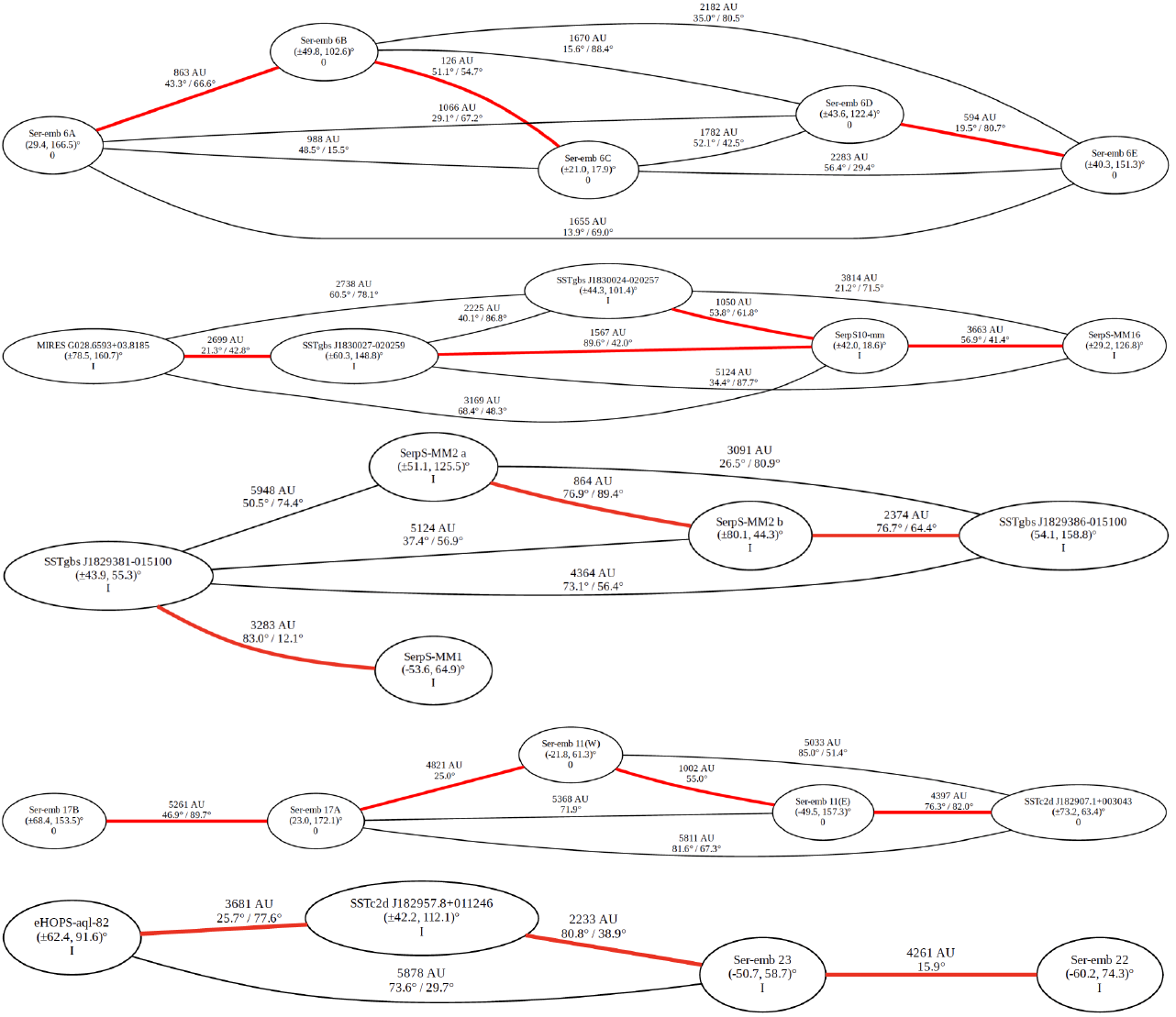}
    \caption{The protostellar disks in the quadruples and quintuples in the Serpens and Aquila molecular cloud. We represent each protostellar disk as a node with the source name, inclination,  position angle, and evolutionary class labeled. We add an edge if the disks are within a projected distance of 6000$\,$au. Each edge is labeled by the separation between the disks as well as the three-dimensional angle between the disk orientations. For each disk (node), we highlighted the edges connecting the nearest neighbor in red. Data from the CAMPOS survey \citep{2024ApJ...973..138H}.} 
\label{fig:Flow_chart_Aql_Serp2} 
\end{figure*} 

\begin{figure*}[tbh!]
    \includegraphics[width=.99\textwidth]{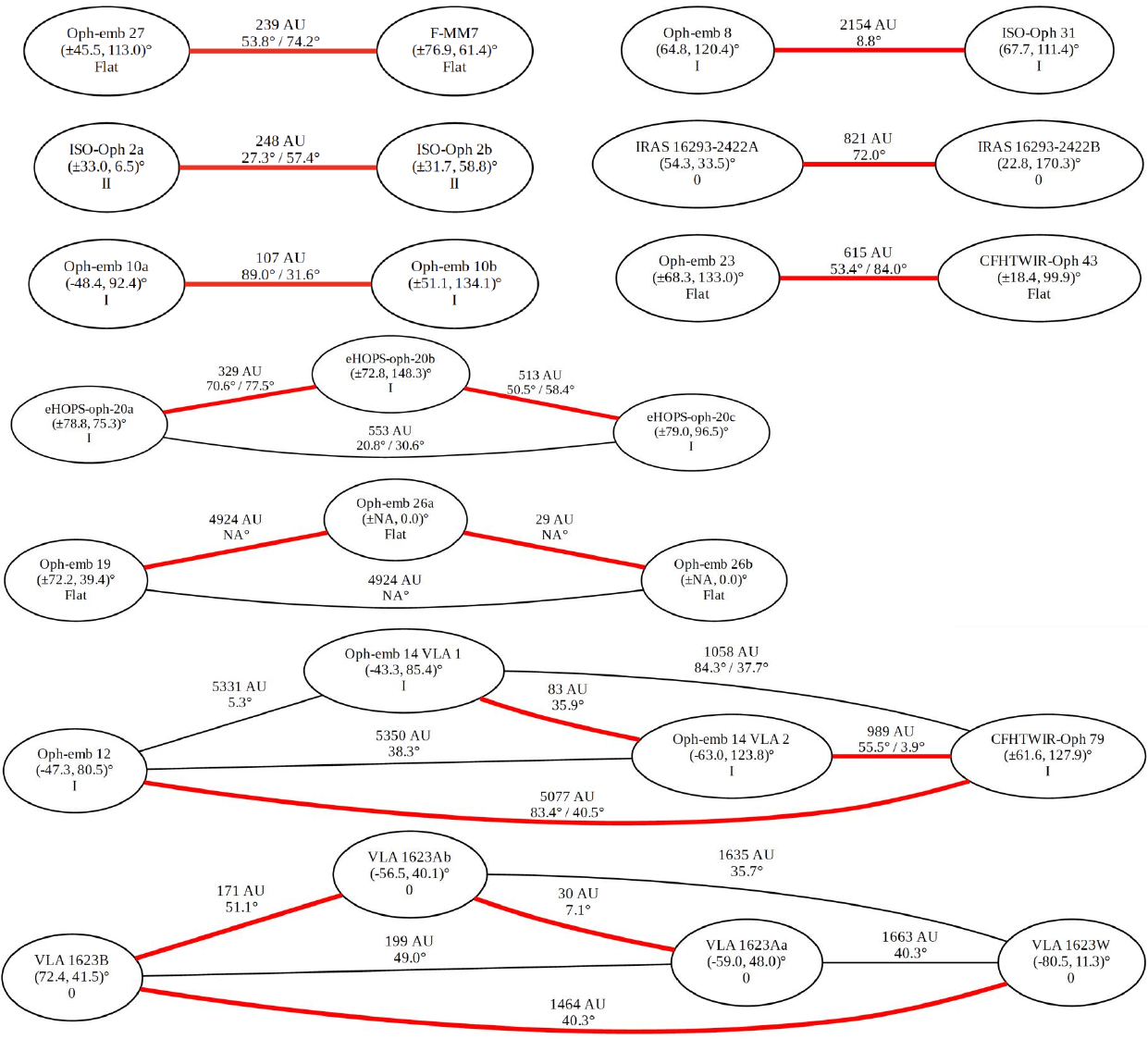}
    \caption{The protostellar disks in the multiple systems in the $\rho$ Ophiuchus molecular cloud. For each disk (node), we highlighted the edges connecting the nearest neighbor. Data from the CAMPOS survey \citep{2024ApJ...973..138H}.} 
\label{fig:Flow_chart_Oph}
\end{figure*}

\begin{figure*}[tbh!]
    \includegraphics[width=.99\textwidth]{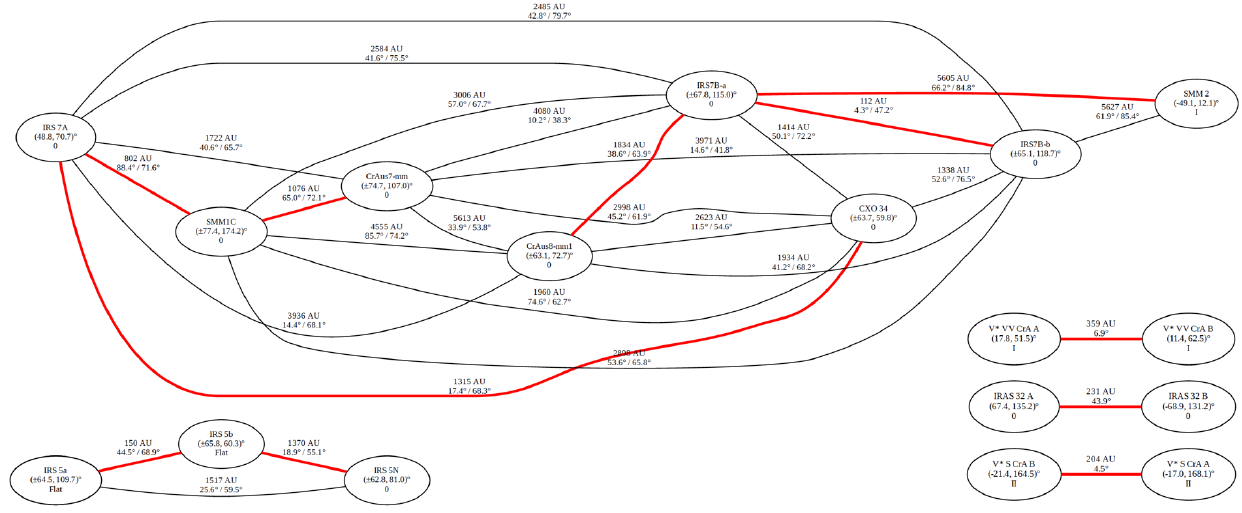}
    \caption{ The protostellar disks in the multiple systems in the Corona Australis molecular cloud. Data from the CAMPOS survey \citep{2024ApJ...973..138H}.} 
\label{fig:Flow_chart_CrAus}
\end{figure*}

\begin{figure*}[tbh!]
\centering
    \includegraphics[width=.95\textwidth]{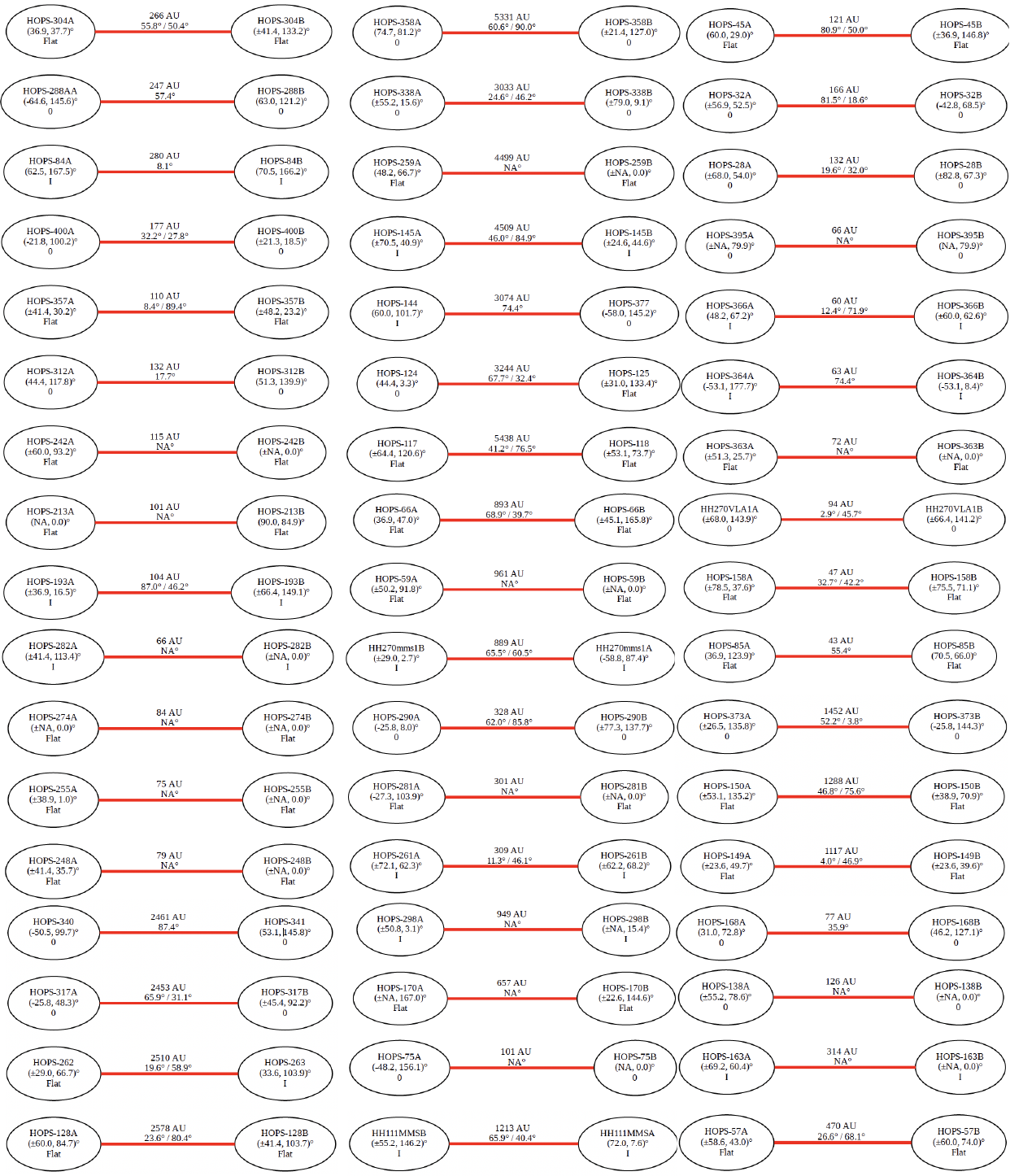}
    \caption{ Protostellar disks in the binary systems in the Orion molecular cloud. Data from the VANDAM Orion survey \citep{2020ApJ...890..130T}.} 
\label{fig:Flow_chart_Orion1}
\end{figure*}

\begin{figure*}[tbh!]
\centering
    \includegraphics[width=.99\textwidth]{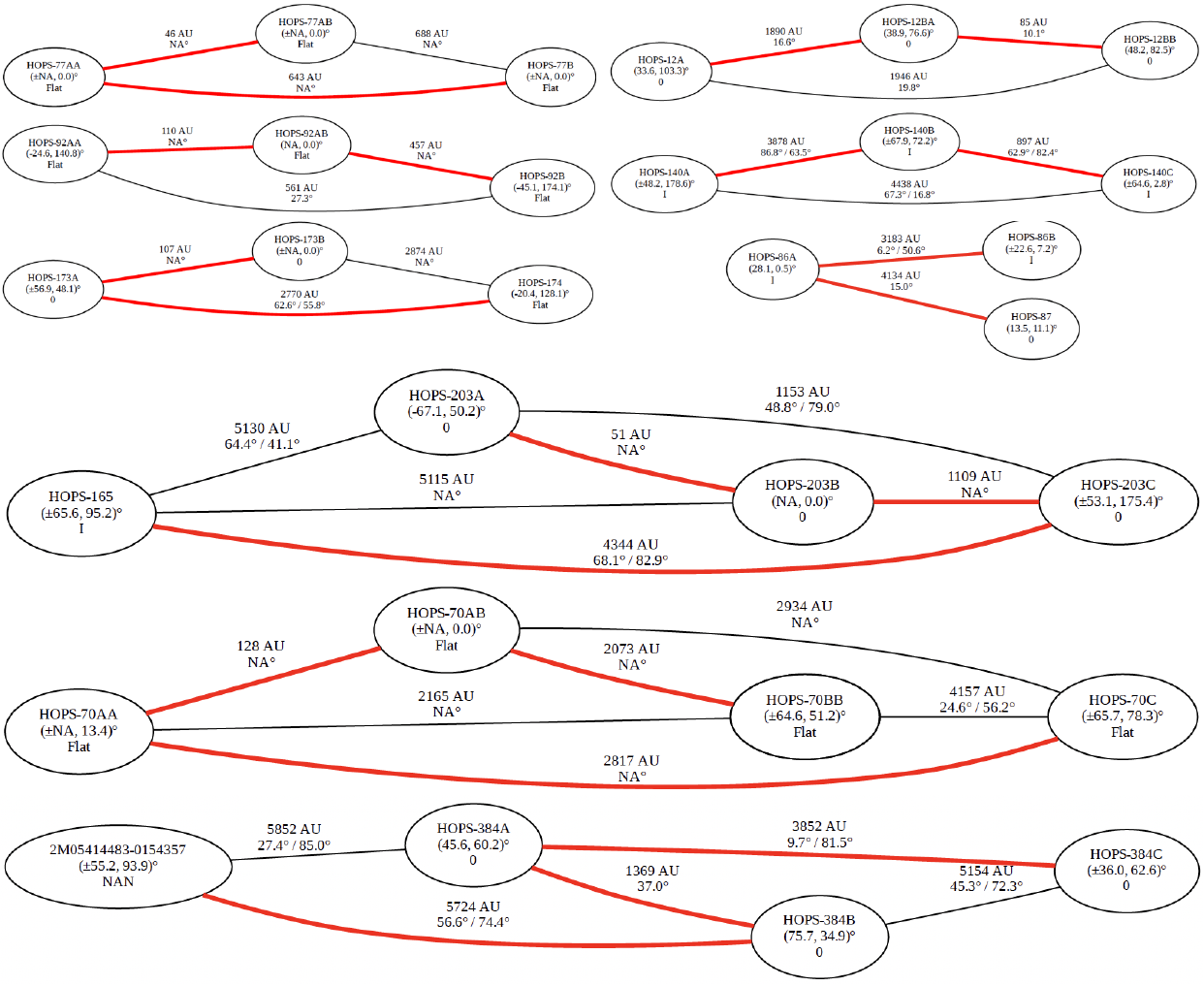}
    \caption{ Protostellar disks in the triple and quadruple systems in the Orion molecular cloud. Data from the VANDAM Orion survey \citep{2020ApJ...890..130T}.} 
\label{fig:Flow_chart_Orion2}
\end{figure*}

\begin{figure*}[tbh!]
\centering
    \includegraphics[width=.99\textwidth]{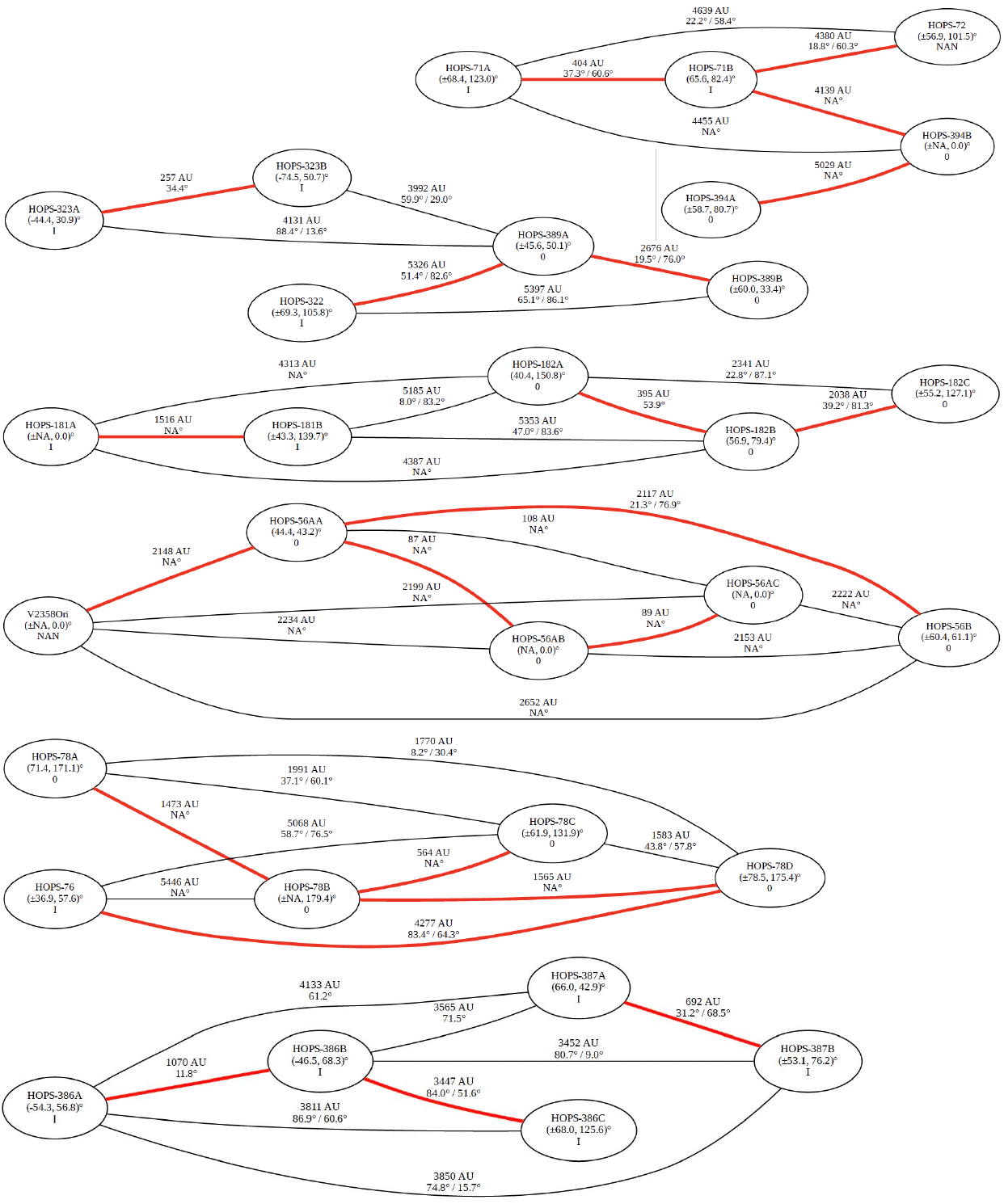}
    \caption{Protostellar disks in the quintuple systems in the Orion molecular cloud. Data from the VANDAM Orion survey \citep{2020ApJ...890..130T}.} 
\label{fig:Flow_chart_Orion3}
\end{figure*}

\begin{figure*}[tbh!]
\centering
    \includegraphics[width=.99\textwidth]{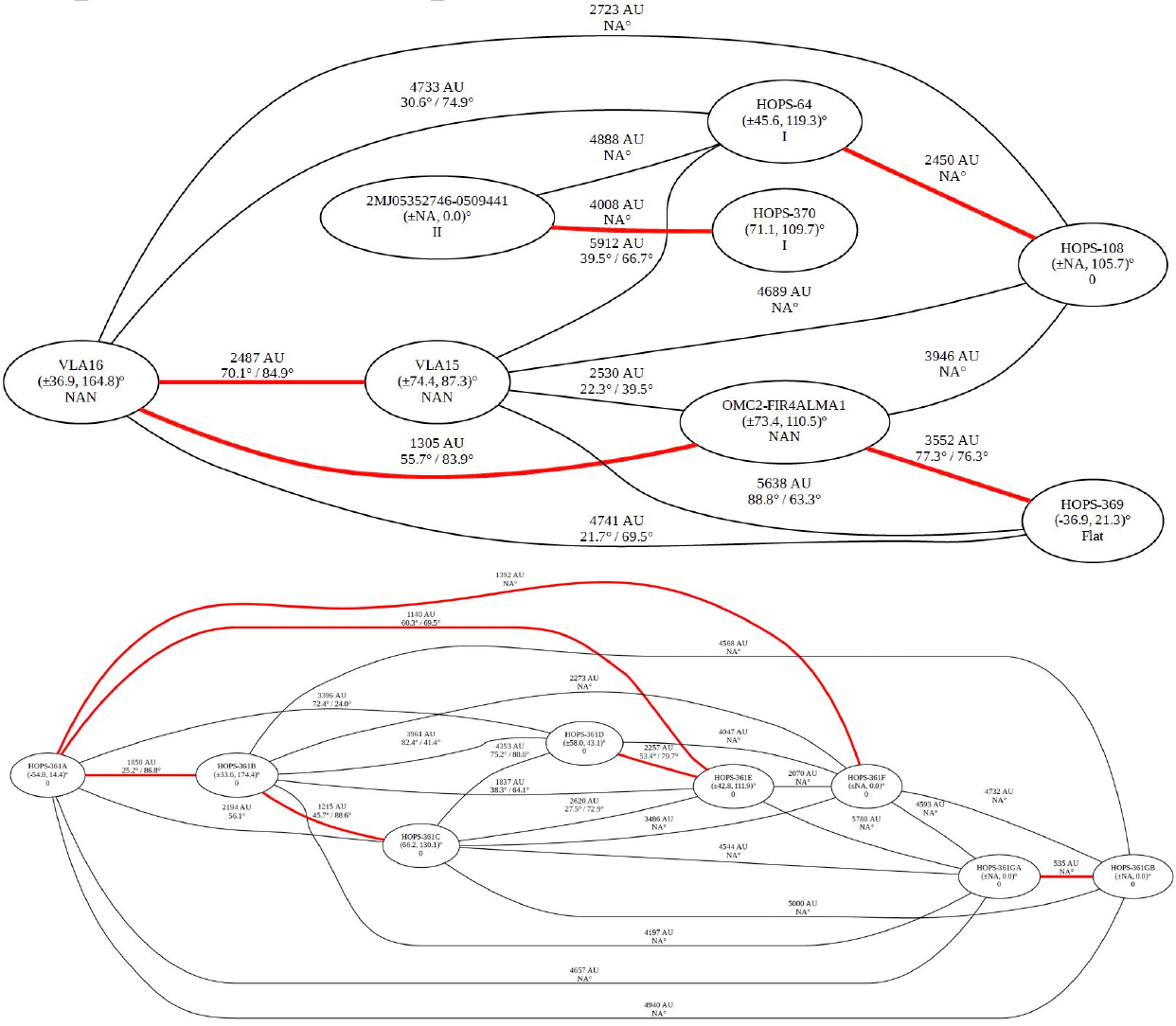}
    \caption{ Protostellar disks in the octuple systems in the Orion molecular cloud. Data from the VANDAM Orion survey \citep{2020ApJ...890..130T}.} 
\label{fig:Flow_chart_Orion4}
\end{figure*}

This Appendix presents the complete dataset of binary and higher-order multiple systems analyzed in this study, excluding those in Chamaeleon, which are shown in \autoref{fig:Flow_chart_Cham}.  In the Figures, each protostellar disk is represented as a node labeled with the source name, inclination, position angle, and evolutionary class. An edge is drawn between two nodes if their projected separation is less than 6000 au. Each edge is labeled with both the projected separation and the three-dimensional angle between the disk orientations. For each disk (node), the edge connecting to its nearest neighbor is highlighted in red.

\autoref{fig:Flow_chart_Aql_Serp} and  \autoref{fig:Flow_chart_Aql_Serp2} illustrate the protostellar disks in multiple systems within the Serpens and Aquila molecular clouds. 
\autoref{fig:Flow_chart_Oph} presents the corresponding diagram for $\rho$ Ophiuchus and $\rho$ Ophiuchus North, and \autoref{fig:Flow_chart_CrAus} for Corona Australis. \autoref{fig:Flow_chart_Orion1} shows the binary systems in Orion A and Orion B, while \autoref{fig:Flow_chart_Orion2} displays the triple and quadruple systems. \autoref{fig:Flow_chart_Orion3} presents the quintuple systems, and \autoref{fig:Flow_chart_Orion4} shows the octuple systems in Orion A and Orion B.

In Table \ref{table:CAMPOS_Tbol}, we present the disk orientation of the nearest neighbor pairs in the protostellar binaries, and high-order systems with projected separation less than 6000 au across the Corona Australis, Chamaeleon I, Chamaeleon II, Ophiuchus, Ophiuchus North, Orion A, Orion B, Serpens, and Aquila molecular clouds. 

\FloatBarrier

\onecolumn

\begin{landscape}

\begin{longtable}{lrrrrrrrr}
\caption{Disk orientation for the nearest neighbor of protostellar binaries and high-order multiple systems with separation less than 6000 au in the Corona Australis, Chamaeleon I, Ophiuchus, Ophiuchus North, Orion A, Orion B, Serpens, and Aquila molecular clouds} \\ 
\label{table:CAMPOS_Tbol} \\
\hline
Source & Closest Neighbor & Source & Neighbor &  $a$ &  $\Delta\phi_1$ & $\Delta\phi_2$ & Source  & Neighbor \\
& & Class & Class & [au] & [$^\circ$] & [$^\circ$] &  inc, pa [$^\circ$] & inc, pa [$^\circ$] \\
\hline

\hline
\endfirsthead

\multicolumn{5}{c}
{{\tablename\ \thetable{} -- Continued from previous page}} \\
\hline
Source & Closest Neighbor & Source & Neighbor &  $a$ &  $\Delta\phi_1$ & $\Delta\phi_2$ & Source  & Neighbor  \\
& & Class & Class & [au] & [$^\circ$] & [$^\circ$] &  inc, pa [$^\circ$] & inc, pa [$^\circ$] \\
\hline
\endhead
\hline
\multicolumn{5}{r}{\emph{Continued on next page}} \\
\endfoot

\hline
\endlastfoot

\multicolumn{8}{p{0.5\linewidth}} {Serpens and Aquila (Binary)}\\
\hline
Ser-emb 1A & Ser-emb 1B & 0 & 0 & 3089 & $40.4^{+7.7}_{-7.3}$ & $27.7^{+6.9}_{-7.6}$ & $(-56.7,-98.5)$ & $({\pm}83.5,106.0)$ \\ [1ex]
Ser-emb 4 & Ser-emb 4(E) & 0 & 0 & 1721 & $24.0^{+25.8}_{-25.2}$ & $24.0^{+25.8}_{-25.2}$ & $(65.8,-103.0)$ & $(87.2,-114.2)$ \\ [1ex]
Ser-emb 9B & Ser-emb 9A & 0 & 0 & 353 & $62.0^{+19.0}_{-18.4}$ & $62.0^{+19.0}_{-18.4}$ & $(-71.7,-103.6)$ & $(50.9,-78.5)$ \\ [1ex]
eHOPS-aql-123 & eHOPS-aql-122 & 0 & 0 & 3046 & $59.7^{+23.8}_{-24.0}$ & $75.6^{+18.8}_{-20.1}$ & $(67.2,-143.4)$ & $({\pm}70.8,79.0)$ \\ [1ex]
eHOPS-aql-86B & eHOPS-aql-86A & 0 & 0 & 350 & $23.1^{+21.2}_{-18.0}$ & $69.0^{+19.6}_{-23.0}$ & $({\pm}44.0,64.4)$ & $(67.0,-61.7)$ \\ [1ex]
eHOPS-aql-96 b & eHOPS-aql-96 a & 0 & 0 & 140 &  &  & (NA, NA) & $({\pm}76.0,63.2)$ \\ [1ex]
Ser-emb 10B & Ser-emb 10A & I & I & 909 & $10.2^{+19.8}_{-19.6}$ & $60.1^{+21.7}_{-22.3}$ & $(60.5,-71.0)$ & $({\pm}60.5,59.2)$ \\ [1ex]
SerpS-MM6a & SerpS-MM6b & I & I & 813 & $74.6^{+15.8}_{-15.0}$ & $46.4^{+15.7}_{-18.0}$ & $(-51.3,-70.9)$ & $({\pm}40.2,5.8)$ \\ [1ex]
V* V370 Ser & Ser-emb 20 & I & I & 5139 & $74.4^{+19.6}_{-18.1}$ & $74.4^{+19.6}_{-18.1}$ & $(-28.4,-81.0)$ & $(46.2,-73.6)$ \\ [1ex]
eHOPS-aql-138 & SSTgbs J1830469-015651 & I & I & 2455 & $80.2^{+14.2}_{-14.1}$ & $80.2^{+14.2}_{-14.1}$ & $(-71.3,-21.0)$ & $(-66.3,-108.3)$ \\ [1ex]
eHOPS-aql-139 b & eHOPS-aql-139 a & I & I & 116 & $4.7^{+25.2}_{-24.0}$ & $4.7^{+25.2}_{-24.0}$ & $(-67.4,-128.1)$ & $(-63.7,-124.9)$ \\ [1ex]
IRAS 18278-0158 a & IRAS 18278-0158 b & Flat & Flat & 458 & $8.4^{+26.5}_{-25.0}$ & $8.4^{+26.5}_{-25.0}$ & $(-38.5,-35.0)$ & $(-42.6,-23.8)$ \\ [1ex]
Ser-emb 24A & Ser-emb 24B & Flat & Flat & 601 & $23.5^{+15.6}_{-16.3}$ & $23.5^{+15.6}_{-16.3}$ & $(48.6,-85.8)$ & $(62.6,-108.8)$ \\ [1ex]
Ser-emb 34B & Ser-emb 34A & Flat & Flat & 1046 & $6.5^{+24.7}_{-24.3}$ & $67.3^{+18.0}_{-18.5}$ & $({\pm}59.5,95.2)$ & $({\pm}53.2,93.1)$ \\ [1ex]
Ser-emb 15B & Ser-emb 15A & II & II & 19 & $30.1^{+27.9}_{-22.8}$ & $30.1^{+27.9}_{-22.8}$ & $(34.7,-108.8)$ & $(64.4,-115.8)$ \\ [1ex]

\hline
\multicolumn{8}{p{0.5\linewidth}} {Serpens and Aquila (Triple)}\\
\hline
Ser-emb 8B & Ser-emb 8A & 0 & 0 & 612 & $30.0^{+23.8}_{-23.2}$ & $30.0^{+23.8}_{-23.2}$ & $(58.6,-77.6)$ & $(29.4,-87.9)$ \\ [1ex]
Ser-emb 8C & Ser-emb 8A & 0 & 0 & 491 & $27.1^{+24.3}_{-23.2}$ & $27.1^{+24.3}_{-23.2}$ & $(47.6,-121.3)$ & $(29.4,-87.9)$ \\ [1ex]
Ser-emb 12A & Ser-emb 12B & I & I & 1828 & $7.7^{+22.9}_{-22.0}$ & $75.8^{+22.1}_{-21.3}$ & $({\pm}35.6,91.1)$ & $(40.5,-81.5)$ \\ [1ex]
Ser-emb 12C & Ser-emb 12B & I & I & 45 & $19.4^{+24.6}_{-23.8}$ & $19.4^{+24.6}_{-23.8}$ & $(43.1,-52.5)$ & $(40.5,-81.5)$ \\ [1ex]
Ser-emb 30A & Ser-emb 30B & II & II & 679 & $83.4^{+24.2}_{-29.1}$ & $51.7^{+30.3}_{-32.3}$ & $(-16.3,-121.2)$ & $({\pm}67.5,107.5)$ \\ [1ex]
Ser-emb 30B & Ser-emb 30C & II & II & 2088 & $71.7^{+21.6}_{-22.6}$ & $27.6^{+26.9}_{-25.1}$ & $({\pm}67.5,107.5)$ & $(-41.7,-95.1)$ \\ [1ex]

\hline
\multicolumn{8}{p{0.5\linewidth}} {Serpens and Aquila (Quadruple and above)}\\ [0.5ex]
\hline
SSTc2d J182907.1+003043 & Ser-emb 11(E) & 0 & 0 & 4397 & $76.3^{+14.7}_{-15.4}$ & $82.0^{+16.9}_{-17.9}$ & $({\pm}73.2,63.4)$ & $(-49.5,-157.3)$ \\ [1ex]
Ser-emb 11(E) & Ser-emb 11(W) & 0 & 0 & 1002 & $55.0^{+25.3}_{-24.5}$ & $55.0^{+25.3}_{-24.5}$ & $(-49.5,-157.3)$ & $(-21.8,-61.3)$ \\ [1ex]
Ser-emb 11(W) & Ser-emb 17A & 0 & 0 & 4821 & $25.0^{+20.1}_{-19.7}$ & $25.0^{+20.1}_{-19.7}$ & $(-21.8,-61.3)$ & $(23.0,-172.1)$ \\ [1ex]
Ser-emb 17A & Ser-emb 17B & 0 & 0 & 5261 & $46.9^{+24.2}_{-25.2}$ & $89.7^{+19.0}_{-20.3}$ & $(23.0,-172.1)$ & $({\pm}68.4,153.5)$ \\ [1ex]
Ser-emb 6B & Ser-emb 6A & 0 & 0 & 863 & $43.3^{+14.3}_{-14.3}$ & $66.6^{+14.5}_{-13.5}$ & $({\pm}49.8,102.6)$ & $(29.4,-166.5)$ \\ [1ex]
Ser-emb 6C & Ser-emb 6B & 0 & 0 & 126 & $51.1^{+17.5}_{-17.5}$ & $54.7^{+18.4}_{-17.0}$ & $({\pm}21.0,17.9)$ & $({\pm}49.8,102.6)$ \\ [1ex]
Ser-emb 6D & Ser-emb 6A & 0 & 0 & 1066 & $29.1^{+22.9}_{-17.6}$ & $67.2^{+21.1}_{-24.5}$ & $({\pm}43.6,122.4)$ & $(29.4,-166.5)$ \\ [1ex]
Ser-emb 6E & Ser-emb 6D & 0 & 0 & 594 & $19.5^{+27.9}_{-26.5}$ & $80.7^{+24.0}_{-25.8}$ & $({\pm}40.3,151.3)$ & $({\pm}43.6,122.4)$ \\ [1ex]
SSTc2d J182957.8+011246 & eHOPS-aql-82 & I & I & 3681 & $25.7^{+23.0}_{-22.8}$ & $77.6^{+18.0}_{-19.1}$ & $({\pm}42.2,112.1)$ & $({\pm}62.4,91.6)$ \\ [1ex]
SSTgbs J1829386-015100 & SerpS-MM2 b & I & I & 2374 & $76.7^{+18.0}_{-19.2}$ & $64.4^{+17.7}_{-17.3}$ & $(54.1,-158.8)$ & $({\pm}80.1,44.3)$ \\ [1ex]
SSTgbs J1830024-020257 & SSTgbs J1830027-020259 & I & I & 2225 & $40.1^{+26.1}_{-25.3}$ & $86.8^{+19.9}_{-20.9}$ & $({\pm}44.3,101.4)$ & $({\pm}60.3,148.8)$ \\ [1ex]
SSTgbs J1830027-020259 & MIRES G028.6593+03.8185 & I & I & 2699 & $21.3^{+25.6}_{-26.7}$ & $42.8^{+22.1}_{-23.5}$ & $({\pm}60.3,148.8)$ & $({\pm}78.5,160.7)$ \\ [1ex]
Ser-emb 22 & Ser-emb 23 & I & I & 4261 & $15.9^{+14.1}_{-14.5}$ & $15.9^{+14.1}_{-14.5}$ & $(-60.2,-74.3)$ & $(-50.7,-58.7)$ \\ [1ex]
Ser-emb 23 & SSTc2d J182957.8+011246 & I & I & 2233 & $80.8^{+17.3}_{-16.2}$ & $38.9^{+18.0}_{-18.9}$ & $(-50.7,-58.7)$ & $({\pm}42.2,112.1)$ \\ [1ex]
SerpS-MM1 & SSTgbs J1829381-015100 & I & I & 3283 & $83.0^{+12.6}_{-13.0}$ & $12.1^{+13.0}_{-13.1}$ & $(-53.6,-64.9)$ & $({\pm}43.9,55.3)$ \\ [1ex]
SerpS-MM16 & SerpS10-mm & I & I & 3663 & $56.9^{+25.3}_{-23.4}$ & $41.4^{+23.7}_{-23.7}$ & $({\pm}29.2,126.8)$ & $({\pm}42.0,18.6)$ \\ [1ex]
SerpS-MM2 a & SSTgbs J1829381-015100 & I & I & 5948 & $50.5^{+11.0}_{-10.8}$ & $74.4^{+12.3}_{-12.0}$ & $({\pm}51.1,125.5)$ & $({\pm}43.9,55.3)$ \\ [1ex]
SerpS-MM2 b & SerpS-MM2 a & I & I & 864 & $76.9^{+13.1}_{-13.6}$ & $89.4^{+9.3}_{-9.4}$ & $({\pm}80.1,44.3)$ & $({\pm}51.1,125.5)$ \\ [1ex]
SerpS10-mm & SSTgbs J1830024-020257 & I & I & 1050 & $53.8^{+25.6}_{-24.9}$ & $61.8^{+22.0}_{-22.3}$ & $({\pm}42.0,18.6)$ & $({\pm}44.3,101.4)$ \\ [1ex]

\hline
\multicolumn{6}{p{0.5\linewidth}} {Ophiuchus and Ophiuchus North (Binary)}\\
\hline
IRAS 16293-2422B & IRAS 16293-2422A & 0 & 0 & 821 & $72.0^{+19.3}_{-20.5}$ & $72.0^{+19.3}_{-20.5}$ & $(22.8,-170.3)$ & $(54.3,-33.5)$ \\ [1ex]
ISO-Oph 31 & Oph-emb 8 & I & I & 2154 & $8.8^{+16.9}_{-16.1}$ & $8.8^{+16.9}_{-16.1}$ & $(67.7,-111.4)$ & $(64.8,-120.4)$ \\ [1ex]
Oph-emb 10b & Oph-emb 10a & I & I & 107 & $89.0^{+13.8}_{-14.6}$ & $31.6^{+16.5}_{-18.5}$ & $({\pm}51.1,134.1)$ & $(-48.4,-92.4)$ \\ [1ex]
CFHTWIR-Oph 43 & Oph-emb 23 & Flat & Flat & 615 & $53.4^{+12.9}_{-20.1}$ & $84.0^{+12.2}_{-7.4}$ & $({\pm}18.4,99.9)$ & $({\pm}68.3,133.0)$ \\ [1ex]
F-MM7 & Oph-emb 27 & Flat & Flat & 239 & $53.8^{+8.5}_{-8.3}$ & $74.2^{+9.0}_{-8.7}$ & $({\pm}76.9,61.4)$ & $({\pm}45.5,113.0)$ \\ [1ex]
ISO-Oph 2b & ISO-Oph 2a & II & II & 248 & $27.3^{+27.9}_{-24.8}$ & $57.4^{+27.6}_{-24.5}$ & $({\pm}31.7,58.8)$ & $({\pm}33.0,6.5)$ \\ [1ex]

\hline
\multicolumn{6}{p{0.5\linewidth}} {Ophiuchus and Ophiuchus North (Triple)}\\
\hline
eHOPS-oph-20b & eHOPS-oph-20a & I & I & 329 & $70.6^{+24.7}_{-25.0}$ & $77.5^{+20.3}_{-22.0}$ & $({\pm}72.8,148.3)$ & $({\pm}78.8,75.3)$ \\ [1ex]
eHOPS-oph-20c & eHOPS-oph-20b & I & I & 513 & $50.5^{+25.3}_{-25.2}$ & $58.4^{+22.6}_{-23.7}$ & $({\pm}79.0,96.5)$ & $({\pm}72.8,148.3)$ \\ [1ex]
Oph-emb 26a & Oph-emb 19 & Flat & Flat & 4924 &  &  & (NA, NA) & $({\pm}72.2,39.4)$ \\ [1ex]
Oph-emb 26b & Oph-emb 26a & Flat & Flat & 29 &  &  & (NA, NA) & (NA, NA) \\ [1ex]
\hline
\multicolumn{6}{p{0.5\linewidth}} {Ophiuchus and Ophiuchus North (Quadruple and above)}\\
\hline
VLA 1623Aa & VLA 1623Ab & 0 & 0 & 30 & $7.1^{+11.5}_{-12.0}$ & $7.1^{+11.5}_{-12.0}$ & $(-59.0,-48.0)$ & $(-56.5,-40.1)$ \\ [1ex]
VLA 1623Ab & VLA 1623B & 0 & 0 & 171 & $51.1^{+14.3}_{-14.1}$ & $51.1^{+14.3}_{-14.1}$ & $(-56.5,-40.1)$ & $(72.4,-41.5)$ \\ [1ex]
VLA 1623W & VLA 1623B & 0 & 0 & 1464 & $40.3^{+1.2}_{-1.2}$ & $40.3^{+1.2}_{-1.2}$ & $(-80.5,-11.3)$ & $(72.4,-41.5)$ \\ [1ex]
CFHTWIR-Oph 79 & Oph-emb 14 VLA 2 & I & I & 989 & $55.5^{+14.2}_{-14.0}$ & $3.9^{+7.8}_{-7.6}$ & $({\pm}61.6,127.9)$ & $(-63.0,-123.8)$ \\ [1ex]
Oph-emb 14 VLA 1 & Oph-emb 12 & I & I & 5331 & $5.3^{+25.7}_{-24.6}$ & $5.3^{+25.7}_{-24.6}$ & $(-43.3,-85.4)$ & $(-47.3,-80.5)$ \\ [1ex]
Oph-emb 14 VLA 2 & Oph-emb 14 VLA 1 & I & I & 83 & $35.9^{+21.9}_{-22.0}$ & $35.9^{+21.9}_{-22.0}$ & $(-63.0,-123.8)$ & $(-43.3,-85.4)$ \\ [1ex]

\hline
\multicolumn{6}{p{0.5\linewidth}} {Orion A and Orion B (Binary)}\\
\hline
HH270VLA1B & HH270VLA1A & 0 & 0 & 94 & $2.9^{+8.0}_{-7.7}$ & $45.7^{+13.7}_{-13.8}$ & $({\pm}66.4,141.2)$ & $({\pm}68.0,143.9)$ \\ [1ex]
HOPS-138B & HOPS-138A & 0 & 0 & 126 &  &  & (NA, NA) & $({\pm}55.2,78.6)$ \\ [1ex]
HOPS-168B & HOPS-168A & 0 & 0 & 77 & $35.9^{+12.0}_{-12.1}$ & $35.9^{+12.0}_{-12.1}$ & $(46.2,-127.1)$ & $(31.0,-72.8)$ \\ [1ex]
HOPS-288B & HOPS-288AA & 0 & 0 & 247 & $57.4^{+5.3}_{-5.3}$ & $57.4^{+5.3}_{-5.3}$ & $(63.0,-121.2)$ & $(-64.6,-145.6)$ \\ [1ex]
HOPS-28B & HOPS-28A & 0 & 0 & 132 & $19.6^{+27.7}_{-23.0}$ & $32.0^{+20.7}_{-19.3}$ & $({\pm}82.8,67.3)$ & $({\pm}68.0,54.0)$ \\ [1ex]
HOPS-290B & HOPS-290A & 0 & 0 & 328 & $62.0^{+10.3}_{-10.3}$ & $85.8^{+15.5}_{-17.7}$ & $({\pm}77.3,137.7)$ & $(-25.8,-8.0)$ \\ [1ex]
HOPS-312B & HOPS-312A & 0 & 0 & 132 & $17.7^{+16.9}_{-17.4}$ & $17.7^{+16.9}_{-17.4}$ & $(51.3,-139.9)$ & $(44.4,-117.8)$ \\ [1ex]
HOPS-317B & HOPS-317A & 0 & 0 & 2453 & $65.9^{+13.8}_{-13.5}$ & $31.1^{+11.1}_{-12.8}$ & $({\pm}45.4,92.2)$ & $(-25.8,-48.3)$ \\ [1ex]
HOPS-32B & HOPS-32A & 0 & 0 & 166 & $81.5^{+8.9}_{-9.5}$ & $18.6^{+9.8}_{-9.8}$ & $(-42.8,-68.5)$ & $({\pm}56.9,52.5)$ \\ [1ex]
HOPS-338B & HOPS-338A & 0 & 0 & 3033 & $24.6^{+23.6}_{-22.6}$ & $46.2^{+21.9}_{-20.8}$ & $({\pm}79.0,9.1)$ & $({\pm}55.2,15.6)$ \\ [1ex]
HOPS-341 & HOPS-340 & 0 & 0 & 2461 & $87.4^{+3.7}_{-3.7}$ & $87.4^{+3.7}_{-3.7}$ & $(53.1,-145.8)$ & $(-50.5,-99.7)$ \\ [1ex]
HOPS-358B & HOPS-358A & 0 & 0 & 5331 & $60.6^{+1.9}_{-1.9}$ & $90.0^{+1.2}_{-1.2}$ & $({\pm}21.4,127.0)$ & $(74.7,-81.2)$ \\ [1ex]
HOPS-373B & HOPS-373A & 0 & 0 & 1452 & $52.2^{+6.9}_{-6.9}$ & $3.8^{+4.2}_{-4.2}$ & $(-25.8,-144.3)$ & $({\pm}26.5,135.8)$ \\ [1ex]
HOPS-395B & HOPS-395A & 0 & 0 & 66 &  &  & (NA,79.9) & (NA,79.9) \\ [1ex]
HOPS-400B & HOPS-400A & 0 & 0 & 177 & $32.2^{+17.8}_{-17.4}$ & $27.8^{+15.9}_{-15.7}$ & $({\pm}21.3,18.5)$ & $(-21.8,-100.2)$ \\ [1ex]
HOPS-75B & HOPS-75A & 0 & 0 & 101 &  &  & (NA, NA) & $(-48.2,-156.1)$ \\ [1ex]
HOPS-377 & HOPS-144 & 0 & I & 3074 & $74.4^{+7.2}_{-7.0}$ & $74.4^{+7.2}_{-7.0}$ & $(-58.0,-145.2)$ & $(60.0,-101.7)$ \\ [1ex]
HH111MMSA & HH111MMSB & I & I & 1213 & $65.9^{+22.5}_{-27.0}$ & $40.4^{+21.0}_{-22.6}$ & $(72.0,-7.6)$ & $({\pm}55.2,146.2)$ \\ [1ex]
HH270mms1A & HH270mms1B & I & I & 889 & $65.5^{+14.9}_{-9.9}$ & $60.5^{+10.6}_{-16.4}$ & $(-58.8,-87.4)$ & $({\pm}29.0,2.7)$ \\ [1ex]
HOPS-145B & HOPS-145A & I & I & 4509 & $46.0^{+17.0}_{-18.0}$ & $84.9^{+11.2}_{-10.4}$ & $({\pm}24.6,44.6)$ & $({\pm}70.5,40.9)$ \\ [1ex]
HOPS-163B & HOPS-163A & I & I & 314 &  &  & (NA, NA) & $({\pm}69.2,60.4)$ \\ [1ex]
HOPS-193B & HOPS-193A & I & I & 104 & $87.0^{+25.1}_{-24.6}$ & $46.2^{+19.3}_{-19.8}$ & $({\pm}66.4,149.1)$ & $({\pm}36.9,16.5)$ \\ [1ex]
HOPS-261B & HOPS-261A & I & I & 309 & $11.3^{+22.0}_{-21.4}$ & $46.1^{+23.1}_{-23.2}$ & $({\pm}62.2,68.2)$ & $({\pm}72.1,62.3)$ \\ [1ex]
HOPS-282B & HOPS-282A & I & I & 66 &  &  & (NA, NA) & $({\pm}41.4,113.4)$ \\ [1ex]
HOPS-298B & HOPS-298A & I & I & 949 &  &  & (NA,15.4) & $({\pm}50.8,3.1)$ \\ [1ex]
HOPS-364B & HOPS-364A & I & I & 63 & $74.4^{+28.6}_{-29.7}$ & $74.4^{+28.6}_{-29.7}$ & $(-53.1,-8.4)$ & $(-53.1,-177.7)$ \\ [1ex]
HOPS-366B & HOPS-366A & I & I & 60 & $12.4^{+18.4}_{-18.3}$ & $71.9^{+16.5}_{-16.9}$ & $({\pm}60.0,62.6)$ & $(48.2,-67.2)$ \\ [1ex]
HOPS-84B & HOPS-84A & I & I & 280 & $8.1^{+14.7}_{-16.9}$ & $8.1^{+14.7}_{-16.9}$ & $(70.5,-166.2)$ & $(62.5,-167.5)$ \\ [1ex]
HOPS-263 & HOPS-262 & I & Flat & 2510 & $19.6^{+13.3}_{-13.8}$ & $58.9^{+16.7}_{-16.2}$ & $(33.6,-103.9)$ & $({\pm}29.0,66.7)$ \\ [1ex]
HOPS-125 & HOPS-124 & Flat & 0 & 3244 & $67.7^{+21.9}_{-14.2}$ & $32.4^{+21.9}_{-13.4}$ & $({\pm}31.0,133.4)$ & $(44.4,-3.3)$ \\ [1ex]
HOPS-118 & HOPS-117 & Flat & Flat & 5438 & $41.2^{+16.7}_{-18.2}$ & $76.5^{+14.4}_{-14.3}$ & $({\pm}53.1,73.7)$ & $({\pm}64.4,120.6)$ \\ [1ex]
HOPS-128B & HOPS-128A & Flat & Flat & 2578 & $23.6^{+12.6}_{-13.0}$ & $80.4^{+11.2}_{-11.3}$ & $({\pm}41.4,103.7)$ & $({\pm}60.0,84.7)$ \\ [1ex]
HOPS-149B & HOPS-149A & Flat & Flat & 1117 & $4.0^{+10.8}_{-10.9}$ & $46.9^{+15.1}_{-15.7}$ & $({\pm}23.6,39.6)$ & $({\pm}23.6,49.7)$ \\ [1ex]
HOPS-150B & HOPS-150A & Flat & Flat & 1288 & $46.8^{+21.7}_{-21.4}$ & $75.6^{+17.2}_{-17.0}$ & $({\pm}38.9,70.9)$ & $({\pm}53.1,135.2)$ \\ [1ex]
HOPS-158B & HOPS-158A & Flat & Flat & 47 & $32.7^{+25.6}_{-24.7}$ & $42.2^{+25.8}_{-26.3}$ & $({\pm}75.5,71.1)$ & $({\pm}78.5,37.6)$ \\ [1ex]
HOPS-170B & HOPS-170A & Flat & Flat & 657 &  &  & $({\pm}22.6,144.6)$ & (NA,167.0) \\ [1ex]
HOPS-213B & HOPS-213A & Flat & Flat & 101 &  &  & $(90.0,-84.9)$ & (NA, NA) \\ [1ex]
HOPS-242B & HOPS-242A & Flat & Flat & 115 &  &  & (NA, NA) & $({\pm}60.0,93.2)$ \\ [1ex]
HOPS-248B & HOPS-248A & Flat & Flat & 79 &  &  & (NA, NA) & $({\pm}41.4,35.7)$ \\ [1ex]
HOPS-255B & HOPS-255A & Flat & Flat & 75 &  &  & (NA, NA) & $({\pm}38.9,1.0)$ \\ [1ex]
HOPS-259B & HOPS-259A & Flat & Flat & 4499 &  &  & (NA, NA) & $(48.2,-66.7)$ \\ [1ex]
HOPS-274B & HOPS-274A & Flat & Flat & 84 &  &  & (NA, NA) & (NA, NA) \\ [1ex]
HOPS-281B & HOPS-281A & Flat & Flat & 301 &  &  & (NA, NA) & $(-27.3,-103.9)$ \\ [1ex]
HOPS-304B & HOPS-304A & Flat & Flat & 266 & $55.8^{+25.7}_{-24.2}$ & $50.4^{+23.6}_{-23.7}$ & $({\pm}41.4,133.2)$ & $(36.9,-37.7)$ \\ [1ex]
HOPS-357B & HOPS-357A & Flat & Flat & 110 & $8.4^{+26.4}_{-24.6}$ & $89.4^{+20.7}_{-20.4}$ & $({\pm}48.2,23.2)$ & $({\pm}41.4,30.2)$ \\ [1ex]
HOPS-363B & HOPS-363A & Flat & Flat & 72 &  &  & (NA, NA) & $({\pm}51.3,25.7)$ \\ [1ex]
HOPS-45B & HOPS-45A & Flat & Flat & 121 & $80.9^{+22.1}_{-25.9}$ & $50.0^{+17.8}_{-16.6}$ & $({\pm}36.9,146.8)$ & $(60.0,-29.0)$ \\ [1ex]
HOPS-57B & HOPS-57A & Flat & Flat & 470 & $26.6^{+17.3}_{-17.2}$ & $68.1^{+16.2}_{-17.4}$ & $({\pm}60.0,74.0)$ & $({\pm}58.6,43.0)$ \\ [1ex]
HOPS-59B & HOPS-59A & Flat & Flat & 961 &  &  & (NA, NA) & $({\pm}50.2,91.8)$ \\ [1ex]
HOPS-66B & HOPS-66A & Flat & Flat & 893 & $68.9^{+16.1}_{-13.6}$ & $39.7^{+15.9}_{-16.5}$ & $({\pm}45.1,165.8)$ & $(36.9,-47.0)$ \\ [1ex]
HOPS-85B & HOPS-85A & Flat & Flat & 43 & $55.4^{+22.5}_{-21.9}$ & $55.4^{+22.5}_{-21.9}$ & $(70.5,-66.0)$ & $(36.9,-123.9)$ \\ [1ex]

\hline
\multicolumn{6}{p{0.5\linewidth}} {Orion A and Orion B (Triple)}\\
\hline
HOPS-12BA & HOPS-12A & 0 & 0 & 1890 & $16.6^{+9.2}_{-9.6}$ & $16.6^{+9.2}_{-9.6}$ & $(38.9,-76.6)$ & $(33.6,-103.3)$ \\ [1ex]
HOPS-12BB & HOPS-12BA & 0 & 0 & 85 & $10.1^{+5.9}_{-5.9}$ & $10.1^{+5.9}_{-5.9}$ & $(48.2,-82.5)$ & $(38.9,-76.6)$ \\ [1ex]
HOPS-173B & HOPS-173A & 0 & 0 & 107 &  &  & (NA, NA) & $({\pm}56.9,48.1)$ \\ [1ex]
HOPS-87 & HOPS-86A & 0 & I & 4134 & $15.0^{+20.2}_{-19.1}$ & $15.0^{+20.2}_{-19.1}$ & $(13.5,-11.1)$ & $(28.1,-0.5)$ \\ [1ex]
HOPS-140B & HOPS-140A & I & I & 3878 & $86.8^{+10.6}_{-16.3}$ & $63.5^{+15.6}_{-9.7}$ & $({\pm}67.9,72.2)$ & $({\pm}48.2,178.6)$ \\ [1ex]
HOPS-140C & HOPS-140B & I & I & 897 & $62.9^{+16.3}_{-13.7}$ & $82.4^{+10.6}_{-9.1}$ & $({\pm}64.6,2.8)$ & $({\pm}67.9,72.2)$ \\ [1ex]
HOPS-86B & HOPS-86A & I & I & 3183 & $6.2^{+19.4}_{-19.1}$ & $50.6^{+22.6}_{-21.5}$ & $({\pm}22.6,7.2)$ & $(28.1,-0.5)$ \\ [1ex]
HOPS-174 & HOPS-173A & Flat & 0 & 2770 & $62.6^{+19.1}_{-19.1}$ & $55.8^{+21.9}_{-22.2}$ & $(-20.4,-128.1)$ & $({\pm}56.9,48.1)$ \\ [1ex]
HOPS-77AB & HOPS-77AA & Flat & Flat & 46 &  &  & (NA, NA) & (NA, NA) \\ [1ex]
HOPS-77B & HOPS-77AA & Flat & Flat & 643 &  &  & (NA, NA) & (NA, NA) \\ [1ex]
HOPS-92AB & HOPS-92AA & Flat & Flat & 110 &  &  & (NA, NA) & $(-24.6,-140.8)$ \\ [1ex]
HOPS-92B & HOPS-92AB & Flat & Flat & 457 &  &  & $(-45.1,-174.1)$ & (NA, NA) \\ [1ex]

\hline
\multicolumn{6}{p{0.5\linewidth}} {Orion A and Orion B (Quadruple and above)}\\
\hline
HOPS-182B & HOPS-182A & 0 & 0 & 395 & $53.9^{+5.0}_{-4.9}$ & $53.9^{+5.0}_{-4.9}$ & $(56.9,-79.4)$ & $(40.4,-150.8)$ \\ [1ex]
HOPS-182C & HOPS-182B & 0 & 0 & 2038 & $39.2^{+18.9}_{-20.2}$ & $81.3^{+15.2}_{-15.0}$ & $({\pm}55.2,127.1)$ & $(56.9,-79.4)$ \\ [1ex]
HOPS-203B & HOPS-203A & 0 & 0 & 51 &  &  & (NA, NA) & $(-67.1,-50.2)$ \\ [1ex]
HOPS-203C & HOPS-203B & 0 & 0 & 1109 &  &  & $({\pm}53.1,175.4)$ & (NA, NA) \\ [1ex]
HOPS-361B & HOPS-361A & 0 & 0 & 1050 & $25.2^{+24.3}_{-22.8}$ & $86.8^{+27.5}_{-30.7}$ & $({\pm}33.6,174.4)$ & $(-54.8,-14.4)$ \\ [1ex]
HOPS-361C & HOPS-361B & 0 & 0 & 1215 & $45.7^{+23.0}_{-22.6}$ & $88.6^{+13.7}_{-12.6}$ & $(66.2,-130.1)$ & $({\pm}33.6,174.4)$ \\ [1ex]
HOPS-361D & HOPS-361A & 0 & 0 & 3396 & $72.4^{+7.2}_{-7.2}$ & $24.0^{+4.9}_{-4.8}$ & $({\pm}58.0,43.1)$ & $(-54.8,-14.4)$ \\ [1ex]
HOPS-361E & HOPS-361A & 0 & 0 & 1140 & $60.3^{+21.1}_{-22.4}$ & $69.5^{+25.5}_{-27.0}$ & $({\pm}42.8,111.9)$ & $(-54.8,-14.4)$ \\ [1ex]
HOPS-361F & HOPS-361A & 0 & 0 & 1392 &  &  & (NA, NA) & $(-54.8,-14.4)$ \\ [1ex]
HOPS-361GA & HOPS-361B & 0 & 0 & 4197 &  &  & (NA, NA) & $({\pm}33.6,174.4)$ \\ [1ex]
HOPS-361GB & HOPS-361GA & 0 & 0 & 535 &  &  & (NA, NA) & (NA, NA) \\ [1ex]
HOPS-384B & HOPS-384A & 0 & 0 & 1369 & $37.0^{+2.7}_{-2.7}$ & $37.0^{+2.7}_{-2.7}$ & $(75.7,-34.9)$ & $(45.6,-60.2)$ \\ [1ex]
HOPS-384C & HOPS-384A & 0 & 0 & 3852 & $9.7^{+12.2}_{-12.2}$ & $81.5^{+12.4}_{-13.3}$ & $({\pm}36.0,62.6)$ & $(45.6,-60.2)$ \\ [1ex]
HOPS-389B & HOPS-389A & 0 & 0 & 2676 & $19.5^{+11.0}_{-11.1}$ & $76.0^{+11.8}_{-12.1}$ & $({\pm}60.0,33.4)$ & $({\pm}45.6,50.1)$ \\ [1ex]
HOPS-56AB & HOPS-56AA & 0 & 0 & 87 &  &  & (NA, NA) & $(44.4,-43.2)$ \\ [1ex]
HOPS-56AC & HOPS-56AB & 0 & 0 & 89 &  &  & (NA, NA) & (NA, NA) \\ [1ex]
HOPS-56B & HOPS-56AA & 0 & 0 & 2117 & $21.3^{+15.6}_{-17.5}$ & $76.9^{+13.7}_{-13.3}$ & $({\pm}60.4,61.1)$ & $(44.4,-43.2)$ \\ [1ex]
HOPS-78B & HOPS-78A & 0 & 0 & 1473 &  &  & (NA,179.4) & $(71.4,-171.1)$ \\ [1ex]
HOPS-78C & HOPS-78B & 0 & 0 & 564 &  &  & $({\pm}61.9,131.9)$ & (NA,179.4) \\ [1ex]
HOPS-78D & HOPS-78B & 0 & 0 & 1565 &  &  & $({\pm}78.5,175.4)$ & (NA,179.4) \\ [1ex]
HOPS-108 & HOPS-64 & 0 & I & 2450 &  &  & (NA,105.7) & $({\pm}45.6,119.3)$ \\ [1ex]
HOPS-182A & HOPS-181A & 0 & I & 4313 &  &  & $(40.4,-150.8)$ & (NA, NA) \\ [1ex]
HOPS-203A & HOPS-165 & 0 & I & 5130 & $64.4^{+1.7}_{-1.7}$ & $41.1^{+1.3}_{-1.3}$ & $(-67.1,-50.2)$ & $({\pm}65.6,95.2)$ \\ [1ex]
HOPS-389A & HOPS-323B & 0 & I & 3992 & $59.9^{+14.8}_{-14.5}$ & $29.0^{+13.3}_{-13.8}$ & $({\pm}45.6,50.1)$ & $(-74.5,-50.7)$ \\ [1ex]
HOPS-394B & HOPS-71B & 0 & I & 4139 &  &  & (NA, NA) & $(65.6,-82.4)$ \\ [1ex]
HOPS-384A & 2M05414483-0154357 & 0 & NA & 5852 & $27.4$ & $85.0$ & $(45.6,-60.2)$ & $({\pm}55.2,93.9)$ \\ [1ex]
HOPS-56AA & V2358Ori & 0 & NA & 2148 &  &  & $(44.4,-43.2)$ & (NA, NA) \\ [1ex]
HOPS-181B & HOPS-181A & I & I & 1516 &  &  & $({\pm}43.3,139.7)$ & (NA, NA) \\ [1ex]
HOPS-323B & HOPS-323A & I & I & 257 & $34.4^{+9.3}_{-9.4}$ & $34.4^{+9.3}_{-9.4}$ & $(-74.5,-50.7)$ & $(-44.4,-30.9)$ \\ [1ex]
HOPS-386B & HOPS-386A & I & I & 1070 & $11.8^{+7.4}_{-7.4}$ & $11.8^{+7.4}_{-7.4}$ & $(-46.5,-68.3)$ & $(-54.3,-56.8)$ \\ [1ex]
HOPS-386C & HOPS-386B & I & I & 3447 & $84.0^{+6.1}_{-6.2}$ & $51.6^{+7.3}_{-7.1}$ & $({\pm}68.0,125.6)$ & $(-46.5,-68.3)$ \\ [1ex]
HOPS-387A & HOPS-386B & I & I & 3565 & $71.5^{+9.3}_{-9.1}$ & $71.5^{+9.3}_{-9.1}$ & $(66.0,-42.9)$ & $(-46.5,-68.3)$ \\ [1ex]
HOPS-387B & HOPS-387A & I & I & 692 & $31.2^{+5.8}_{-5.8}$ & $68.5^{+7.7}_{-7.9}$ & $({\pm}53.1,76.2)$ & $(66.0,-42.9)$ \\ [1ex]
HOPS-71B & HOPS-71A & I & I & 404 & $37.3^{+11.4}_{-11.5}$ & $60.6^{+17.5}_{-16.7}$ & $(65.6,-82.4)$ & $({\pm}68.4,123.0)$ \\ [1ex]
HOPS-370 & 2MJ05352746-0509441 & I & II & 4008 &  &  & $(71.1,-109.7)$ & (NA, NA) \\ [1ex]
HOPS-64 & VLA16 & I & NA & 4733 & $30.6^{+18.4}_{-19.1}$ & $74.9^{+17.2}_{-18.0}$ & $({\pm}45.6,119.3)$ & $({\pm}36.9,164.8)$ \\ [1ex]
HOPS-70AB & HOPS-70AA & Flat & Flat & 128 &  &  & (NA, NA) & (NA,13.4) \\ [1ex]
HOPS-70BB & HOPS-70AB & Flat & Flat & 2073 &  &  & $({\pm}64.6,51.2)$ & (NA, NA) \\ [1ex]
HOPS-70C & HOPS-70AA & Flat & Flat & 2817 &  &  & $({\pm}65.7,78.3)$ & (NA,13.4) \\ [1ex]
HOPS-369 & OMC2-FIR4ALMA1 & Flat & NA & 3552 & $77.3^{+11.8}_{-11.8}$ & $76.3^{+19.1}_{-18.3}$ & $(-36.9,-21.3)$ & $({\pm}73.4,110.5)$ \\ [1ex]
HOPS-72 & HOPS-71B & NA & I & 4380 & $18.8^{+13.0}_{-13.1}$ & $60.3^{+20.0}_{-19.1}$ & $({\pm}56.9,101.5)$ & $(65.6,-82.4)$ \\ [1ex]
OMC2-FIR4ALMA1 & VLA16 & NA & NA & 1305 & $55.7^{+21.7}_{-21.1}$ & $83.9^{+13.5}_{-13.6}$ & $({\pm}73.4,110.5)$ & $({\pm}36.9,164.8)$ \\ [1ex]
VLA15 & VLA16 & NA & NA & 2487 & $70.1^{+14.9}_{-16.2}$ & $84.9^{+10.6}_{-9.4}$ & $({\pm}74.4,87.3)$ & $({\pm}36.9,164.8)$ \\ [1ex]
\hline
\multicolumn{6}{p{0.5\linewidth}} {Corona Australis (Binary)}\\
\hline
IRAS 32 B & IRAS 32 A & 0 & 0 & 231 & $43.9^{+2.4}_{-2.4}$ & $43.9^{+2.4}_{-2.4}$ & $(-68.9,-131.2)$ & $(67.4,-135.2)$ \\ [1ex]
V* VV CrA B & V* VV CrA A & I & I & 359 & $6.9^{+9.5}_{-9.6}$ & $6.9^{+9.5}_{-9.6}$ & $(11.4,-62.5)$ & $(17.8,-51.5)$ \\ [1ex]
V* S CrA A & V* S CrA B & II & II & 204 & $4.5^{+13.5}_{-13.4}$ & $4.5^{+13.5}_{-13.4}$ & $(-17.0,-168.1)$ & $(-21.4,-164.5)$ \\ [1ex]
\hline
\multicolumn{6}{p{0.5\linewidth}} {Corona Australis (Triple)}\\
\hline
IRS 5N & IRS 5b & 0 & Flat & 1370 & $18.9^{+11.1}_{-11.8}$ & $55.1^{+19.2}_{-18.6}$ & $({\pm}62.8,81.0)$ & $({\pm}65.8,60.3)$ \\ [1ex]
IRS 5b & IRS 5a & Flat & Flat & 150 & $44.5^{+26.3}_{-24.4}$ & $68.9^{+19.2}_{-19.6}$ & $({\pm}65.8,60.3)$ & $({\pm}64.5,109.7)$ \\ [1ex]
\hline
\multicolumn{6}{p{0.5\linewidth}} {Corona Australis (Quadruple and above)}\\
\hline
CXO 34 & IRS 7A & 0 & 0 & 1315 & $17.4^{+15.0}_{-15.6}$ & $68.3^{+16.4}_{-16.6}$ & $({\pm}63.7,59.8)$ & $(48.8,-70.7)$ \\ [1ex]
CrAus7-mm & SMM1C & 0 & 0 & 1076 & $65.0^{+6.9}_{-7.0}$ & $72.1^{+9.5}_{-8.9}$ & $({\pm}74.7,107.0)$ & $({\pm}77.4,174.2)$ \\ [1ex]
CrAus8-mm1 & IRS 7A & 0 & 0 & 3936 & $14.4^{+18.4}_{-18.1}$ & $68.1^{+18.3}_{-18.0}$ & $({\pm}63.1,72.7)$ & $(48.8,-70.7)$ \\ [1ex]
IRS7B-a & CrAus8-mm1 & 0 & 0 & 1834 & $38.6^{+16.4}_{-17.0}$ & $63.9^{+17.8}_{-18.8}$ & $({\pm}67.8,115.0)$ & $({\pm}63.1,72.7)$ \\ [1ex]
IRS7B-b & IRS7B-a & 0 & 0 & 112 & $4.3^{+2.5}_{-2.5}$ & $47.2^{+3.2}_{-3.1}$ & $({\pm}65.1,118.7)$ & $({\pm}67.8,115.0)$ \\ [1ex]
SMM1C & IRS 7A & 0 & 0 & 802 & $88.4^{+9.4}_{-9.7}$ & $71.6^{+13.6}_{-12.8}$ & $({\pm}77.4,174.2)$ & $(48.8,-70.7)$ \\ [1ex]
SMM 2 & IRS7B-a & I & 0 & 5605 & $66.2^{+2.0}_{-2.0}$ & $84.8^{+3.0}_{-2.9}$ & $(-49.1,-12.1)$ & $({\pm}67.8,115.0)$ \\ [1ex]
\hline
\multicolumn{6}{p{0.5\linewidth}} {Chameleon I (Binary)}\\
\hline
V* HO Cha b & V* HO Cha a & Flat & Flat & 155 & $5.2^{+9.8}_{-10.0}$ & $43.9^{+17.9}_{-18.0}$ & $({\pm}66.9,58.6)$ & $({\pm}69.5,63.4)$ \\ [1ex]
ISO-ChaI 204 & ChamI-9 mm & II & II & 5580 &  &  & $(50.8,-35.2)$ & (NA, NA) \\ [1ex]
\end{longtable}
\tablefoot{Protostellar disks in multiple systems in Serpens and Aquila, Ophiuchus and Ophiuchus North, Corona Australis, and Chameleon I molecular cloud were observed in our CAMPOS survey. See also the CAMPOS data paper \citep{2024ApJ...973..138H}. Protostellar disks in Orion A and Orion B were observed by the VANDAM survey \citep{2020ApJ...890..130T}.

\textdagger \; The RA/Dec reported for this source in \citet{2024ApJ...973..138H} are slightly offset. We adopt the corrected coordinates for Oph-emb 26a: RA = 16:27:30.164, Dec = -24:27:44.11. The deconvolution radius from CASA \emph{imfit} is unreliable for this source, and therefore the derived position angle and inclination are not robust.

\textdaggerdbl  \; The RA/Dec reported for this source in \citet{2024ApJ...973..138H} are slightly offset. We adopt the corrected coordinates for Oph-emb 26b: RA = 16:27:30.173, Dec = -24:27:43.96. The deconvolution radius from CASA \emph{imfit} is unreliable for this source, and therefore the derived position angle and inclination are not robust.

}

\end{landscape}

\twocolumn

\clearpage
\section{Spin alignment in \textsc{Starforge} including the closest stellar pairs}
\label{Appendix_B} 
Here we compare the alignments of close pairs (within 400 au projected) in \textsc{Starforge} simulations and observations. Here we include pairs with projected separations below 40 au, even though such close pairs are absent in observations.

\begin{figure}[tbh!]
\includegraphics[width=\columnwidth]{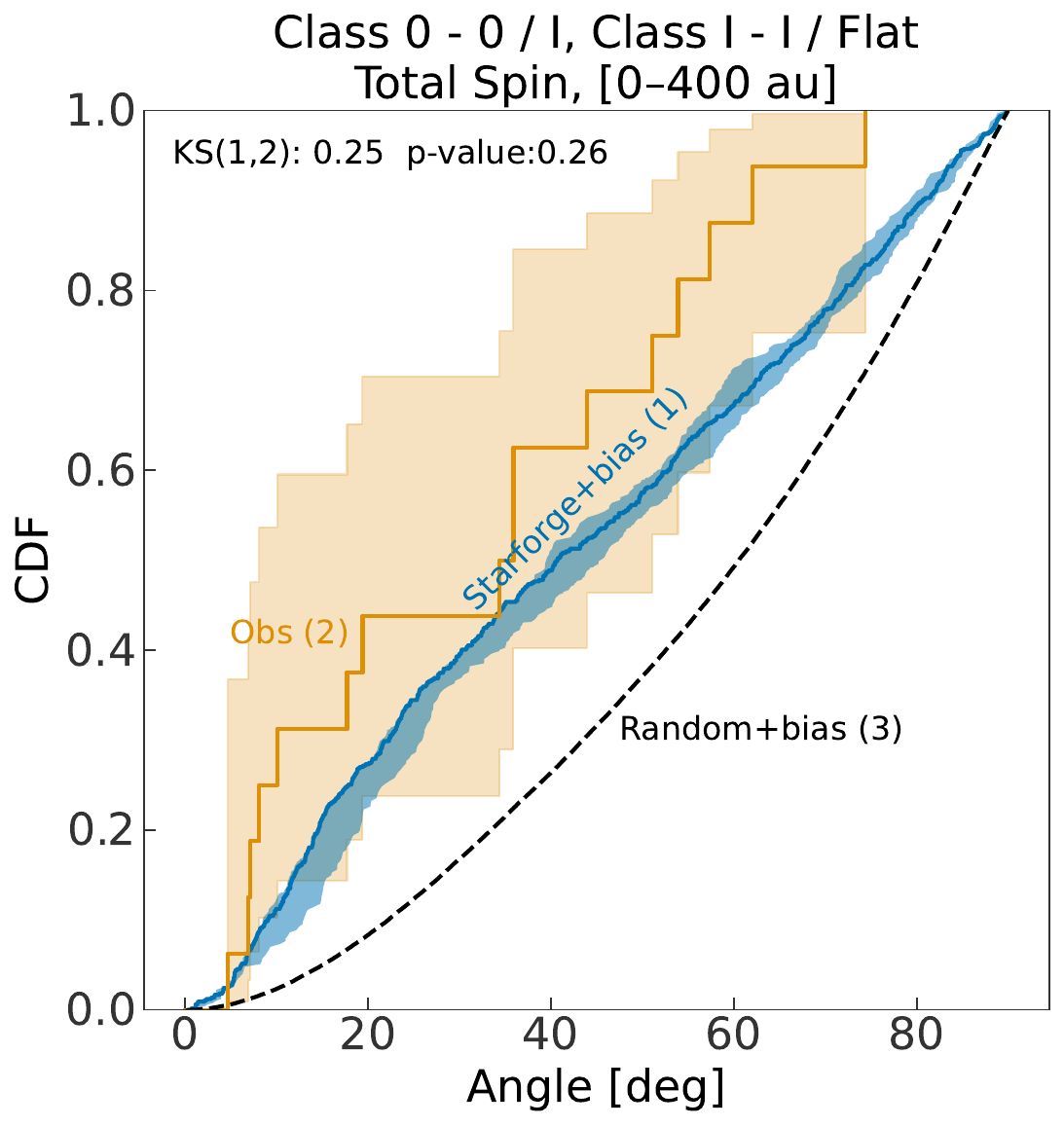}

\caption{Same as \autoref{fig:fidObs1DCut}, except including close pairs within a projected separation of 40 au from \textsc{Starforge}. These very close pairs have a strong alignment signal.} 
\label{fig:theoryNoCut}
\end{figure}

\end{appendix}

\end{document}